\documentclass[prx,twocolumn, showpacs, amsmath,amssymb,superscriptaddress,floatfix,reprint, nofootinbib]{revtex4-2}

\usepackage{graphicx}

\makeatletter
\def\l@subsubsection#1#2{}
\makeatother

\usepackage{macros}
\usepackage{braket}

\normalsize

\usepackage[pagebackref=false,colorlinks=true,linkcolor=blue,urlcolor=blue,filecolor=black,citecolor=blue,pdfstartview=FitV,pdftitle={},pdfauthor={},pdfsubject={},pdfkeywords={},pdfpagemode=None,bookmarksopen=true]{hyperref}
  
\pdfoutput=1 

\usepackage{tensor}
\usepackage{comment}

\usepackage{dsfont}
\usepackage{bbm}
\usepackage{tikz}
\usepackage{upgreek}

\usepackage{amsmath}
\usepackage{enumerate}
\usepackage{epsfig}
\usepackage{mathbbol}

\newcommand{\be}{\begin{equation}}
\newcommand{\ee}{\end{equation}}
\newcommand{\bea}{\begin{eqnarray}}
\newcommand{\eea}{\end{eqnarray}}
\newcommand{\bega}{\begin{gather}}
\newcommand{\eega}{\end{gather}}
\newcommand{\nn}{\nonumber\\}
\newcommand{\bi}{\begin{itemize}}
\newcommand{\ei}{\end{itemize}}
\newcommand{\ben}{\begin{enumerate}}
\newcommand{\een}{\end{enumerate}}
\newcommand{\bca}{\begin{cases}}
\newcommand{\eca}{\end{cases}}
\newcommand{\bln}{\begin{align}}
\newcommand{\eln}{\end{align}}
\newcommand{\bst}{\begin{split}}
\newcommand{\est}{\end{split}}
\def\ie{\begin{equation}\begin{aligned}}
\def\fe{\end{aligned}\end{equation}}
\newcommand{\bma}{\le(\begin{matrix}}
\newcommand{\ema}{\end{matrix}\ri)}
\newcommand{\bwt}{\begin{widetext}}
\newcommand{\ewt}{\end{widetext}}

\def\le{\left}
\def\ri{\right}

\newcommand\sC{{\ensuremath{{\mathcal C}}}}

\newcommand\sE{{\ensuremath{{\mathcal E}}}}

\newcommand\sH{{\ensuremath{{\mathcal H}}}}

\newcommand\sM{{\ensuremath{{\mathcal M}}}}
\newcommand\sN{{\ensuremath{{\mathcal N}}}}

\newcommand\sS{{\mathcal S}}

\newcommand\sZ{{\mathcal Z}}

\usepackage{amsthm}

\begin{document}

\title{Entanglement dynamics  from universal low-lying modes}

\author{Shreya Vardhan}
\email{vardhan@stanford.edu}
\affiliation{Stanford Institute for Theoretical Physics, Stanford University, Stanford, CA 94305, USA}

\author{Sanjay Moudgalya}
\email{sanjay.moudgalya@gmail.com}
\affiliation{School of Natural Sciences, Technische Universit\"{a}t M\"{u}nchen (TUM), James-Franck-Str. 1, 85748 Garching, Germany}
\affiliation{Munich Center for Quantum Science and Technology (MCQST), Schellingstr. 4, 80799 M\"{u}nchen, Germany}

\begin{abstract}
Information-theoretic quantities such as Renyi entropies show a remarkable universality in their late-time behaviour across a variety of chaotic many-body systems.
Understanding how such common features emerge from very different microscopic dynamics remains an important challenge.
In this work, we address this question in a class of Brownian models with random time-dependent Hamiltonians and a variety of different microscopic couplings.
In any such model, the Lorentzian time-evolution of the $n$-th Renyi entropy can be mapped to evolution by a Euclidean Hamiltonian on 2$n$ copies of the system.
We provide evidence that in systems with no symmetries, the low-energy excitations of the Euclidean Hamiltonian are universally given by a gapped quasiparticle-like band.
The eigenstates in this band are plane waves of locally dressed domain walls between ferromagnetic ground states associated with two permutations in the symmetric group $\sS_n$.
These excitations give rise to the membrane picture of entanglement growth, with the membrane tension determined by their  dispersion relation.
We establish this structure in a variety of cases using analytical perturbative methods and numerical variational techniques, and extract the associated dispersion relations and membrane tensions for the second and third Renyi entropies.
For the third Renyi entropy, we  argue that phase transitions in the membrane tension as a function of velocity are needed to ensure that physical constraints on the membrane tension are satisfied.
Overall, this structure provides an understanding of entanglement dynamics in terms of a universal set of gapped low-lying modes, which may also apply to systems with time-independent Hamiltonians.

\end{abstract}

\maketitle

\tableofcontents

\section{Introduction} 
\label{sec:intro}
Chaotic quantum many-body systems show the universal phenomenon of thermalization.
When an arbitrary initial state is evolved to sufficiently late times, it starts to macroscopically resemble a thermal density matrix $\rho^{\rm(eq)}$. 
This process is independent of most details of the initial state and microscopic dynamics of the system.
While thermalization is ubiquitously observed, much remains to be understood both about the mechanism for its robustness across a variety of different microscopic dynamics,  and about  the effective field-theoretic approaches which can capture its essential aspects.
Such an understanding would be valuable not only for quantum many-body physics, but also for understanding the process of black hole formation in quantum gravity, which is an example of thermalization~\cite{banks, trivedi}.    
Thermalization can be probed by using correlation functions of few-body operators in the time-evolved state, as well as information-theoretic quantities such as the Renyi entropies of a subsystem.
For an initial state $\rho_0$ and a unitary time-evolution operator $U$,  the time-evolved $n$-th Renyi entropy of a subsystem $R$ is given by~\footnote{The $n\to 1$ limit is  the von Neumann entropy.}
\begin{align}
&S_{n, R}(t) = - \frac{1}{(n-1)}\log\Tr[\rho_R(t)^n], \nn 
& \rho_R(t) =  \Tr_{\bar R}[U\rho_0 U^{\dagger}],  
\quad n \geq 2  \, . \label{ent}
\end{align}
At late times in chaotic systems, $S_{n, R}(t)$ saturates to a value that depends only on $\rho^{\rm (eq)}$, reflecting the fact that most details of the initial state are forgotten. 
For example, if $\rho_0$ is pure and one of the subsystems is much larger than the other,  we expect the general behaviour 
\be 
\lim_{t\to \infty} S_{n, R}(t) = \text{min} \le(S_{n, R}(\rho^{\rm (eq)}), S_{n, \bar R}(\rho^{\rm (eq)})  \ri) \, . \label{page}
\ee
The late-time value \eqref{page} is intuitively expected based on the behaviour of the Renyi entropies in random pure states~\cite{page, lubkin, pagels, nadal, leut}, and was argued for more systematically in~\cite{eq_approx}.
For the evolution of correlation functions during thermalization, it has long been understood that there are universal behaviours not only in the saturation value, but also in the way in which it is approached at late times.
For example, one expects the late-time behaviour of correlation functions of any conserved charge density to be governed by hydrodynamic modes, which depend on the conservation law but not on  details of the microscopic dynamics. 

A substantial amount of evidence has been gathered for a similar universality in the growth of $S_{n,R}(t)$ in chaotic systems before it approaches its late-time value \eqref{page}, starting with observations of a linear in $t$ regime in a variety of chaotic systems~\cite{kim_huse, liu_suh, hartman_maldacena}.
By synthesizing various observations, \cite{huse} conjectured a ``membrane formula'' to describe entanglement growth  in general chaotic quantum many-body systems.
This formula was found to hold in two very different examples of analytically tractable chaotic quantum many-body systems: random unitary circuits~\cite{vijay_randomcircuit, nahum_randomcircuit, rak_randomcircuit, zhou_nahum_statmech}, which involve a discrete chaotic evolution with random gates, and holographic conformal field theories~\cite{mark_membrane}.
While the result turns out to be the same, the formula is derived in these examples using techniques which are specific to each model. 
There have been attempts to understand the general origin of this universality in non-random chaotic Floquet systems~\cite{zhou_nahum}.
In this work, we identify the origin of the membrane picture across various examples in a large class of chaotic quantum many-body systems from a common set of gapped low-lying modes of an effective Hamiltonian on multiple copies of the system.
These modes resemble hydrodynamic modes that are commonly used to understand the late-time behavior of chaotic systems with conserved quantities, which, in certain kinds of systems, can also been understood as low-lying modes of appropriate effective Hamiltonians~\cite{mcculloch2023full, ogunnaike2023unifying, moudgalya2023symmetries}.
These hydrodynamic modes are universal in the sense that the ground states of the excitations are completely determined by the symmetries of the systems~\cite{moudgalya2023symmetries}, and hence the form of the low-energy excitations is fixed.
The gapped modes that govern the entanglement dynamics are also universal in the same sense, i.e., they are dressed domain-wall excitations on top of ferromagnetic ground states of the effective Hamiltonian, which are in turn completely specified by (the absence of) symmetries of the evolution.
We demonstrate the robustness of this physical picture in  random time-dependent ``Brownian"  Hamiltonians, which, unlike Haar random circuits,  have a variety of tunable parameters that allow us to probe a large class of models.
We provide a precise physical interpretation of the ``membrane tension'' function, the key ingredient of the membrane formula, in terms of the dispersion relation of these modes.
We show that the membrane picture for discrete-time random unitary circuits derived in~\cite{nahum_randomcircuit,rak_randomcircuit, huse, zhou_nahum_statmech, zhou_nahum} is a specific case where this structure applies, but in most of this work, the time-evolutions we consider are continuous. 
The entanglement membrane in such discrete circuits has been related to dressed domain-walls in effective statistical models that appear in their analysis~\cite{zhou_nahum_statmech, zhou_nahum}. 
Our analysis provides a simple physical picture for the appearance of such a dressed domain-wall structure, and relates it to the symmetries of the multiple-copy effective Hamiltonian.
The same set of modes that we find here can in principle be defined in systems with fixed time-independent Hamiltonians, and even in continuum quantum field theories such as holographic CFTs.
It is tempting to speculate that the same modes also govern the late-time evolution of the Renyi entropies in these contexts, and are entanglement analogs of hydrodynamic modes for correlation functions.
In the rest of the introduction, we first briefly review the membrane picture, and then summarize our methods and results. 
\subsection{Review of membrane picture}
\begin{figure}[t]
\includegraphics[width=0.5\textwidth]{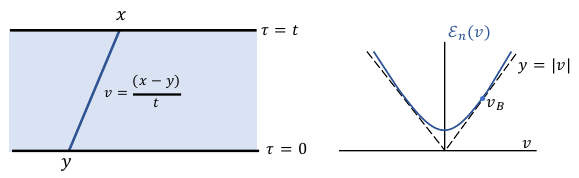}
\caption{Left: Example of a candidate line appearing in the minimization of \eqref{membrane_form}. Right: A cartoon of the membrane tension function. The first two constraints in \eqref{const} are equivalent to the fact the $\sE(v)$ is tangent to the $y=v$ line at some $v=v_B$.} 
\label{fig:formula}
\end{figure}

For simplicity, let us state the membrane formula in the case of one spatial dimension. Consider the  evolution of the Renyi entropy $S_{n}(x, t)$ of a pure or mixed state for the left half-line region ending at $x$.
According to the conjecture of \cite{huse}, this quantity can be expressed as the following minimization problem in any chaotic system.
Let us extend the one-dimensional system  to a two-dimensional slab, with an auxilliary time axis $\tau$ going from $\tau=0$ to $\tau=t$, as shown in Fig.~\ref{fig:formula}.
Then consider all possible lines with different velocities $v$ starting at $x$, and extending into the $\tau$ direction. 
At sufficiently late times, $S_{n}(x,t)$ is given by: 
\be 
S_{n}(x, t) = \min_{v}\left[ s_{n,{\rm eq}} ~ \sE_n(v) ~ t  + S_n(x - vt, t=0) \ri]  \label{membrane_form}
\ee
Here $s_{n, {\rm eq}}$ is the $n$-th Renyi entropy density of the equilibrium state.
The  function  $\sE_n(v)$  is model-dependent, but is conjectured to always be an even convex function, with a minimum at $v=0$.
It is expected to also universally satisfy the following constraints: 
\be 
\sE_n'(v_B) = 1 \, , \quad \sE_n(v_B)=v_B \,   ,  \quad \sE_n(v) \geq v ~ \text{ for all } ~ v \,    \label{const}
\ee
for some velocity $v_B$, which are necessitated by physical conditions we discuss below.
A natural generalization of the formula also holds for multiple intervals and higher dimensions, for details see~\cite{huse}.

To understand the physical consequences of this formula, it is useful to understand its prediction for an initial mixed state with volume law entropy with some coefficient $s$:
\be 
S_n(x, t=0) = s\times \le( x+ L/2\ri) , \quad 0 \leq s \leq s_{n,\rm eq}\, .  \label{initial_s}
\ee
where we have assumed that the system has total length $L$, with positions labelled from $-L/2$ to $L/2$. 
For such states, \eqref{membrane_form} predicts that 
\be 
S_n(x, t)=S_n(x, t=0) + s_{n, \rm eq}~ \Gamma_n(s)~ t \label{gamma_n} 
\ee
where the entropy growth rate $\Gamma_n(s)$ is related to $\sE_n(v)$ through 
\be 
\Gamma_n(s) = \text{min}_v \le(\sE_n(v) - \frac{vs}{s_{n, \rm eq}} \ri) \, . \label{gammadef}
\ee
The constraints \eqref{const} are known to be equivalent to~\cite{huse} the physical condition that the entropy of the initial equilibrium state should not grow, i.e.,
\be 
\Gamma_n(s_{n, \rm eq}) =0\, . \label{gamma_condition}
\ee
$v_B$ is the velocity that minimizes \eqref{gammadef} for $s=s_{n,{\rm eq}}$, and due to the fact that $s \leq s_{n, {\rm eq}}$, only velocities $v\leq v_B$ are physically relevant for the evolution of the entropy.

\subsection{Summary of results}
In this work, we will consider a family of ``Brownian'' time-evolutions in lattice systems, with independent random local Hamiltonians acting at each infinitesimal time-step.
Various specific examples of similar models have been introduced and studied in the literature over the years~\cite{lashkari, swingle, brownian_syk, sunderhauf2019quantum, bauer2017stochastic, bauer2019equilibrium, bernard2021entanglement, bernard2022dynamics,  nahum_freefermion,zhou2019operator, sunderhauf2019quantum, jian2021note, bao2021symmetry,  jian2022linear, leontica2023purification, moudgalya2023symmetries, ogunnaike2023unifying}. These models are less random than Haar-random unitary circuits, and have various tunable microscopic parameters. 
The key simplification of such models is that after averaging over randomness,  the Lorentzian evolution $(U \otimes U^{\ast})^n$ on $2n$ copies of the system, which governs the evolution of the $n$-th Renyi entropy,  can be replaced with a Euclidean evolution on $2n$ copies with a non-negative frustration-free local Hamiltonian $P_{2n}$:\footnote{In Haar random circuits, this average maps to a classical statistical mechanics model~\cite{nahum_randomcircuit, rak_randomcircuit}} 
\be 
\overline{(U \otimes U^{\ast})^{\otimes n}} = e^{- P_{2n}t} \, .  \label{p2n_intro}
\ee
We will explain the precise setup and derive this mapping from Lorentzian to Euclidean evolution in Sec. \ref{sec:setup}.
We will sometimes refer to $P_{2n}$ as the \textit{superhamiltonian} in the discussion below.
Due to the mapping in \eqref{p2n_intro}, the low-energy properties of the superhamiltonian determine the late-time evolution of quantities such as the $n$-th  Renyi entropy.
This allows us to use both physical intuition and precise analytical and numerical techniques from the low-energy physics of quantum many-body systems, and apply them to understanding the physics of thermalization.
Moreover, it turns out that the ground states of $P_{2n}$ can completely be understood in terms of the symmetries of evolutions, independent of the details of the couplings used in the Brownian evolutions.
This structure also allows us to conjecture general statements about the nature of low-energy excitations that control the late-time dynamics of relevant physical quantities.
For $n=1$, correlation functions of few-body operators in the thermal state or a time-evolved state can be written as transition amplitudes under the evolution operator~\eqref{p2n_intro}.
Some hints that the lessons we learn from Brownian models apply more generally come from Refs.~\cite{moudgalya2023symmetries, ogunnaike2023unifying}, which studied Brownian models with a variety of symmetries, and derived the associated hydrodynamic modes using the low-energy spectrum of $P_2$.
For example, in models with global $U(1)$ symmetry, these works used the low-energy gapless modes of $P_2$ to derive diffusive behaviour of two-point functions of the  charge density.
These low-energy modes are the natural excitations on top of the ground states of $P_2$ that are completely specified in terms of the symmetry algebra of $U(1)$ ~\cite{moudgalya2022from, moudgalya2023symmetries}.
The Renyi entropies can be written as transition amplitudes under \eqref{p2n_intro} for $n \geq 2$.
We first show that in general Brownian models with any symmetry, we can use the  ground states of $P_{2n}$ to derive a late-time saturation value of $S_{n,R}(t)$ consistent with the equilibrium approximation of \cite{eq_approx}, and in particular with \eqref{page}.
We then specialize to the case of Brownian models with no symmetries, where $P_{2n}$  has an $n!$-dimensional ground state subspace.
The ground state subspace is spanned by states associated with permutations in $\sS_n$, which have a product form between different sites $i$ of the system:
\be 
\otimes_i\ket{\sigma}_i, \quad \sigma \in \sS_n \, . 
\ee
The precise definition of $\ket{\sigma}$ will be given in Sec. \ref{sec:setup}.
We will provide more intuition for why states associated with permutations should be relevant for the late-time behaviour of the Renyi entropies at the end of the introduction.
In one spatial dimension, we find evidence that the low-energy excitations of $P_{2n}$ in models with no symmetries have the following universal structure.
Let us denote the identity permutation in $\sS_n$ by $e$, and the cyclic permutation $(n ~ n-1 ~ n-2  ... \, 1)$ which sends $n$ to $n-1$, $n-1$ to $n-2$, and so on, by $\eta$.
There exists a band of first excited eigenstates above the ground state that are well-approximated by plane waves of locally dressed domain walls between the states associated with $\eta$ and $e$. 
More explicitly, for $n = 2$ and $n = 3$,  we find that they can be well-approximated as
\be 
\ket{\psi_k} \approx  \sum_x e^{-ikx} \, \ket{\eta} ...  \ket{\eta}_x \ket{\phi}_{x+1, ..., x+\Delta} \ket{ e}_{x+\Delta+1} ... \ket{e} \label{eta_e_modes_mt}
\ee
where $\Delta$ is $O(1)$ in the thermodynamic limit, and $\ket{\phi}$ is an arbitrary state in the full Hilbert space on $2n$ copies of $\Delta$ sites from $x+1$ to $x + \Delta$.
For $n = 2$, we will show that the structure \eqref{eta_e_modes_mt} of the eigenstates in the thermodynamic limit leads to the membrane formula \eqref{membrane_form} for a half-line region, while for $n = 3$ we get the membrane formula along with some additional structures.
These eigenstates have a gapped dispersion relation $E(k)$, which determines the entanglement growth rate $\Gamma_n(s)$ of \eqref{gamma_n} through the relation 
\be 
\Gamma_n(s) = E(is)/s_{n, \rm eq} \, .
\ee
This in turn determines the membrane tension $\sE$ through the inverse of \eqref{gammadef}.
While we mostly focus on infinite systems and the entanglement about a single cut, the natural ``multiparticle''\footnote{Note that these effective ``particles'' which appear in the chaotic systems in this work have an entirely different structure from the quasiparticle picture of Calabrese and Cardy~\cite{calabrese}.
The latter  applies to integrable systems and gives very different results for the evolution of $S_{n, A}(t)$ for multiple intervals from the chaotic case~\cite{leichenauer, spread}.}  versions of these single domain wall excitations  give rise to the membrane picture for subsystems consisting of one or more intervals.\footnote{However, for entanglement in finite systems or finite intervals, one needs to also account for annihilation of domain walls at the boundary or with each other, which would modify the simple plane-wave nature of the excitations.
These processes are important for capturing the saturation of entanglement in such regions at late times.}
We establish the above universal structure of the low-energy eigenstates of $P_{2n}$ by studying the following cases: 
\begin{enumerate}
\item We start with the simplest case of a maximally random Brownian Hamiltonian, where the local coupling operators are drawn from the GUE ensemble and the local Hilbert space dimension $q$ is large.
For the second Renyi entropy in this case, the superhamiltonian  is analytically tractable, and allows us to explicitly see that the low-energy  eigenstates have the form  \eqref{eta_e_modes_mt}.
\item  Next, we  consider the second Renyi entropy in the same model at finite local Hilbert space dimension $q$.
Since the superhamiltonian is no longer analytically tractable, we use a version of the variational approach  used for extracting low-energy excitations of gapped Hamiltonians in~\cite{elementary_excitations, variational_ansatz, scattering_particles}.
This method allows us to both verify that the eigenstates are well-approximated by \eqref{eta_e_modes_mt}, and to extract their dispersion relation $E(k)$.
In this case, the on-site Hilbert space dimension of the superhamiltonian  is sufficiently small that we can also check $E(k)$ obtained from the variational method with results from exact diagonalization, finding good agreement. 
\item We then turn to the case of the higher Renyi entropies in the same model, in particular focusing on the third Renyi entropy. While we can no longer use exact diagonalization due to the large on-site Hilbert space dimension of the superhamiltonian, we again  use the variational approach to check that the low-energy eigenstates have the structure \eqref{eta_e_modes_mt}, and extract the associated $E(k)$.  
\item Finally, we consider the evolution of $S_2$ in a class of Brownian models where the coupling operators are fixed to be those of the mixed field Ising model, and only the coefficients appearing next to the operators have time-dependent randomness. For generic values of the coupling strength, the model is expected to be chaotic, except close to a special integrable (free-fermion) point.
Consistent with this expectation, we find good evidence for the structure \eqref{eta_e_modes_mt} using the variational approach in the general case, and a breakdown of this structure close to the integrable point. 
\end{enumerate}

In cases 1, 2, and 4 above, we find that the membrane tensions for the second Renyi entropy resulting from the dispersion relations $E(k)$ satisfy \eqref{const}, or equivalently $\Gamma_2(s)$ satisfies \eqref{gamma_condition}.
The case of the third Renyi entropy from point 3 turns out to be more subtle. The naive growth rate $\bar \Gamma_3(s)$ from the dispersion relation of modes \eqref{eta_e_modes_mt} appears to be non-zero at $s=s_{3, \rm eq}$, which is unphysical.
However, we conjecture that the evolution of $S_3$ also receives contributions from a second set of modes besides \eqref{eta_e_modes_mt} in this case. 
We argue that beyond some value of $s$, the naive growth rate $\bar\Gamma_3(s)$ should be replaced with the growth rate implied by this second set of modes, which is the same as $\Gamma_2(s)$.
In terms of the true membrane tension $\sE_3(v)$,  we find that this single first-order phase transition in $\Gamma$ leads to two phase transitions in terms of $\sE_3(v)$ at velocities $v_1^{\ast}<v_2^{\ast}<v_B$.
We have a first-order phase transition at $v=v_1^{\ast}$ (where $\sE'_3(v)$ is discontinuous) and a second-order transition at $v=v_2^{\ast}$ (where  $\sE_3''(v)$ is discontinuous but $\sE_3'(v)$ is not).
For $v<v_2^{\ast}$, $\sE_3(v)$ is smaller than $\sE_2(v)$, while for $v> v_2^{\ast}$, $\sE_3(v)=\sE_2(v)$.

\begin{figure}[t]
\includegraphics[width=0.4\textwidth]{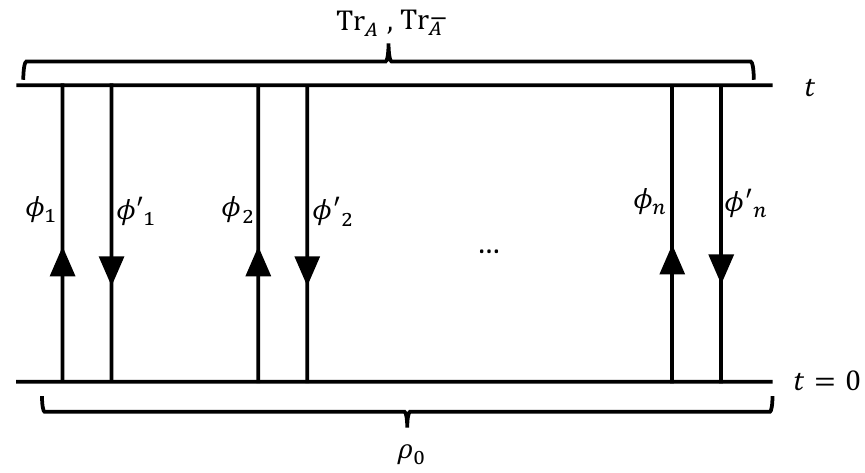}
\caption{$\Tr[\rho_A(t)^n]$ can be represented as a Lorentzian path integral on $2n$ copies of the theory, with $n$ forward and $n$ backward evolutions. The initial conditions of the path integral are determined by $\rho_0$, and the final conditions are determined by the pattern of traces in $\Tr[\rho_A(t)^n]$.}
\label{fig:pathint}
\end{figure}

While the above discussion of $\sE_3(v)$ uses an assumption about the existence of the second set of modes which should be more carefully checked in future work, it allows us to propose a form of $\sE_3(v)$ at finite $q$ and general $v$. 
Hence, we are able to provide a characterization of the phase transitions of $\sE_3(v)$ in a more general regime than previous discussions in random unitary circuits~\cite{zhou_nahum_statmech}, where evidence for a phase transition was found using expansions for large $q$ and large $v$.
Our physical picture for the origin of the phase transition appears to be similar to the one in~\cite{zhou_nahum_statmech}.  
The structure of the modes \eqref{eta_e_modes_mt} can be seen as a simple and precise realization of an insight from~\cite{zhou_nahum} about the crucial role played by permutations in the late-time evolution of the Renyi entropies.
Note that the Lorentzian path integral representation of the quantity $\Tr[\rho_A(t)^n]$, shown schematically in Fig. \ref{fig:pathint}, involves an integrand of the form $e^{i\sum_{j=1}^n  S[\phi_j]- i \sum_{j=1}^n  S[\phi'_j] }$, where $\phi_i, \phi_i'$ represent the dynamical fields of the theory along forward and backward contours respectively.
Ref.~\cite{zhou_nahum}
noted that we get stationary contributions in this path integral from configurations where each $\phi_j$ is equal to some $\phi'_{\sigma(j)}$ for some  $\sigma \in \sS_n$, as the phase in the exponent cancels. From other configurations, in a chaotic system, we should expect rapidly oscillating contributions that cancel among themselves.
Based on this observation,  \cite{eq_approx} developed a systematic approximation for the saturation value of $S_{n, A}(t)$.

At late times before saturation, it is natural to expect that the dominant configurations should be such that each $\phi_i$ is {\it locally} equal to some $\phi'_{\sigma(j)}$ for some $\sigma \in \sS_n$, but different permutations can appear in different regions.
\cite{zhou_nahum} developed a self-consistent numerical scheme based on this idea for evaluating the membrane tension in circuit and Floquet models, including non-random models. 
The scheme involved a sum over space-time diagrams for the propagation of entanglement domain walls, where the contributions from diagrams with large spacetime regions with states orthogonal to the permutation subspace were neglected.
In the random models considered in this work, we can better understand the suppression of such diagrams due to the high energy of the associated configurations in the Euclidean superhamiltonian.
The analog of the summation over diagrams from~\cite{zhou_nahum} is automatically performed by the low-energy dispersion relation of the superhamiltonian.
The structure of low-lying modes in \eqref{eta_e_modes_mt} provides a natural language for generalization to continuum systems, as we discuss further in the final section.   
The plan of this paper is as follows. We introduce the family of models we study and derive the mapping \eqref{p2n_intro} from the Lorentzian to the Euclidean time evolution in Sec.~\ref{sec:setup}.
We discuss the structure of the ground states and derive the equilibrium approximation for these models in Sec.~\ref{sec:eq}.
We then provide a detailed analysis of both the second and third Renyi entropy in the Brownian local GUE model in Sec.~\ref{sec:gue}.
In Sec.~\ref{sec:fixed_op}, we discuss the robustness of the same structure in more general Brownian models, and provide numerical results in a Brownian version of the mixed field Ising model.  

The results up to this point are all for one spatial dimension. For the more challenging case of higher dimensions, we derive the membrane formula in a large $q$, small $v$ limit of the local GUE model in Sec.~\ref{sec:eom}, using a different approach from the one-dimensional case.  We end with a number of open questions in Sec.~\ref{sec:discussion}. Various technical details as well as a few conceptual points are discussed in the appendices.

\section{Setup} \label{sec:setup} 
\begin{figure}[!h]
\includegraphics[width=0.4\textwidth]{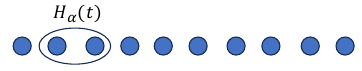}
\caption{We consider a family of random time-dependent Hamiltonians, where individual terms have a spatially local structure.}
\label{fig:structure}
 \end{figure}

In this work, we will consider a class of lattice models one or more spatial  dimensions with ``Brownian'' time-dependent Hamiltonians.
We label one copy of the full Hilbert space $\sH$.
The Hamitonians consist of a sum of local random terms $\{H_{\alpha}(t)\}$ (shown in Fig.~\ref{fig:structure}) which are uncorrelated for different $\alpha$ and $t$:
\begin{align} 
H(t) = \sum_{\alpha} H_{\alpha}(t), \quad \overline{H_{\alpha}(t)}=0, \\
\overline{H_{\alpha}(t)_{ij} H_{\beta}(t')_{kl}} \propto \delta_{\alpha \beta} \delta(t-t')  \label{halpha_gen}
\end{align}
We can analyze the dynamics under this setup by formally discretizing the time-evolution in small steps of size $\epsilon$, and regularizing the delta function between different times by replacing it with $\frac{1}{\epsilon} \delta_{tt'}$, so that 
\be 
U(t) = \prod_{j = 1}^{t/\epsilon} e^{- i\epsilon \, H(t_j) }, \quad t_j = j \epsilon \, . \label{u_op} 
\ee

One simple choice, which we will discuss in Sec. \ref{sec:gue}, will be to take the matrices $\{H_{\alpha}(t)\}$ themselves to be random.
A less random class of models is one where we fix  some set of local Hermitian operators $\{B_{\alpha}\}$, and take the coefficients appearing next to them to be random:  
\be 
H(t) = \sum_{\alpha} J_{\alpha}(t) B_{\alpha}, ~  \label{htdef}
\ee
where $\{J_{\alpha}(t)\}$ are random i.i.d. real numbers drawn from a Gaussian distribution, such that 
\be 
\overline{J_{\alpha}(t)}=0, \quad \overline{J_{\alpha}(t) J_{\beta}(t')} = \frac{2\,  g_{\alpha} \,  \delta_{\alpha \beta}\, \delta_{tt'}}{\epsilon} \label{jab_av}
\ee
for some arbitrary positive numbers $g_{\alpha}$. 
We can make a variety of choices of $\{B_{\alpha}\}$, where cases with different symmetries will correspond to different dynamical universality classes~\cite{moudgalya2023symmetries}.
For example, in a spin-1/2 system in $d$ spatial dimensions with sites labelled by $i, j$, we could consider a case where  $\{B_{\alpha}\}= \{X_i , Z_i, Z_i Z_{j}\}$ for nearest neighbours $i, j$.
In this case, the time-evolution does not have any symmetry. 
Another choice is to take $\{B_{\alpha}\} = \{Z_j, Z_i Z_{j}, X_i X_{j} + Y_iY_{j}\}$.
 These operators commute with the total charge $\sum_i Z_i$, so that the time-evolution has  a $U(1)$ symmetry.~\footnote{We can  see the symmetries in each case by computing the \textit{commutant} of the operators $\{B_{\alpha}\}$~(i.e., the algebra of operators that commute with these terms)~\cite{moudgalya2022from, moudgalya2023symmetries}.}

\begin{figure}[!ht]
    \centering
    \includegraphics[width=0.45\textwidth]{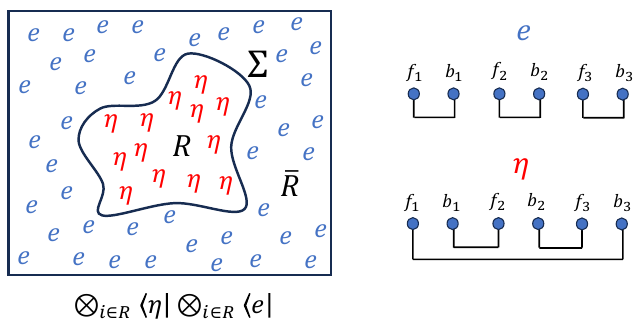}
    \caption{Left: Example of a domain wall final state relevant for the evolution of the $n$-th Renyi entropy for region $R$ in two spatial dimensions. Right: We show the pattern of entanglement between different copies in the states $\ket{e}$ and $\ket{\eta}$ for $n=3$.}
    \label{fig:domain_wall}
\end{figure}

The dynamical quantities of interest in this work are the $n$-th  Renyi entropies $S_{n, R}(t)$ of a subsystem $R$.
These  can be expressed as transition amplitudes on $2n$ copies of the system $\sH^{\otimes 2n}$,
\begin{align}
&e^{-(n-1)S_{n, R}(t)} = \Tr[\rho_R(t)^n] \nn
&= 
\braket{\eta_R \otimes e_{\bar R}|(U(t) \otimes U(t)^{\ast})^{\otimes n}| \rho_0, e} ~\label{renyi_ev}
\end{align}
where $\ast$ denotes complex conjugation, and in the bra and the ket we have introduced compact notations for certain states in $\sH^{\otimes 2n}$.
These states are associated with an operator $O$ acting on $\sH$, and permutations $\sigma \in \sS_{n}$, which are defined as follows.
Let $\{\ket{i}\}$ be the basis states for one copy of the Hilbert space, i.e., $\sH$.
We define 
\be 
\braket{i_1 i_1' ... i_n i_n'|O, \sigma} = O_{i_1 i_{\sigma(1)}'} ... O_{i_n i_{\sigma(n)}'} \, . \label{o15}
\ee
In cases where $O$ is the identity operator, it will be convenient to label the corresponding states simply by the permutation, that is, 
\be 
\braket{i_1 i_1' ... i_n i_n'| \sigma} = \delta_{i_1 i_{\sigma(1)}'} ... \delta_{i_n i_{\sigma(n)}'}. \label{eq:sigmadefn}
\ee
In \eqref{renyi_ev}, $e$ refers to the identity permutation, and $\eta$ refers to the single-cycle permutation $(n~ n-1 ~...1)$.
Further, we can also consider such states $\ket{\sigma_R}$ with a fixed permutation on the Hilbert space of some subsystem $R$; in lattice systems with sites labelled by $i$, we have $\ket{\sigma_R} = \otimes_{i \in R} \ket{\sigma}_i$.
The final state in the bra in \eqref{renyi_ev} therefore has a domain wall at the boundary $\Sigma$ between the regions, see Fig. \ref{fig:domain_wall}.
From \eqref{renyi_ev}, the evolution of the Renyi entropy of any initial state can  be understood in terms of the backward time-evolution of this domain wall final state under $(U \otimes U^{\ast})^{\otimes n}$.
In the rest of this work, we will make use of this perspective, which appears often  in the literature on Haar-random circuits, e.g.  \cite{nahum_randomcircuit, rak_randomcircuit}, 
Let us label the $2n$ copies of the system in \eqref{renyi_ev} by $f_i$, $b_i$ for $i = 1, ..., n$, corresponding respectively to the $n$ forward evolutions by $H(t)$ and $n$ backward evolutions by $H(t)^T$.
On expanding \eqref{u_op} for a step of size $\epsilon$ and using the averages in \eqref{jab_av}, we find  
\begin{align} 
&\overline{\le(U(\epsilon)\otimes U(\epsilon)^{\ast}\ri)^{\otimes n}} \approx 1 - \epsilon P_{2n} + O(\epsilon^2)\label{p2ndef}\\
&P_{2n} = \sum_{\alpha} \,  J_{\alpha} \, P_{2n, \alpha}, \, \;\; P_{2n, \alpha} = [\sum_{j=1}^n (B_{\alpha, f_j} - 
 B_{\alpha, b_j}^T)]^2 \, \nonumber. 
\end{align}
By re-exponentiating \eqref{p2ndef}, we find   
\be 
\overline{(U(t) \otimes U(t)^{\ast})^{\otimes n}}  = e^{- P_{2n} t} \, .  \label{euclidean}
\ee
In particular,  the average over the random $J_{\alpha}(t)$ allows us to replace the original Lorentzian time-evolution on $2n$ copies of the system with a Euclidean evolution, with a non-negative ``superhamiltonian'' $P_{2n}$.
From \eqref{renyi_ev}, the expression for the randomness-averaged Renyi entropy then reads
\begin{equation}
\overline{e^{-(n-1)S_{n, R}(t)}} = 
\braket{\eta_R \otimes e_{\bar R}|e^{-P_{2n}}| \rho_0, e} 
\label{eq:renyi_avg}
\end{equation}
Note that throughout this work, we assume that 
\begin{equation}
    \overline{S_n(x, t)} \approx -\log(\overline{e^{-S_n(x, t)}}),
\label{eq:assumption}
\end{equation}
hence we will use Eq.~(\ref{eq:renyi_avg}) to evaluate the randomness averaged $S_n(x, t)$.
Various earlier works~\cite{rak_randomcircuit,nahum_randomcircuit} have showed numerically that that the difference between the two quantities in negligible in similar random unitary circuit models.
Analytic justifications for this approximation in other similar models are presented in \cite{zhou_nahum_statmech, zhou_nahum, fisher2023random}.
We will henceforth also be suppressing the use of $\overline{\cdots}$, and all the quantities we compute are randomness averaged with the assumption of Eq.~(\ref{eq:assumption}).

We will first briefly discuss the structure of the zero energy ground states of $P_{2n}$ in Sec.~\ref{sec:eq}, and then discuss the structure of its low energy eigenstates in models without conserved quantities in the later sections. 

\section{Late-time saturation value} \label{sec:eq} 
The superhamiltonian \eqref{p2ndef} consists of a sum of positive semidefinite operators, so that its eigenvalues are all non-negative.  
Any ground state must be \textit{frustration-free}, meaning that it is annihilated by each term: 
\be 
P_{2n}\ket{\psi}=0  \quad \Leftrightarrow \quad P_{2n, \alpha}\ket{\psi}= 0 \quad \text{for each} \quad \alpha \, .  
\ee
One explicit set of zero energy eigenstates of $P_{2n}$ can be constructed as follows.
Let $\sC$ be the set of all operators which commute with all elements of $\{B_{\alpha}\}$, which is also known as its commutant algebra~\cite{moudgalya2022from}, and characterizes the symmetries of the time-evolution.
Let $\{Q_m\}$ be an orthonormal basis of operators for $\sC$,  
i.e., they satisfy $\Tr[Q_m^{\dagger} Q_{m'}] = \delta_{m m'}$.
Now for any choice of a sequence $(m_1 , ..., m_n)$ and a permutation $\sigma \in \sS_{n}$, let us define a state   $\ket{m_1, m_2, ..., m_n;~ \sigma}$ in $\sH^{\otimes 2n}$: 
\begin{align} 
 &\braket{i_1 i'_1 ... i_n i'_n \, |~ m_1, m_2, ..., m_n;~ \sigma}\nn 
 &= (Q_{m_1})_{i_1 i'_{\sigma(1)}}  (Q_{m_2})_{i_2 i'_{\sigma(2)}}... (Q_{m_n})_{i_n i'_{\sigma(n)}} \label{basis_2n}
\end{align}
Here each $i_p, i'_{p}$ labels basis states in $\sH$ like in \eqref{o15}, and runs from 1 to $D$, the total Hilbert space dimension, and each $m_k$ goes from 1 to $d_{\sC}$, the dimension of $\sC$, and labels an element of $\sC$. 
Since each $Q_m$ commutes with all $\{B_{\alpha}\}$, the above states are  zero energy eigenstates of $P_{2n}$.
Assuming that these states span the ground state subspace of $P_{2n}$,\footnote{This is true if this ensemble of evolutions forms an exact $n$-design in the $\to \infty$ limit~~\cite{harrow2009random, brandao2016local, hunterjones2019unitary}, which should be the case for generic choices of $B_{\alpha}$.
This is easily provable in many cases for $n = 1$~\cite{moudgalya2023symmetries} but there are simple exceptions for $n \geq 2$, e.g., when $\{B_\alpha\}$ can be written as quadratic operators in fermions~\cite{bao2021symmetry, nahum_freefermion, lastres2024nonuniversality}, which do not form $2$-designs.
We also discuss this in Section \ref{sec:fixed_op}.}~we have the following late-time limit of the Euclidean time-evolution operator: 
\begin{align}
&\lim_{t \to \infty} e^{- P_{2n}t}  \nn  &\approx\sum_{\substack{m_1, ..., m_n\, ;\\ \, \sigma \in \sS_n}}\ket{m_1, m_2, ..., m_n;~ \sigma} \bra{m_1, m_2, ..., m_n;~ \sigma } \label{late_t_proj}
\end{align}
The above expression  assumes that the states in \eqref{basis_2n} can be treated as approximately orthonormal, which is true for the purpose of the expression \eqref{renyi_ev} when the initial state $\rho_0$ can access a large effective Hilbert space dimension.\footnote{See the example in Appendix \ref{app:eq} for a more explicit discussion of this point.}
More generally, \eqref{late_t_proj} would also involve various Weingarten functions of the permutations on $n$ copies, see e.g. \cite{hearth2023unitary}.
The main idea, however, is that the ground state of the superhamiltonian, and hence the associated late-time saturation values, can be determined by the symmetries of the evolution.
Putting this projector into \eqref{eq:renyi_avg}, in the case where the initial state  $\rho_0$ is pure, we find
\begin{align} \label{eqa_mt} 
\lim_{t\to\infty}e^{-(n-1)S_{n,R}(t)} = \sum_{\sigma \in \sS_n}\braket{\eta_R \otimes e_{\bar R}| \rho^{\rm (eq)}, \sigma } 
\end{align}
where 
\be \label{rho_eq_mt}
\rho^{\rm (eq)} = \sum_m \Tr[Q_m^{\dagger}\rho_0] Q_m \, . 
\ee
$\ket{\rho^{(\rm eq)}, \sigma}$ is defined as in \eqref{o15}.
See Appendix \ref{app:eq} for details of the derivation.  
It is natural to think of $\rho^{\rm (eq)}$ defined above as an equilibrium density matrix which coarse-grains over all details of $\rho_0$ other than the information about the conserved charges.
We discuss an explicit example in Appendix \ref{app:eq} for the case where the time-evolution has a $U(1)$ symmetry,  which makes this interpretation clearer.   
It was previously argued in the context of general chaotic quantum many-body systems in \cite{eq_approx} that the expression \eqref{eqa_mt} gives the saturation value of the $n$-th Renyi entropy in an equilibrated pure state which macroscopically resembles some equilibrium state $\rho^{\rm (eq)}$. The Brownian models we consider in this work provide one explicit confirmation of this general argument, with a precise form of $\rho^{\rm (eq)}$ given by \eqref{rho_eq_mt}. The general properties of the expression \eqref{eqa_mt} are discussed in \cite{eq_approx}. In particular, the sum over permutations ensures that the $n$-th Renyi entropy in $R$ is equal to that in $\bar R$ at late times, as required by the unitarity of the dynamics. In the thermodynamic limit, it is explained in \cite{eq_approx} that the dominant permutation in \eqref{eqa_mt} is always either $\sigma=e$ or $\sigma = \eta$, leading to the physically expected result in \eqref{page}.
\section{Brownian local GUE model}
\label{sec:gue}
Let us consider a $d$-dimensional lattice, with a $q$-dimensional Hilbert space at each site.
As a first simple model, we take the $\{H_{\alpha}(t)\}$ to be random Hermitian $q^2 \times q^2$ matrices acting on pairs of nearest neighbours $i, j$ on a lattice, and drawn from the GUE ensemble, so that\footnote{For  recent discussions of the spectrum of this time-dependent Hamiltonian in the case without spatial locality, see \cite{guo2024complexity, haifeng}.}  
\be  \label{gue}
\overline{(H_{i, j}(t))_{\alpha \beta}} = 0, \quad  \overline{(H_{i, j}(t))_{\alpha \beta} (H_{i, j}(t))_{\delta\gamma}} = \frac{1}{2\epsilon q^2} \delta_{\alpha \gamma} \delta_{\beta \delta}
\ee
By similar steps to the discussion around \eqref{p2ndef}, for this model we obtain the following superhamiltonian on $\sH^{\otimes n}$ (again labelling copies with forward evolution $f_i$ and those with backward evolution $b_i$, with $i = 1, ..., n$): \begin{align} 
&P_{2n} = \sum_{\braket{ij}}P_{2n, ij}, \nn 
&P_{2n, ij} = \frac{1}{2} \bigg[n \, I -  \sum_{k, l=1}^n  M^i_{f_k, b_l} M^j_{f_k, b_l}   \nn 
& \quad \quad \quad \quad \quad + \frac{1}{q^2}\sum_{1\leq k<l\leq n} (S^i_{f_k,f_l} S^j_{f_k,f_l} + S^i_{b_k, b_l} S^j_{b_k b_l} )\bigg] \label{p2n_gue}
\end{align}
where $I$ is the identity operator in $\sH^{\otimes 2n}$,  $M_{r,s}^i$ is the projector onto the maximally entangled state  between the copies $r$ and $s$ at site $i$, defined as
\begin{equation}
\ket{\rm MAX}_{i_r, i_s} = \frac{1}{\sqrt{q}} \sum_{a=1}^q\ket{a}_{i_r}\ket{a}_{i_s},
\label{eq:MAXdefn}
\end{equation}
and $S_{r,s}^i$ is the swap operator between copies $r$ and $s$ at site $i$, which has the action $S^i_{rs} \ket{a}_{i_r} \ket{b}_{i_s} =\ket{b}_{i_r} \ket{a}_{i_s}$.
$P_{2n}$ has a large symmetry group which includes  $\sS_n \times \sS_n$ corresponding to permuting the forward and backward copies independently; see \cite{bao2021symmetry} for a detailed symmetry analysis of Hamiltonians of this kind.

For $n=1$, this superhamiltonian has a unique ground state given by $\otimes_i \ket{e}_i$, where $\ket{e}$ is two-copy state associated with the identity permutation, see Eq.~(\ref{eq:sigmadefn}).
Alternately, this can just be understood as the vectorization of the identity operator, which is the only ``symmetry" of the evolution on the doubled Hilbert space~\cite{moudgalya2023symmetries}.
Moreover, it is easy to check that it is composed of commuting terms, and is therefore exactly solvable, with a gapped spectrum with discretely spaced energy levels. 
This leads to a simple exponential decay of infinite-temperature autocorrelation functions of any operator $A$, which can be written as 
\be 
\braket{A(t) A}_{\beta =0} = \braket{A, e|~ e^{-P_2 \, t}~| A, e} \, . 
\ee
This is consistent with the physical expectation from the lack of any symmetries in the time-evolution.

Since Eq.~(\ref{gue}) does not have any symmetries, for general $n$, as discussed in Sec.~\ref{sec:eq}, its zero energy ground states are the $n!$ product states with the same permutation at each site:
\be 
\otimes_i \le(\frac{1}{q^{n/2}}\ket{\sigma}_i\ri), \quad \sigma \in \sS_n \, . \label{nth_gs_mt}
\ee
These can be thought of as ferromagnets of $\sS_n$ degrees of freedom.
We hence expect $P_{2n}$ to be gapped, and the low-energy excitations to be domain-walls between the different ferromagnetic ground states.
The structure of the low-energy excitations leads to the membrane picture of entanglement.
$P_{2n}$ also has the nice property that the subspace spanned by states of the form $\otimes_i \ket{\sigma_i}_i$ for $\sigma_i \in \sS_n$ is closed.
While this closure is indeed special to the choice of local terms drawn from the GUE ensemble in Eq.~(\ref{gue}), and does not generally hold for all Brownian evolutions of the form of Eq.~(\ref{htdef}), it simplifies analytical and numerical computations discussed later.
However, we emphasize that the ground states are always of the form of (\ref{nth_gs_mt}) for general Brownian models as well, which leads to the universality of the physics we discuss below.

The saturation value of the $n$-th Renyi entropy in this model obtained from the ground states \eqref{nth_gs_mt} is the special case of \eqref{eqa_mt} with $\rho^{\rm (eq)}= I/D$, where $D$ is the total Hilbert space dimension. As discussed in \cite{eq_approx}, Eq.~\eqref{rho_eq_mt} for this case is equal to the average value in random product states~\cite{page, lubkin, pagels, nadal, leut}.
The equilibrium entropy density for each of the Renyi entropies is therefore 
 \be 
s_{n,{\rm eq}} = s_{{\rm eq}} =\log q \, . 
 \ee
\subsection{Second Renyi entropy }\label{subsec:secondrenyi}
Let us now focus on the structure of the superhamiltonian 
 \eqref{p2n_gue} for the case $n=2$.
In this case, following the discussion around Eq.~(\ref{nth_gs_mt}), we have two degenerate ground states,
\begin{equation}
    \ket{G_\up} =  \ket{\uparrow \cdots \uparrow},\;\;\;\ket{G_\dn} = \ket{\dn \cdots \dn}.
\label{eq:P4gs}
\end{equation} where
 \begin{align} 
 &\ket{\up} = \frac{1}{q} \ket{e} =     \ket{\rm MAX}_{f_1 b_1} \ket{\rm MAX}_{f_2 b_2}, \nn ~  &\ket{\down} = \frac{1}{q} \ket{\eta}  = \ket{\rm MAX}_{f_1 b_2} \ket{\rm MAX}_{f_2 b_1},\label{spin_def}
\end{align}
where $\ket{\rm MAX}$ is defined in Eq.~(\ref{eq:MAXdefn}).
It will be useful to express $P_4$ in terms of these spin states.
Note that  $\ket{\up}$ and $\ket{\down}$ are normalized, but do not form an orthonormal basis as $\braket{\uparrow|\downarrow}= 1/q$.
It will be convenient to work in a bi-orthogonal system and introduce  the  following two states   
\begin{align} 
\ket{\bar \up} = \frac{q^2}{q^2-1} \le( \ket{\up} - \frac{1}{q}\ket{\down}\ri) , \quad \ket{\bar \down} = \frac{q^2}{q^2-1} \le( \ket{\down} - \frac{1}{q}\ket{\up}\ri)  \label{bar_spin_def}
\end{align} 
which have the property that 
$\braket{\bar{s'}| s} = \delta_{s s'}$ for $s, s'\in \{\up, \down\}$. We can then write $P_4$ in the subspace spanned by $\ket{\up}, \ket{\down}$ at each site as follows:
\begin{align} 
& P_4 = A_0 + A_1, \nn 
&A_0 =\sum_{\braket{ij}}\bigg[  I  - \ket{\bar \up\bar \up} \bra{\up \up} -   \ket{\bar \down \bar \down} \bra{\down \down} \nn 
 &\quad \quad\quad \quad- \frac{1}{q} (\ket{\bar \up \bar \down} +\ket{\bar \down \bar \up} )(\bra{\up \up} + \bra{\down \down})\bigg]_{i,j}   \nn 
 & A_1 =
 \sum_{\braket{ij}}\bigg[   \frac{1}{q^2} (\ket{\bar \up \bar \down} \bra{\down \up} +\ket{\bar \down \bar \up} \bra{\up \down} )\bigg]_{i,j} \label{aleft_def_mt}
\end{align}
where $\braket{ij}$ denotes nearest neighboring sites on a lattice, and for instance $\ket{\up\up}$ denotes $\ket{\up}_i \ket{\up}_j$.
This representation will be useful as the final state in the expression for the second  Renyi entropy in \eqref{eq:renyi_avg} is a domain wall state of the form 
\begin{equation}
    \bra{\eta_R \otimes e_{\bar R}} = q^L\otimes_{i\in R} \bra{\down}_i \otimes_{i \in \bar R} \bra{\up}_i
\end{equation} 
Note that the matrix expression for $P_4$ in this basis is non-Hermitian, due to the non-orthonormality of the basis. 
These non-Hermitian matrices can be expressed in terms of Pauli matrices, and these have been referred to as \textit{entanglement feature} Hamiltonians in earlier literature~\cite{you2018entanglement, kuo2020markovian}. 

While the above  representation will be convenient for some of our later analysis,  we can also represent $A$ in the following orthonormal basis for one site:  
\be 
\ket{+} = \frac{1}{\sqrt{2(1+\frac{1}{q})}}(\ket{\up}+\ket{\down}), \quad \ket{-} =  \frac{1}{\sqrt{2(1-\frac{1}{q})}}(\ket{\up}-\ket{\down})
\label{eq:orthobasis}
\ee
Defining the Pauli matrices with respect to these states, where $\ket{+}$ and $\ket{-}$ are the eigenstates of the  $Z$ operator, we obtain the following representation:
\begin{align}
P_4 =  \frac{1}{2}\sum_{\braket{i,j} }[& 1 - X_i X_{j} - \frac{1}{q}(Z_i + Z_{j}) + \frac{1}{q^2}(Z_i Z_{j} + X_i X_{j})],
\label{eq:PEspin_mt}
\end{align}
In one spatial dimension,  $P_4$ lies on a  parameter line of the spin-1/2 XYZ spin chain in an external magnetic field known as the ``Peschel-Emery'' line~\cite{peschel1981calculation, katsura2015exact}.
On this line, the Hamiltonian is known to lie within the Ising ferromagnetic ($Z_2$ symmetry broken) phase, and was previously noted to have two frustration-free product ground states, which are of the form~(\ref{eq:P4gs}).
We discuss this mapping  in Appendix \ref{app:num}. 

In the rest of this section, we specialize to one spatial dimension. 
We will consider higher dimensions in Sec. \ref{sec:eom}. 
\begin{figure*}[!ht]
    \centering
    \includegraphics[scale=0.95]{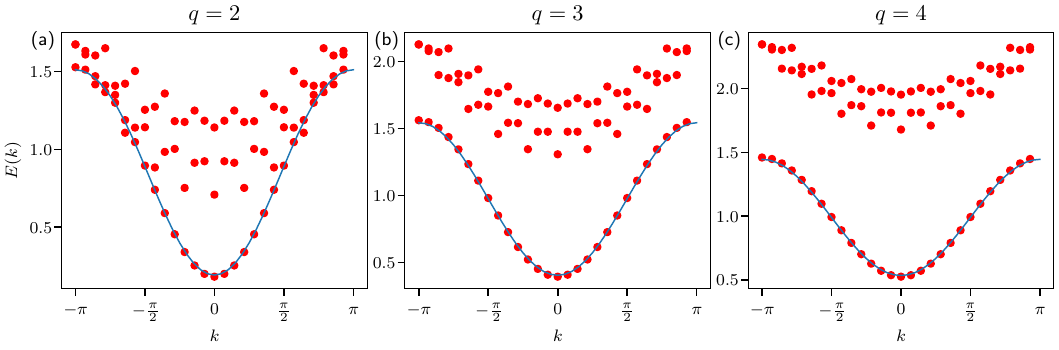}
    \caption{The red data points show the momentum-resolved low-energy spectra for $P_4$ in \eqref{aleft_def_mt} or \eqref{eq:PEspin_mt} for  $L = 14$  for (a) $q = 2$, (b) $q = 3$, and (c) $q = 4$.
    These energies are found using  exact diagonalization   with the twisted boundary conditions described in Appendix \ref{app:num}. The ground states are at $E = 0$ and the low-energy spectrum is almost unchanged between $L=12$ and $L=14$.
    These are compared to the blue curves, which are the dispersion relations obtained from the minimization of the expectation value of $P_4$ in the states  \eqref{phikdef} for $\Delta=4$, $L=40$ with open boundary conditions.
    }
    \label{fig:PEspectra}
\end{figure*}
\subsubsection{Large $q$ limit }\label{sec:large_q}
Note that without the $O(1/q^2)$ term, Eq.~(\ref{eq:PEspin_mt}) is simply the transverse field Ising model. 
Hence in the large $q$ limit, the low-energy physics is expected to be similar to that of the Ising model, although we will not use this for any of the analytical analysis.
Instead, we choose to take a large $q$ limit in the bi-orthogonal basis by  ignoring the $A_1$ term in \eqref{aleft_def_mt}, which is $O(1/q^2)$.
In one spatial dimension, the left eigenstates of the non-Hermitian Hamiltonian $A_0$ will turn out to be analytically tractable.
We assume that the system has $L$ sites labelled from $x=-L/2$ to $x=L/2-1$ and open boundary conditions (OBC), but we will take $L\to \infty$ for most purposes, and the left eigenstates of $A_0$ will be exactly solvable in this limit.
Note that the zero energy left-eigenstates of $A_0$ are simply $\bra{G_{\up}}$ and $\bra{G_{\dn}}$ given by (\ref{eq:P4gs}), which are the same as the zero energy eigenstates of the full Hamiltonian $P_4$.
Due to this feature, $A_0$ turns out to be a better first approximation for understanding the structure of the first excited states than the transverse field Ising model obtained by keeping the $O(1)$ and $O(1/q)$ terms in \eqref{eq:PEspin_mt}.    
To find the low-energy left-eigenstates, note that $A_0$ has a particularly simple left-action on the domain wall states defined as 
\be 
\bra{D_x} = \bra{\down \,  \down ... \down_x \, \up_{x+1} \,  \up ... \up} 
\label{eq:Dxdefn}
\ee
given by
\be 
\bra{D_x} A_0 = \bra{D_x} - \frac{1}{q}(\bra{D_{x+1}} + \bra{D_{x-1}})  \,,\label{dshift}
\ee
where we have implicitly defined $\bra{D_{-\frac{L}{2} - 1}} = \bra{G_{\up}}$ and  $\bra{D_{\frac{L}{2} - 1}} = \bra{G_{\dn}}$ at the boundaries.
This is thus a process where the domain wall can hop freely in the bulk of the system, and can be absorbed at the boundaries.\footnote{Note that since $A_0$ is non-Hermitian, and because we are working in a non-orthogonal basis,  domain-walls cannot be created at the boundaries.}
From \eqref{dshift}, we can observe that in the $L \rightarrow \infty$ limit, the lowest band of left-eigenstates of $A_0$ above the ground state is given by\footnote{Strictly speaking, if we are working with a finite but large system size $L$ with open boundary conditions, the $e^{ikx}$ should be replaced by a $\sin(kx)$, see App.~\ref{app:exactDW} for an exact solution.}
\be 
\bra{\psi_k} = \sum_{x}{e^{i k x} \bra{D_x}}, \quad E(k) = 1- \frac{2}{q}\cos k  \, .  \label{plane_waves}
\ee
We expect that this should be a good approximation for the eigenstates for finite but large $L$ as well, and the spectrum is gapped as we are in the regime $q \gg 1$. 
As we illustrate in detail in App.~\ref{app:exactDW}, the complete spectrum of $A_0$ is solvable at finite system sizes $L$, which we can use to further verify the structure of these left-eigenstates. 
In fact, as we illustrate there, a Hamiltonian that resembles $A_0$ with $q \rightarrow q + q^{-1}$ also turns out to be Hermitian (evident when expressed in the basis of Eq.~(\ref{eq:orthobasis})), and can hence be used as a solvable toy model for understanding the essential physics of superhamiltonians that appear in this work.

Let us now apply this structure to understand the evolution of $S_2$ for a half-line region to the left of $x$,
\be 
e^{-S_2(x, t)} = q^L\braket{D_x|e^{-P_4 t}| \rho_0, e} \, . \label{purity}
\ee
It is useful to introduce the states 
\begin{equation}
    \ket{\bar D_x} = \ket{\bar \down ... \bar \down_x \bar \up_{x+1} ... \bar \up},
\label{eq:Dxbardefn}
\end{equation}
and a ``domain wall propagator'' (considered for instance in \cite{zhou_nahum})
\be 
G(x, y, t) = \braket{D_x| e^{-A_0 t}|\bar D_y} \label{g0def}
\ee
in terms of which \eqref{purity} can be written as
\be 
e^{-S_2(x,t)} =
\sum_y G(x, y, t) e^{-S_2(y, t=0)}\, . \label{entlarge}
\ee
Here we have used the fact that the action of $A_0$ does not create additional domain walls, and we can hence insert a resolution of identity $\sum_{x}\ket{\bar D_x}\bra{D_x}$ restricted to the single domain wall subspace in \eqref{purity}.
Then using \eqref{plane_waves} and working in the $L \rightarrow \infty$ limit, we can express (\ref{g0def})  as
\be \label{gprop_2}
G(x, y, t) =  \int_{-\pi}^{\pi} \frac{dk}{2\pi} \, e^{-(E(k) + i k v)t}, \quad v= \frac{x-y}{t}\, .
\ee
For $E(k)$ of \eqref{plane_waves}, this can be evaluated exactly for integer values of $vt$ (which is what we are interested in) to be
\begin{equation}
    G(x, y, t) = e^{-t}I_{vt}\left(\frac{2t}{q}\right)\;\;\textrm{if }\;\;v t \in \mathbb{Z},
\label{eq:Gexact}
\end{equation}
where $I_{vt}(\bullet)$ is the modified Bessel function of the first kind of integer order $vt$, and we have used the identity of Eq.~(\ref{eq:besselidentity}). 
While Eq.~(\ref{eq:Gexact}) is exact, we are interested in the late-time limit $t \gg 1$ with $v$ being $\mathcal{O}(1)$ in this limit.
This can in principle be obtained using a standard Debye expansion of the modified Bessel function, see the discussion around Eq.~(\ref{eq:Debye}) in App.~\ref{app:membraneextract}.
However, here we illustrate a more general method to evaluate (\ref{gprop_2}) in this limit, that should work for arbitrary dispersion relations $E(k)$. 

Then the integral over $k$ in Eq.~(\ref{gprop_2}) can alternatively be evaluated using the saddle-point approximation for large $t$.
The saddle-point value $k_v$  lies on the imaginary axis, and by deforming  the contour to pass through its steepest descent contour we get
\be 
G(x, y, t) \sim  e^{- s_{\rm eq} \sE(v) t}   , \quad v = \frac{x-y}{t}  \label{436}
\ee
where 
\be 
\sE(v) = \frac{E(k_v) + i k_v v}{s_{\rm eq}} \,, \, \quad E'(k_v)=-iv \, .  \label{mem_0}
\ee 
Now putting \eqref{436} into \eqref{entlarge}, we obtain precisely the membrane formula for the second Renyi entropy in the large $t$ limit:
\begin{align}
e^{-S_2(x,t)} &\approx \sum_v{e^{-s_{\rm eq} \sE(v) t - S_2(x - vt, t = 0)}}\nonumber\\
\implies S_2(x, t) &\ \approx \  \text{min}_v \le[\, s_{\rm eq}\, \sE(v) \, t + S_2(x-vt, t=0) \, \ri], \label{1dmem}
\end{align}
where in the second step we have used that $v \sim O(1)$ and $t \gg 1$, and ignored contributions that are smaller than linear in $t$.

The  membrane tension from the dispersion relation \eqref{plane_waves}  is~\footnote{\label{ft:toy}Note that \eqref{mem_gue} by itself does not satisfy the constraints \eqref{const} except in the strict large $q$ limit \eqref{largeq_abs}. It does satisfy the constraints at finite $q$ if we make some replacements of $q \to q + q^{-1}$, which can be seen as a simple analytical ``toy model'' for the membrane tension function, see Apps.~\ref{sec:hermitiantoy} and App.~\ref{subsec:varcalc}. 
This is not an accident, and there is a different choice of ensemble of random circuits for which a toy dispersion relation~\cite{znidaric2008exact, suzuki2024more} is relevant, see discussion in Sec.~\ref{subsec:haarcompare}}
\be 
 \sE^{\rm (large~ q) }(v) = \frac{1}{\log q} \le[1 - \frac{2}{q} \sqrt{1 + \le(\frac{qv}{2}\ri)^2}+ v \, \text{arcsinh}\le(\frac{v q}{2}\ri)\ri] \label{mem_gue} \, . 
 \ee 
Its strict large $q$ limit is 
\be 
\lim_{q \to \infty}\sE(v) = |v|  \, \label{largeq_abs}
\ee
which satisfies the constraints \eqref{const} in a somewhat degenerate way.
As we will discuss in the following sections, the expressions \eqref{entlarge}-\eqref{1dmem} will turn out to apply to more general cases than the large $q$ limit of the current model.
The modified dispersion relations $E(k)$ in these cases will lead to other, more general expressions for $\sE(v)$. 
Finally, we would also like to point readers to Apps.~\ref{app:exactDW} and \ref{app:membraneextract} for exact computations of the propagator and membrane tension for a toy superhamiltonian where the domain wall excitations are exact for any finite size $L$, where the final results closely resemble to ones discussed above.

Before discussing these more general cases, let us briefly discuss the form of $S_2$ for a region consisting of one or more intervals in the large $q$ limit.
Since the ``final state" in this case would contain more than one domain wall,   we now need to consider ``multi-particle'' excitations of $A_0$ involving more than one domain wall. 
For a  state with two or more domain walls, we can check from \eqref{aleft_def_mt} that the left-action of $A_0$ can cause domain walls to annihilate in pairs. It is useful to divide $A_0$ into two parts, 
\be 
A_0 = A_{f} + A_{c}
\label{eq:A0divide}
\ee
where $A_{f}$ keeps the number of domain walls fixed, while $A_{c}$  causes the annihilation of pairs of domain walls.
$A_{f}$ has a block-diagonal structure, 
while the whole matrix $A_0$ has a lower-triangular structure due to $A_{c}$.
The energy eigenvalues of $A_0$ are therefore the same as those of $A_{f}$. 
Denoting the multiple domain wall states by  $\bra{D_{x_1, ..., x_k}}$,~$x_1<x_2<...<x_n$, it is easy to check that the domain walls are ``non-interacting" under the action of $A_f$.
Hence, the eigenstates of $A_{f}$ take a simple  free fermion-like Slater determinant form:
\begin{align} 
&\bra{ \psi_{k_1, ..., k_n}} = \sum_{\sigma \in \sS_n}  \text{sgn}(\sigma) \bra{\phi_{k_{\sigma(1)}, ..., k_{\sigma(n)}}}, \nn
&\bra{\phi_{k_1, ..., k_n}} = \sum_{x_1< x_2<  ...<  x_n} e^{i \sum_{j=1}^n k_j x_j} \bra{D_{x_1, ..., x_n}}   \label{psin_def_mt}
\end{align}
with energies $E(k_1,..., k_n)$ given by the sum of the one-particle energies $E(k_i)$ in \eqref{plane_waves}.
Hence, the energies of the multiparticle states under $A_0$ are also sums of one-particle energies,\footnote{This is also evident if one performs the basis change of \eqref{eq:orthobasis} on $A_0$ in \eqref{aleft_def_mt} and then performs a Jordan-Wigner transformation of \eqref{eq:JWtransforms} -- the resulting Hamiltonian is a non-Hermitian non-interacting Hamiltonian.} although the eigenstates are more complicated superpositions of \eqref{psin_def_mt}.
As we discuss in more detail in Appendix \ref{app:multiple}, this structure leads to the membrane picture for multiple intervals. 
\subsubsection{General structure}
The key physical properties of $A_0$ that give rise to the membrane picture in the above discussion are:
\begin{enumerate}
\item[(i)] The ground states are simply ferromagnetic states of the form of Eq.~(\ref{eq:P4gs}).
\item[(ii)] The spectrum is  gapped, so that $e^{-S_2(t)}$ has an exponential decay for any initial state, leading to linear growth of $S_2(t)$;
\item[(iii)] The one-particle eigenstates of the $A_0$ are plane waves of domain walls between the two ferromagnetic ground states.
\item[(iv)] Domain walls are non-interacting other than the possibility of  pair-wise annihilation and annihilation at the boundaries.
\end{enumerate}
In the following, we will show that properties (i)-(iii) remain robust for the full Hamiltonian $P_4$ at finite $q$.
(i) just follows from the symmetries of the system, and always holds for all the systems we consider in this work.
Given that ground state structure, we expect that it is generically gapped, so (ii) holds, and we expect that its excitations are domain-walls, so (iii) should be robust up to some local ``dressing" of the domain walls.
In Sec. \ref{sec:gue}, we will further show that all these properties remain robust for Brownian models with fixed coupling operators.
We also expect that property (iv) is robust, but this is harder to show explicitly, and we do not comment further on it until the Discussion section  ~\ref{sec:discussion}. 
Due to the robustness of (iii), the formulas \eqref{entlarge} and \eqref{gprop_2} will also apply to the second Renyi entropy in all remaining examples we consider.
It is useful at this point to summarize some general consequences of these formulas, which we will use in the later discussion. 
For an initial mixed state of the form \eqref{initial_s} with entropy density $s$, the evolution of $S_2$ is given by
\be 
e^{-S_2(x,t)} = e^{-s \le(x+ \frac{L}{2}\ri)}\int_{-\infty}^{\infty} dv \int_{-\pi}^{\pi} \frac{dk}{2\pi} e^{-(E(k) + i k v  - s v)t}  \label{fullint_2ren}
\ee
The saddle-point equations for $v$ and $k$ in the above integral are:
\be 
k= -is, \quad E'(k) = -iv \label{k2}
\ee
which imply \eqref{gamma_n} with the entropy growth rate\footnote{Note that initial states of the form \eqref{initial_s} should grow at this rate for any finite $t$ in the $L \rightarrow \infty$ limit, although for a finite system this would slow down due to the gradual increase of entropy density in the system.}
\begin{align}
\Gamma_2(s)\equiv\defn \frac{\partial S_2}{\partial t} = E(is)/s_{\rm eq} \, . \label{growthrate}  
\end{align}
where we have used the fact that $E(k)$ is an even function. 
Eq.~\eqref{mem_0} is then  equivalent to the statement that $\sE(v)$ is the Legendre transform of $-\Gamma(s)$, 
\be 
\sE_2(v) = \text{max}_s\le(\frac{v s}{s_{\rm eq}} + \Gamma_2(s)\ri) \,  \label{leg} 
\ee
which also follows from \eqref{gammadef} and 
was previously noted in \cite{huse}.
In particular, from Eq.~(\ref{growthrate}), we obtain that for an initial pure product state, which has entropy density $s = 0$, the entanglement velocity of the second Renyi entropy $v_{E, 2}$ is proportional to the gap in the spectrum: 
\be 
S_2(x, t) = s_{\rm eq}  \, v_{E, 2} \,  t, \quad v_{E, 2} = \frac{E(0)}{s_{\rm eq}} \, .  \label{ve_disp} 
\ee
From \eqref{growthrate}, the constraints \eqref{const} or \eqref{gamma_condition}  are equivalent to the fact that 
\be
E(i s_{\rm eq}) =0 \, .   \label{econd}
\ee
Using \eqref{k2}, in terms of the dispersion relation, $v_B$ is given by 
\be 
v_B = -i E'(-i s_{\rm eq}) \,. \label{vbdef}
\ee

Note that the condition \eqref{econd} does not need to be imposed as an external input.  The evolution of the entanglement entropy of $S_2$ for a maximally mixed initial state can be expressed as 
 \be 
e^{-S_{2}^{\rm (max)}(x, t)} = q^L\bra{D_x} e^{-P_{4}t} \le(\otimes_i \frac{1}{q} \ket{\up}_i\ri) \, .  \label{459}
 \ee
By acting with $e^{-P_{4}t}$ on the left, we get \eqref{fullint_2ren} with $s=s_{\rm eq}$ due to the structure of the low-energy excitations, and by acting on the right, we get a time-independent result due to the fact that $\otimes_i \ket{\up}_i$ is a zero energy eigenstate of $P_4$. \eqref{econd} must always be true to ensure consistency between these results.\footnote{There is a subtlety here when working with finite system sizes, see App.~\ref{sec:commentsmembrane} for a discussion.}

One interesting aspect of the condition \eqref{econd} is that it is sensitive to the UV behaviour of the dispersion relation $E(k)$, as it involves an $O(1)$ imaginary value of $k$.
We will return to the implications of this UV sensitivity in the Discussion section.
\subsubsection{Finite $q$}  \label{sec:finite_q}
At finite $q$, the $A_1$ term in  \eqref{aleft_def_mt} 
can send a single domain wall state $\bra{D_x}$ to a three domain wall state $\bra{D_{x-1, x, x+1}}$ and vice versa, so that the exact low-energy eigenstates are no longer plane waves of single domain walls as in \eqref{plane_waves}, and the exact energies are also modified.
Nevertheless, since the ground states of $P_4$ are still ferromagnetic states of the form of (\ref{eq:P4gs}), we expect the low-energy excitations to be gapped (dressed) domain walls between the two ground states, similar to (\ref{plane_waves}).
To numerically determine the momentum-resolved dispersion of the low-energy eigenstates, we study $P_4$  with symmetry-twisted (antiperiodic) boundary conditions, as discussed  in Appendix \ref{app:num}.\footnote{We thank Tibor Rakovszky for useful discussions on this.}$^{,}$\footnote{In summary, the momentum resolution cannot be obtained directly with a finite-size OBC Hamiltonian due to lack of translation-invariance. On the other hand, periodic boundary conditions (PBC) does not allow for an odd number of domain walls. 
By inserting a symmetry twist at the boundary, one domain wall gets pinned at the boundary while the other can disperse, and we can obtained the momentum-resolved dispersion of a single domain wall.
This is expected to match the low-energy spectrum to match that of the OBC Hamiltonian for large system sizes.}

 The low-lying  spectrum as a function of $k$ is shown in Fig. \ref{fig:PEspectra} for $q=2, 3, 4$.
In all cases, we find that $P_4$ is gapped, consistent with expectations in the ferromagnetic phase. The gap leads to a linear growth of entanglement for a product state, according to \eqref{ve_disp}.
To verify the robustness of the membrane picture, we need to further address the following questions: 
\begin{enumerate}
\item For $q \geq 3, 4$, we find a single-particle band in the spectrum well-separated from the multi-particle continuum for all $k$. It is natural to expect that the eigenstates in this band have a quasiparticle  structure~\cite{elementary_excitations}. Can these quasiparticle states be understood as locally dressed versions of the domain wall states in  \eqref{plane_waves} in a precise sense?
\item For $q=2$, the gap between the single-particle states and the multi-particle continuum vanishes beyond some value of $k$. Are there still well-defined quasiparticle states within the continuum at large $k$? 

\end{enumerate} 
Both questions can be simultaneously addressed using a  technique for obtaining low-energy dispersion relations along the lines of  \cite{elementary_excitations, variational_ansatz, scattering_particles}.
These references introduced a general variational ansatz for low-energy excitations of gapped spin chain systems, starting from the assumption that the ground state is well-approximated by a matrix product state.
In the case of the Hamiltonian $P_4$, since we know that the exact zero energy eigenstates are the product states $\ket{\up...\up}$ and $\ket{\down...\down}$, we can use a particularly simple version of the general ansatz:
\be 
\ket{\psi_k} = \sum_{x} e^{-ikx} \ket{\down ...\down_{x}} \ket{\phi_{x+1,  ..., x+\Delta}} \ket{\up_{x+\Delta+1} ... \up} \label{phikdef}
\ee
where $\ket{\phi_{x+1,  ..., x+\Delta}}$ is an arbitrary state in the subspace spanned by  $\ket{\up}, \ket{\down}$ on $\Delta$ sites,   which has total Hilbert space dimension $ 2^{\Delta}$.~\footnote{The actual dimension of the subspace spanned by the states in \eqref{phikdef} is slightly smaller than $2^{\Delta}$ due to the redundancy between certain choices of $\ket{\phi}$ in the thermodynamic limit. For example, for $n=1$, there is only one linearly independent choice $\ket{\phi}=\ket{\down}$, and for $n =2$, we can consider an arbitrary superposition of the form  $\ket{\phi}= \alpha \ket{\down}\ket{\up}+ \beta \ket{\up} \ket{\down}$.}  
 \begin{figure*}
\includegraphics[scale=0.9]{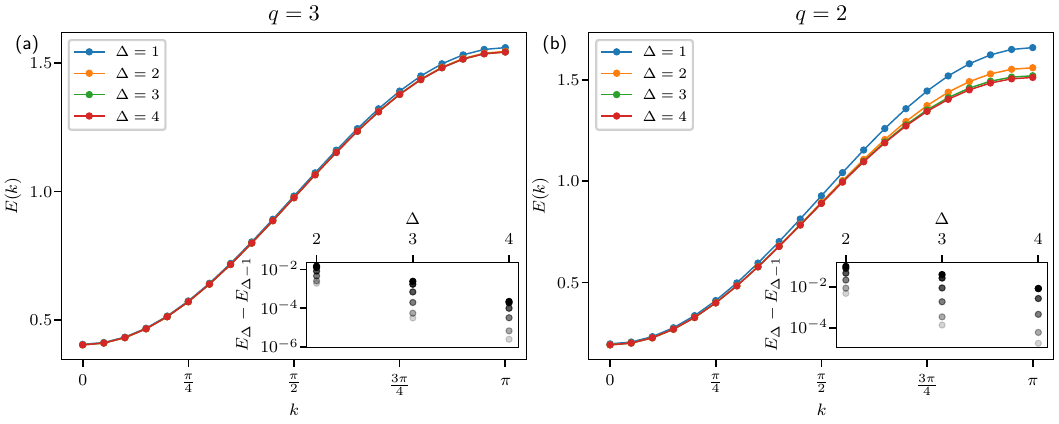}
\caption{
The energies obtained by minimizing the expectation value of $P_4$ for the Brownian local GUE model, \eqref{aleft_def_mt} within states of the form \eqref{phikdef} for increasing values of $\Delta$ from 1 to 4, for $L = 40$ and (a) $q=3$ and (b) $q=2$.
The insets plot $E_{\Delta}-E_{\Delta+1}$ as a function of $\Delta$ for a few values of $k$ from $k = 0$ (lightest) to $k = \pi$ (darkest), which shows that the dispersion relations are converging rapidly with $\Delta$ in both cases, indicating that the domain wall excitations \eqref{phikdef} are a good approximation to the true eigenstates.} 
\label{fig:variational_gue_2renyi}
\end{figure*}

We increase the value of $\Delta$ starting from $0$, and minimize the expectation value 
$\braket{\psi_k|P_4|\psi_k}$ over all choices of $\ket{\phi}$ for a given $\Delta$.
We explain  details of the variational optimization in Appendix \ref{app:var}.
As discussed in \cite{elementary_excitations, variational_ansatz, scattering_particles}, rapid convergence of the dispersion relation on increasing $\Delta$ indicates that the  eigenstates of $P_4$ are well-approximated by \eqref{phikdef} for $O(1)$ $\Delta$, and this interpretation also holds when the dispersion relation lies within a multi-particle continuum.
The results are shown in Fig.~\ref{fig:variational_gue_2renyi}, where we find rapid convergence of the dispersion relation with $\Delta$ for all values of $k$ for both $q=2$ and $q=3$.
We also compare the variational  dispersion relations to the  energies obtained from exact diagonalization in Fig.~\ref{fig:PEspectra}, finding good agreement in cases where the latter show a well-defined single-particle band.  The expectation value with $\Delta = 0$ already gives a good approximation for the numerically observed dispersion relation, see Appendix~\ref{subsec:varcalc} for details.

These results confirm that the low-energy eigenstates relevant for the evolution of $e^{-S_2(x, t)}$  have the structure of localized domain wall-like states at finite $q$, including in the case $q=2$ at all $k$.
Putting this structure \eqref{phikdef} of the eigenstates into \eqref{purity}, we have:~\footnote{Note in particular that in the $q=2$ case, even though the energies of the multi-particle continuum are comparable to those of the single-particle band, the final state $\bra{D_x}$ has significant overlap only with the eigenstates of the single particle band, so that the approximation in \eqref{expansion} is valid.}
\begin{align}
&e^{-S_2(x, t)} \approx \sum_k \sum_{y_1, y_2} e^{ik(y_2-y_1)} e^{-E(k) t}  \nn &\quad \bra{D_x}\le(\ket{\down ...\down_{y_1}} \ket{\phi_{y_1+1,  ..., y_1+\Delta}} \ket{\up_{y_1+\Delta+1} ... \up}\ri)\nn 
&\quad \le(\bra{\down ...\down_{y_2}} \bra{\phi_{y_2+1,  ..., y_2+\Delta}} \bra{\up_{y_2+\Delta+1} ... \up}\ri)\ket{\rho_0, e} \label{expansion}
\end{align}
Since $\ket{\phi_{\cdots}}$ is some superposition of configurations of $\ket{\uparrow}$ and $\ket{\downarrow}$ spins, for $\Delta$ of $O(1)$, the factor in the last line of \eqref{expansion} contains a term proportional to $e^{-S_2(y_2, t=0)}$, as well as terms proportional to $e^{-S_2([-\frac{L}{2}, y_2+\Delta_1]\cup [y_2+\Delta_2, y_2 + \Delta_3]\cup ... \cup [y_2+\Delta_{m-1}, y_2 + \Delta_{m}], t=0)}$ for all odd $m\geq 3$,  for some $O(1)$ $\Delta_i$.
Since $\Delta$ is $O(1)$, in the scaling limit of late time and large system size, the differences between these terms can be ignored, and they can be combined into $c\,   e^{-S_2(x,t=0)}$ for some $O(1)$ number $c$.
Similarly, the factor in the first line of \eqref{expansion} can be replaced with $\delta_{xy_1}$ in this limit.
Hence, the analysis of \eqref{g0def}-\eqref{1dmem} also applies to this case, with the change that $E(k)$ should be used to denote the numerically obtained exact dispersion relation at finite $q$ from Fig.~\ref{fig:PEspectra}, instead of 
 the large-$q$ dispersion relation  of \eqref{plane_waves}.  

\begin{figure}[t]
      \centering
      \includegraphics[width=0.5\textwidth]{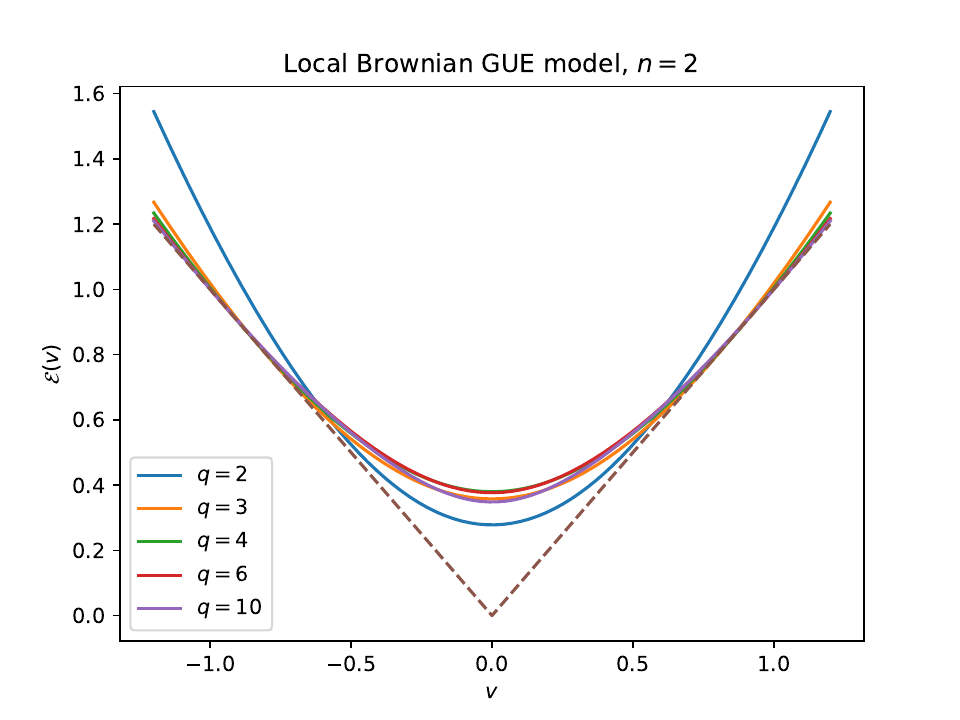} \includegraphics[width=0.5\textwidth]{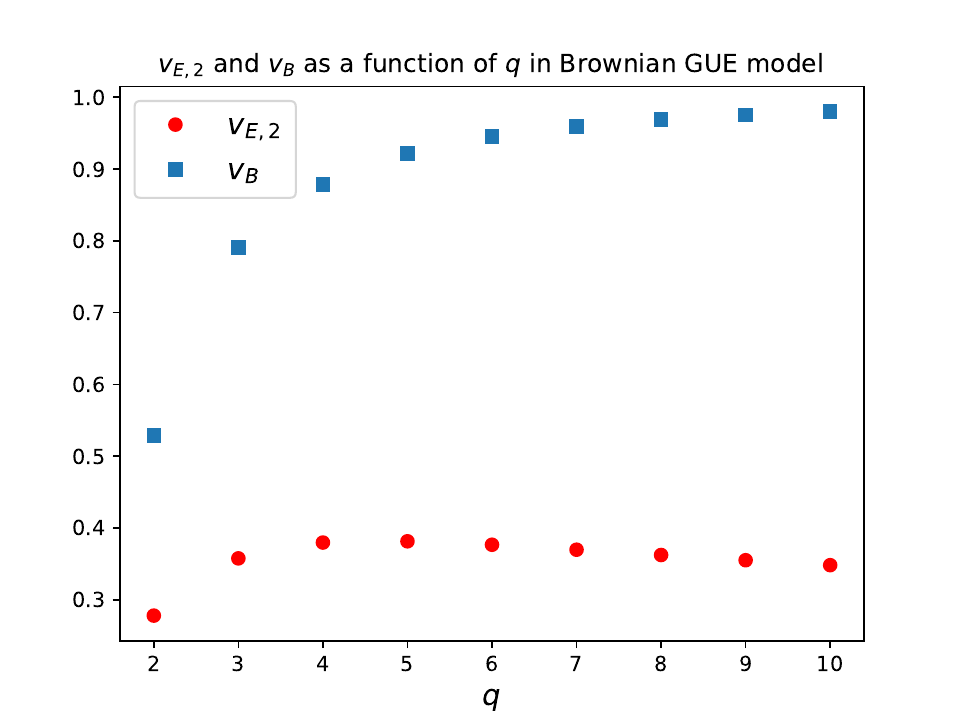}
      \caption{Top: We show the membrane tension curves for the second Renyi entropy for various $q$ from the fitting of $E(k)$ to \eqref{disp_fit} with $N_{\rm max} = 4$. ($N_{\rm max}=2, 3$ give similar curves.) For $q\geq 3$, we use $E(k)$ from exact diagonalization, while for $q=2$, we use $E(k)$ from the variational method for $\Delta=4$, so that \eqref{const} is only approximately satisfied. Bottom: We show $v_{E,2}$ and $v_B$ as a function of $q$.}
      \label{fig:mem_tensions}
  \end{figure}

We determine $\sE(v)$ by  fitting $E(k)$ to the general form 
\be 
E(k) = \sum_{n=0}^{N_{\rm max}} c_n \cos(n k), \label{disp_fit}
 \ee
for some finite $N_{\rm max}$, 
and then numerically solving the equation for $k_v$ in \eqref{mem_0}.
This procedure gives us the membrane tensions in Fig.~\ref{fig:mem_tensions}.
Let us make a few observations about these results: 
\begin{enumerate}
\item 
For each $q$, $\sE(v)$ is convex and satisfies the general constraints \eqref{const} for  $v_B$ given by \eqref{vbdef}.
\item 
The entanglement velocity $v_{E,2}$ from \eqref{ve_disp} is non-monotonic as a function of $q$, increasing up to $q=5$ and then decreasing. The eventual decreasing behaviour is consistent with the large $q$ limit \eqref{largeq_abs}, where $v_{E,2}\to 0$.
Note, however, that the quantity  $s_{\rm eq} v_{E,2}$, which determines the coefficient of the linear growth of $S_2$ for a product state with time, increases  monotonically with $q$. Moreover, by choosing a different overall normalization of the Brownian GUE hamiltonian, we can obtain a finite $v_{E,2}$ in the $q \to \infty$ limit, similar to previous calculations in the SYK chain~\cite{syk_chain}. 
\item 
The butterfly velocity $v_B$ from \eqref{vbdef} monotonically increases with $q$. 
\end{enumerate}

We note one subtlety of the above discussion.
For any choice of $\Delta$ at which we choose to truncate the approximation \eqref{phikdef}, there will be small corrections in the exact eigenstate  proportional to $\ket{\down ...\down_{x}} \ket{\phi_{x+1,  ..., x+r}} \ket{\up_{x+r+1} ... \up}$ 
 for $r> \Delta$.
In principle, there could be initial states $\rho_0$ with entanglement structures  that would lead to a non-trivial competition in \eqref{purity} between the suppression of such components in the eigenstate, and an enhancement of the corresponding overlap factor $\bra{\down ...\down_{x}} \bra{\phi_{x+1,  ..., x+r}} \bra{\up_{x+\Delta+1} ... \up}\ket{\rho_0, e}$.
We argue in Appendix \ref{app:competition} using a somewhat different approach that such corrections are not important.
This argument makes use of the convexity of the numerically obtained membrane tensions in Fig. \ref{fig:mem_tensions}, together with the structure of the interaction picture diagrams we get from treating $A_1$ as a perturbation. 

\subsection{Comparison to Haar random unitary circuits}\label{subsec:haarcompare}

\begin{figure}[!h]
\includegraphics[width=\columnwidth]{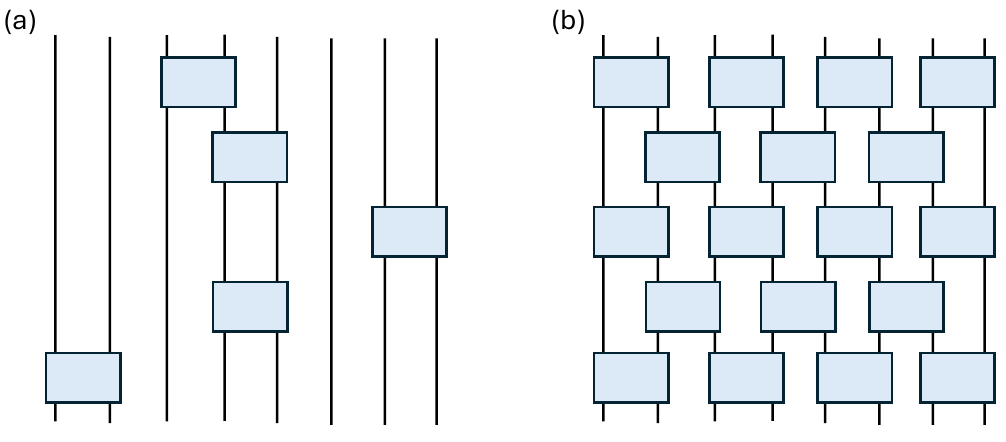}
\caption{Random unitary circuits with (a) randomly applied two-site gates and (b) a brickwork pattern.}
\label{fig:randomcircuit}
\end{figure}
It is instructive to rephrase the evolution of the second Renyi entropy in  Haar random circuits in the above language of low-energy modes of a one-dimensional quantum Hamiltonian, as an alternative to the standard discussion in terms of a mapping to a two-dimensional classical statistical mechanics problem.
In these models, we apply two-site random unitaries drawn with the Haar measure in some pattern, which can either be applied to a random pair of nearest-neighbor qubits at each time step as shown in Fig.~\ref{fig:randomcircuit}~(a), or in a brickwork pattern shown in Fig. \ref{fig:randomcircuit}~(b).
For the circuit where Haar random gates are applied randomly (say with open boundary conditions), the average of the unitary $\overline{(U(t)\otimes U(t)^{\ast})^{\otimes 2}}$ for integer times $t$ can be written as~\cite{suzuki2024more}
\begin{gather}
    \overline{(U(t)\otimes U(t)^{\ast})^{\otimes 2}} = M_R^{(L-1)t},\nonumber\\
    M_R = \frac{1}{L-1}\sum\nolimits_{i}{M_{i,i+1}}, 
\label{eq:randomapplication}
\end{gather}
where we have also averaged over the positions of application of the random gates, defined ``time" $t$ with a factor of $(L-1)$ to account for the fact that there is a single gate in each layer.
Further, $M_{i,i+1}$ is the four copy average of \textit{each} Haar random unitary, which is given by\footnote{The expression in terms of kets and bras of $\up$ and $\down$ spins (without involving $\overline{\up}$ and $\overline{\dn}$) can be derived using standard Weingarten calculus, e.g., see \cite{vijay_randomcircuit}, and this expression is obtained by applying similarity transformations between the different bases discussed in App.~\ref{subsec:mapping}.}
\begin{gather}
    M_{i,i+1} =  \big[\ket{\bar \up \bar \up} \bra{\up \up} + \ket{\bar \down \bar \down} \bra{\down \down} \nonumber \\
\quad + \frac{q}{q^2+1} \le(\ket{\bar \up \bar \down} + \ket{\bar \down \bar \up} \ri)  \le(\bra{\up \up} + \bra{\down \down}\ri) \big]_{i, i+1}
\end{gather}
Note that $M_{j,j+1}$ is Hermitian, and this can be seen by expressing it the orthonormal basis of Eq.~(\ref{eq:orthobasis}), where it reads
\begin{equation}
    M_{i,i+1} = \frac{1}{2} + \frac{1}{2(q + q^{-1})}[q X_j X_{j+1} + q^{-1} Y_j Y_{j+1}  + Z_j + Z_{j+1}].
\label{eq:momentopdef}
\end{equation}
Interestingly, this is a completely free-fermion operator under the Jordan-Wigner transform [of the form of Eq.~(\ref{eq:JWtransforms})], hence the spectrum of $M_R$ is completely solvable.
As we discuss below, a single domain wall $\bra{D_x}$ is only allowed to hop around under the action of such a moment operators, and hence unlike in the GUE model, there is no splitting of domain wall.
From these expressions, it is easy to verify that $\bra{\uparrow \cdots \uparrow}$ and $\bra{\downarrow \cdots \downarrow}$ are eigenstates of $M_R^{(L-1)}$ of Eq.~(\ref{eq:randomapplication}) with eigenvalue $1$.
Further, the action of $M_{i,i+1}$ on each domain wall reads
\begin{equation}
    \bra{\up\dn}M_{i,i+1} = \frac{q}{q^2 + 1}(\bra{\up\up} + \bra{\dn\dn}).
\end{equation}
$M_R$ of Eq.~(\ref{eq:randomapplication}) then has precisely the same structure as the toy Hamiltonians discussed in App.~\ref{app:exactDW} (up to overall factors).
For large $L$, the plane waves of domain walls of the form $\bra{\psi_k}$ of Eq.~(\ref{plane_waves}) can then be shown to be exact eigenstates of $M_R^{(L-1)}$ with eigenvalue $e^{-E_R(k)}$, where 
\begin{align} 
E_{R}(k) &= -(L-1)\log\left[1 - \frac{1}{L-1}\left(1 - \frac{2}{q + q^{-1}}\cos(k)\right)\right] \nonumber \\
&\approx \left(1 - \frac{2}{q + q^{-1}}\cos(k)\right).
\label{eq:haarrandomdispersion}
\end{align}
Hence the averaged evolution of the domain walls under this Haar random circuit is simply equivalent to the imaginary time evolution under a superhamiltonian with a dispersion relation $E_R(k)$, and the membrane tension can be derived using Eq.~(\ref{mem_0}).
This dispersion relation is also very close to the one obtained for the Peschel-Emery model (see App.~\ref{subsec:varcalc}) and such a membrane tension also satisfies the membrane constraints (see Ft.~\ref{ft:toy}).
On the other hand, for the brickwall circuit, the average $\overline{(U(t)\otimes U(t)^{\ast})^{\otimes 2}}$ for a time $t$ can be written in the notation of the previous subsections as  
\begin{align}
& \overline{(U(t) \otimes U(t)^{\ast})^{\otimes 2}}  = M^{t}_B,\nn
&M_B = \prod_{i=\text{odd}} M_{i, i+1}\prod_{i=\text{even}} M_{i, i+1} ,  \label{mdef}
\end{align}
Note that $M_B$ is not Hermitian due to the brick-wall geometry of non-commuting Hermitian operators $\{M_{i,i+1}\}$.
Here we can see that $\bra{\up ... \up}$, $\bra{\down ... \down}$ are left eigenstates of $M$ with eigenvalue 1, and that $\bra{\psi_k}$ in 
\eqref{plane_waves}~(with the sum restricted to odd $x$, hence the momentum restricted to $-\frac{\pi}{2} < k < \frac{\pi}{2}$) is an {\it exact} left eigenstate of $M$ for any value of $q$, with eigenvalue $e^{-2E_{\rm Haar}(k)}$, where $E_{\rm Haar}(k)$ is defined as
\begin{equation} 
E_{\rm Haar}(k) = \log \frac{q^2+1}{2q} - \log \cos k.
\label{eq:haardispersion}
\end{equation}
Applying the relation \eqref{mem_0} for this dispersion relation gives a simple derivation of the membrane tension for this case, found previously in \cite{huse}.
Unlike the membrane tensions in the Brownian models in the rest of this work, $\sE(v)$ in this model diverges for $v=1$ and is not well-defined for $v>1$, indicating the sharp light-cone in Haar-random circuits.

We would like to emphasize again that the main difference between the four-copy evolution by \eqref{mdef} for the Haar random circuit and the evolution by $e^{-P_4t}$ for the local GUE model is the fact that a single domain wall can split into multiple domain walls under the latter but not the former.
From the discussion of the previous subsection and Appendix \ref{app:competition}, we learn that this splitting of the domain walls does not have a qualitative effect on the membrane picture for the second Renyi entropy, and only serves to renormalize the membrane tension.
This is equivalent to the fact that the exact low-energy excitations in the GUE model are not bare domain walls of the form of Eq.~(\ref{plane_waves}), but rather dressed domain walls of the form of Eq.~(\ref{phikdef}).
However, this can lead to differences in certain physical quantities, and in Appendix \ref{app:op_growth}, we give one such example related to operator growth which do shows a qualitative difference between Haar random circuits and the GUE model due to this domain wall splitting. 
We also note that \cite{suzuki2024more} recently obtained a large class of random evolutions that have solvable or free-fermion second-moment operators, and it would be interesting to examine the domain wall dynamics in those cases.
An even earlier work \cite{znidaric2008exact} also studied the dynamics of purity using the low-energy excitations of Hamiltonians that closely resemble Eq.~(\ref{eq:momentopdef}), and it would be interesting to establish a more concrete correspondence.

\subsection{Higher Renyi entropies} \label{sec:higher_renyi}
Let us now discuss the structure of the superhamiltonian $P_{2n}$ of \eqref{p2n_gue} in the more general case where $n\geq 3$.
Recall from the discussion of Sec.~\ref{sec:eq} that due to the absence of symmetries of the model, the ground states are given by ferromagnetic states of the form (\ref{nth_gs_mt}).
Further, as discussed around \eqref{nth_gs_mt}, the Hilbert space at each site that has dimension $n!$, i.e., spanned by the states $\frac{1}{q^{n/2}}\ket{\sigma}_i$, is closed under the action of $P_{2n}$, which simplifies the analysis although this is special to the GUE model.
For studying the excitations, it is convenient to introduce the notion of the Cayley distance  $d(\sigma, \tau)$ between two permutations $\sigma, \tau \in \sS_n$, which is the minimum number of transpositions (swaps) $(i \, j)$ needed to go from $\sigma$ to $\tau$.
For any $\sigma, \tilde \sigma \in \sS_n$ such that $d(\sigma, \tilde \sigma)= 1$, we have  
\be 
\bra{\sigma}\bra{\tilde \sigma} P_{2n} = \bra{\sigma}\bra{\tilde \sigma}  - \frac{1}{q} (\bra{\sigma}\bra{ \sigma} + \bra{\tilde \sigma}\bra{\tilde \sigma}) +  \frac{1}{q^2} \bra{\tilde \sigma}\bra{\sigma}
\label{eq:P2naction}
\ee
Comparing to \eqref{aleft_def_mt}, we see that the action of $P_{2n}$ on states constructed only from any such pair $\bra{\tilde \sigma}, \bra{\sigma}$ is identical to the action of $P_4$ on $\bra{\up}, \bra{\down}$.
Indeed, this reduction is needed to ensure that we get consistent results on  computing the quantity $\Tr[\rho(t)]^{n-2}\Tr[\rho(t)^2] = \Tr[\rho(t)^2]$ using a general $P_{2n}$.
 Hence, the subspaces spanned by configurations of such pairs $\{\sigma, \tilde{\sigma}\}$ are closed under the action of $P_{2n}$.
Following the discussion in the $n = 2$ case in Sec.~\ref{subsec:secondrenyi}, we have eigenstates of $P_{2n}$ approximately given by plane waves of domain walls between each $\sigma$ and $\tilde \sigma$, with the same dispersion relations as those  found in Fig. \ref{fig:PEspectra}. 
However, the action of $P_{2n}$ on permutations $\sigma$, $\tilde{\sigma}$ with $d(\sigma, \tilde{\sigma}) > 1$ is not as simple as \eqref{eq:P2naction}, and this complicates their analysis, as we discuss below.

Recall from \eqref{eq:renyi_avg} that the final state that appears in the expression for $e^{-(n-1)S_n(x, t)}$ is $\bra{\eta ... \eta_x e_{x+1} ... e }$, where $\eta= (n ~ n-1~ 
 n-2 ~ ... 1)$.
Since $d(\eta, e) > 1$ for $n>2$, it turns out that the subspace spanned by configurations of $\bra{\eta}$ and $\bra{e}$ is not closed, and there is no exact eigenstate of $P_{2n}$ consisting only of $\bra{\eta}$ and $\bra{e}$, even in the large $q$ limit.
See Eq.~(\ref{eq:eetaaction}) in App.~\ref{app:superhamhigher} for the explicit expressions for $n = 3$.
\subsubsection{Naive entanglement growth rate from domain wall excitations}
\begin{figure*}[t]
\centering
\includegraphics[scale=0.9]{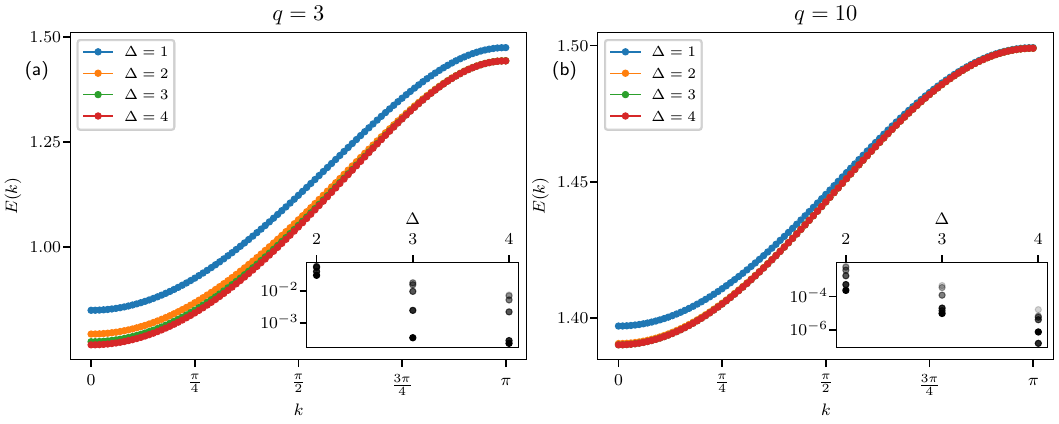}
\caption{Dispersion relation $E_n(k)$ for the third Renyi entropy ($n=3$) in the local GUE model, for $L = 100$ and (a) $q=3$, and (b) $q=10$.
The insets show $|E_{\Delta+1} - E_{\Delta}|$ for various values of $k$, indicating the convergence (see caption of Fig.~\ref{fig:variational_gue_2renyi} for the conventions).
For $q=10$, the curves from $\Delta=2$ to $\Delta=4$ are almost coincident.}
\label{fig:renyi_dispersionplots}
\end{figure*}

To analyze this case, let us start with the following variational ansatz for the excitations   of $P_{2n}$ in the  sector with $\ket{\eta} ... \ket{\eta}$  towards the left boundary and  $\ket{e} ... \ket{e}$ states towards the right boundary\footnote{We look for eigenstates which asymptotically have this form so that they have non-negligible overlap with the final state  in the expression for $e^{-(n-1)S_n(x,t)}$.}  
\be 
\ket{\psi_k} \approx \sum_x e^{-ikx} \ket{\eta...\eta_x} \ket{\phi}_{x+1, ... x+\Delta} \ket{e_{x+\Delta+1}...e} \, ,  \label{psik_renyi}
\ee
 where $\ket{\phi}_{x+1, ... x+\Delta}$ is some state in the $(n!)^{\Delta}$-dimensional Hilbert space consisting of all possible permutation states on $\Delta$ sites.
Like in Sec. \ref{sec:gue},  we minimize the expectation value $\braket{\psi_k|P_{2n}|\psi_k}$ over all possible choices of $\ket{\phi}$ for increasing values of $\Delta$, and check whether the resulting estimate for the dispersion relation converges with $\Delta$.
We show the results of this procedure for $n=3$ with $\Delta$ from 0 to 4 in Fig.~\ref{fig:renyi_dispersionplots}.
We find rapid convergence for $\Delta\geq 2$ for all values of $q$.
Like in the case of $P_2$, the convergence with $\Delta$ is increasingly fast for larger $q$. We therefore learn that the lowest energy excitations in the relevant sector are well-approximated by \eqref{psik_renyi} for $O(1)$ $\Delta$. 

Let us now discuss the consequences of \eqref{psik_renyi} for the time-evolution of the $n$-th Renyi entropy given by \eqref{eq:renyi_avg}.
Here we are interested in the evolution of a domain wall state of the form $\bra{\eta...\eta_x e_{x+1}...e}$, which we will refer to as an \textit{entanglement domain wall}.
For general $n$, it is useful to define the domain wall propagator (similar to \eqref{g0def}) as 
\be 
G_n(x, y, t) = \braket{\eta...\eta_x e_{x+1}...e| e^{-P_{2n}t}|\bar \eta...\bar\eta_y \bar e_{y+1}...\bar e} \label{n_prop}
\ee
where $\ket{\bar\sigma}$ is defined such that (see App.~\ref{app:superhamhigher} for an explicit example with $n = 3$) 
\be 
\braket{\tau|\bar \sigma} = \delta_{\sigma \tau} \, . 
\ee
We assume that the $n$-th Renyi entropy in the scaling limit is  well-approximated by
\be 
e^{-(n-1)S_{n}(x,t)} = \sum_y G_n(x, y, t) e^{-(n-1)S_n(y, t=0)}\, , \label{nth_renyi} 
\ee
which can be justified by arguments analogous to those below (\ref{expansion}).
Putting \eqref{psik_renyi} and the corresponding dispersion relation $E_n(k)$ into \eqref{nth_renyi}, for an initial mixed state with entropy density $s$ for the $n$-th Renyi entropy, analogous to (\ref{fullint_2ren}), we find 
\begin{align} 
&e^{-(n-1)S_{n}(x,t)}=\nn
& e^{-(n-1)s \le(x+\frac{L}{2}\ri)}\int_{-\infty}^{\infty}dv\int_{-\pi}^{\pi} \frac{dk}{2\pi}~  e^{-[E_n(k)+ikv - (n-1)s v] t} 
\label{saddle_point}
\end{align}
The saddle-point equations for $v$ and $k$ are 
\be
k = - i(n-1) s~, \quad E'_n(k) = -i v \, 
\ee
which naively lead to the following growth rate for the $n$-th Renyi entropy: 
\be 
\bar \Gamma_n(s) = \frac{E_n\le(k= - i (n-1)s\ri)}{s_{\rm eq}(n-1)} \, .  \label{74}
\ee
An example of $\bar \Gamma_n(s)$ from \eqref{74} for $n=3$ and $q=8$ (with $\Delta=4$) is shown in Fig.~\ref{fig:gammasandthirdren}.
Again, we find this function by fitting $E_3(k)$ to the form \eqref{disp_fit}. 
The figure also shows $\Gamma_2(s)$ for the second Renyi entropy from the dispersion relation $E_2(k)$ for the same value of $q$.

\subsubsection{Unphysicality of naive analysis}
However, we find that $\bar \Gamma_3(s)$ cannot be the true growth rate.
We see that while $\Gamma_2(s)$ obeys the condition \eqref{gamma_condition} needed to ensure that $S_2(x,t)$ of the equilibrium state should not grow, the curve for $\bar \Gamma_3(s)$ appears to not obey this condition.
This implies it gives the unphysical prediction that the $S_3(x, t)$ of the maximally mixed equilibrium state increases rapidly.
By similar, reasoning to the discussion around \eqref{459}, we know that this prediction cannot be correct.

In the above discussion, we  made the approximation that in the expression for $e^{-(n-1)S_n}$, only the domain-wall eigenstates of $P_{2n}$, i.e., $\ket{\psi_k}$ of \eqref{psik_renyi} contributed, and this was sufficient for the analysis of the dynamics of entanglement domain walls for $n = 2$.
Assuming Eq.~(\ref{eq:assumption}), the only way to avoid the above unphysical conclusion should be that comparable contributions to $e^{-(n-1)S_n}$ must also come from a different set of energy eigenstates which we have not taken into account.
We now conjecture a possible structure of these other eigenstates which can give a simple resolution of the above issue. 
\subsubsection{Possible resolution: Scattering states}
As discussed above, for the $n$-th Renyi entropy we are interested in the dynamics of a domain wall with Cayley distance $n$, i.e., between the permutation degrees of freedom $\bra{e}$ and $\bra{\eta}$, which we refer to as the \textit{entanglement domain wall}. 
It is easy to show using Eq.~(\ref{p2n_gue}) that at large $q$ this entanglement domain wall splits into smaller domain walls with a smaller Cayley distance, see App.~\ref{app:superhamhigher} for the explicit expressions for $n = 3$. 
This splitting happens until we are ultimately left with a gas of domain walls, each between permutations of Cayley distance 1, which we will refer to as \textit{elementary domain walls}.

Recall from the discussion around \eqref{eq:P2naction} at the beginning of this section that elementary domain walls can still be thought of as low-energy excitations of $P_{2n}$ for $n>2$, although these eigenstates would have a very small overlap with the final state $\bra{\eta ... \eta_x e_{x+1}....e}$.
Then there is a natural guess for \textit{scattering states} formed by multiple  elementary domain walls, which could have significant overlap with the final state, which is a single entanglement domain wall.
These scattering states are expected to be of the form 
\begin{align} 
&\sum_{x_1 \leq ...\leq x_{n-1} }e^{-i(k_1x_1 + ... k_{n-1} x_{n-1})} \nn 
&\ket{\eta ... \eta_{x_1} \, \eta^{(n-1)}_{x_1+1}... \eta^{(n-1)}_{x_2} \, \eta^{(n-2)}_{x_2+1} ...\eta_{x_3}^{(n-2)} \, ... e_{x_{n-1}+1} ...e }\label{free_state}
\end{align}
where we have defined the permutation
\be \eta^{(k)}\equiv (k \, k-1\, ...\, 1).
\ee
We would further expect the energies of the states of the form \eqref{free_state} to be approximately $\sum_{i=1}^{n-1} E_2(k_i)$, where $E_2(k)$ is the dispersion relation found numerically for the second Renyi entropy in Sec.~\ref{sec:finite_q}.
For $n = 3$, we can check numerically that $E_3(k)$ obtained from the variational calculation in Fig. \ref{fig:renyi_dispersionplots} is smaller than the energies of these hypothetical states with free single-transposition domain walls: 
\be 
E_3(k)  < \text{min}_{k_1, k_2 ~\text{s.t.}~k_1+k_2 =k}(E_2(k_1)+E_2(k_2))\, . 
\ee
This is also visible from the expression for the ``bare energy" in the action of $P_{6}$ on these permutation states, see discussion around Eq.~(\ref{eq:eetaaction}) in App.~\ref{app:superhamhigher}.
The entanglement domain wall can be interpreted as a \textit{bound state} of the $n-1$ elementary domain walls, which have an effective attractive interaction between them.
A similar attractive attraction was also argued for using different techniques in random unitary circuits~\cite{zhou_nahum_statmech}.   
Even though the lowest excitations are these \textit{bound states}, scattering states of the form \eqref{free_state} may still be present as higher excited states in the spectrum of $P_{2n}$, and they may have significant overlap with the entanglement domain wall.
Such higher energy states can also contribute to the growth rate $\Gamma_n(s)$ for an initial mixed state with entropy density $s$, since these might contribute more in the saddle point evaluation of integrals such as those in Eq.~(\ref{fullint_2ren}).
We will assume this in the rest of the discussion and discuss a scenario that could resolve the unphysicality discussed in the previous section.\footnote{It would be challenging to numerically check that these states are present in the spectrum by exact diagonalization due to the large on-site Hilbert space dimension of the superhamiltonian, which allows us to access only small system sizes.
However, we comment on other methods that can be used to check this in the Discussion.}
\subsubsection{Route to a consistent membrane picture}
We now assume the existence of contributing scattering states, and conjecture a plausible scenario for how a consistent membrane picture could emerge.
Scattering states such as \eqref{free_state} such as would give a contribution $G_2(x, y, t)^n$ to the entanglement domain wall propagator $G_n(x, y, t)$, which would correspond to the elementary propagating freely and join back to form the entanglement domain wall.
\begin{figure*}[t]
    \centering
    \begin{tabular}{ccc}
    \includegraphics[width=0.33\textwidth]{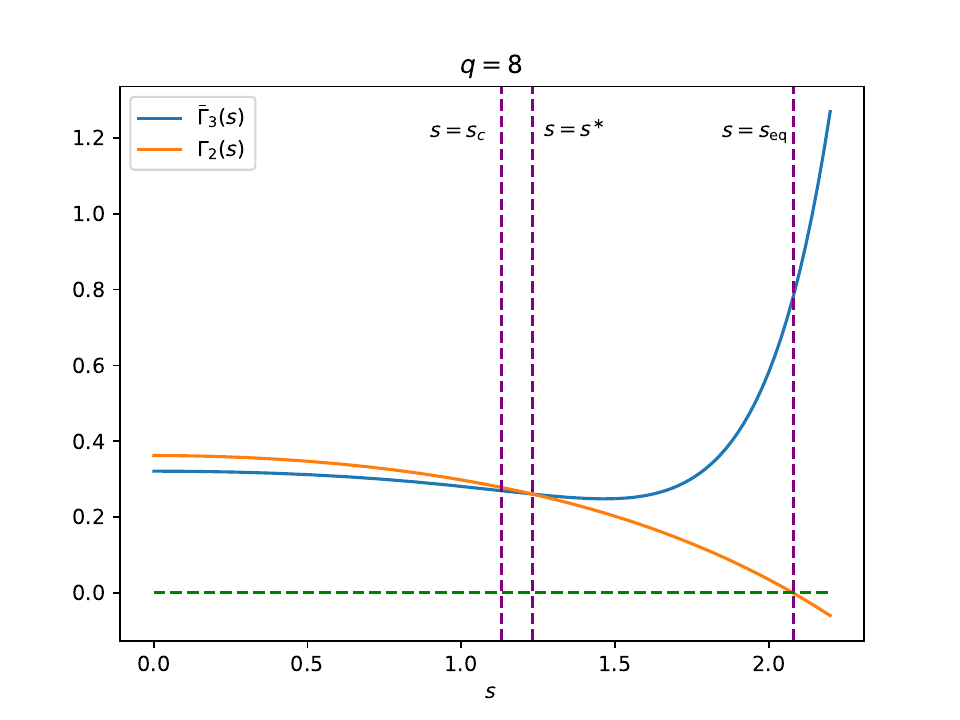}     &  \includegraphics[width=0.33\textwidth]{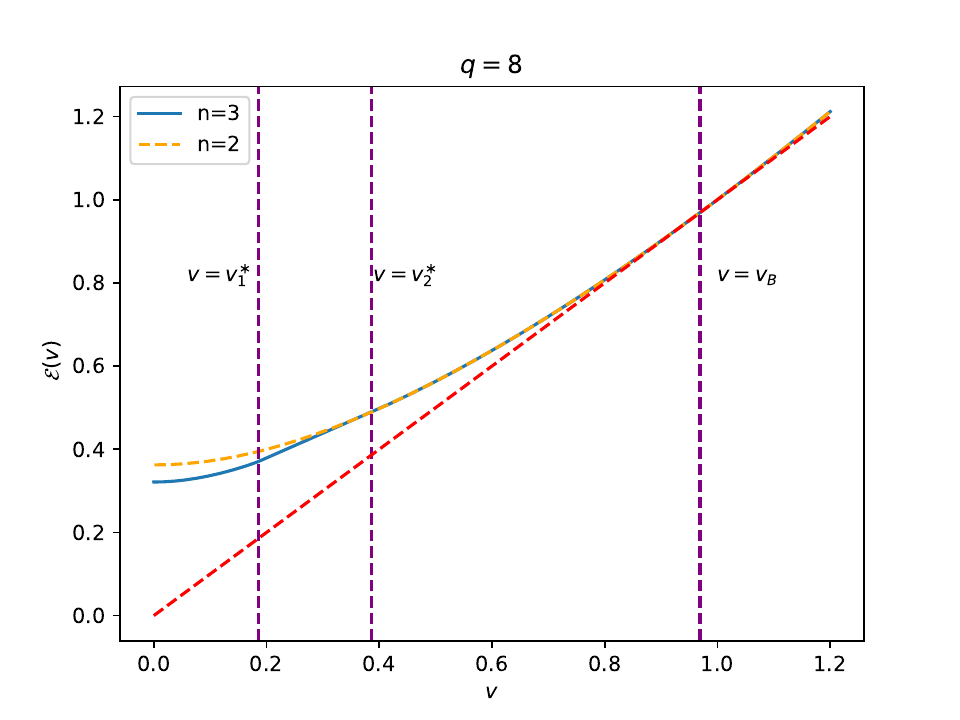}
    &
    \includegraphics[width=0.33\textwidth]{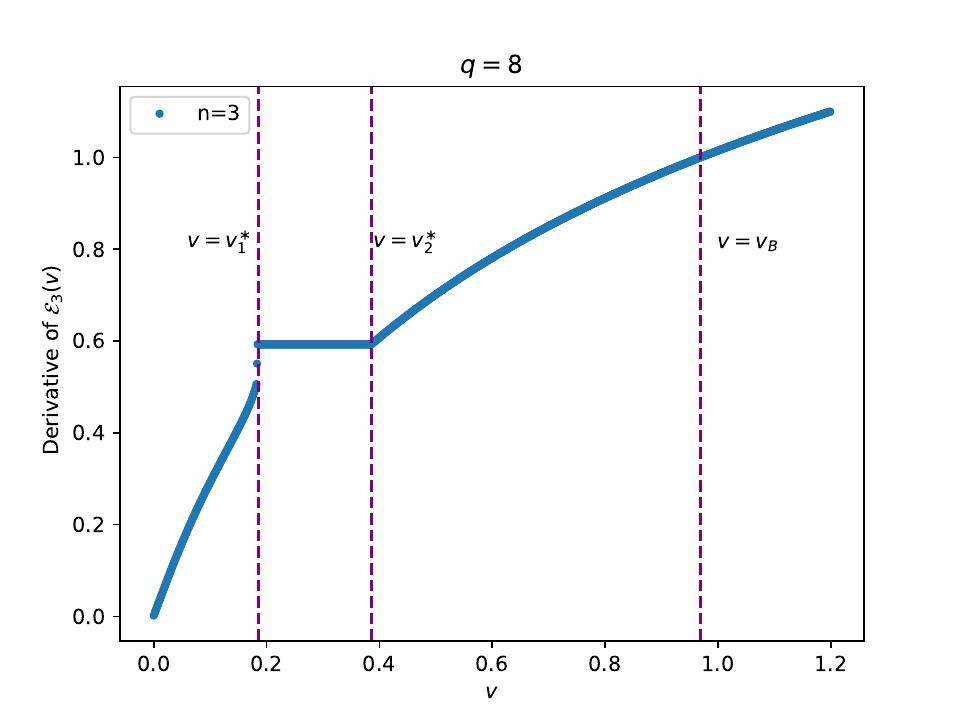}
    \end{tabular}
    \caption{(Left) Naive growth rate $\bar \Gamma_3(s)$ from \eqref{74} alongside $\Gamma_2(s)$ from the dispersion relation for $n=2$. The vertical lines mark the various significant values of $s$ discussed in the text, and the horizontal one marks $\Gamma(s)=0$. In particular, $\bar \Gamma_3(s_{\rm eq}) \neq 0$, which leads to an unphysical prediction of growth of $S_3(x,t)$ in an equilibrium state. 
    (Center) The solid blue curve shows $\sE_3(v)$ from \eqref{mem78} for $q=8$. $\sE_2(v)$ is also shown with the dashed orange curve for comparison.
    (Right) Derivative $\sE_3'(v)$, which has a discontinuity at $v=v_1^{\ast}$ and is continuous but not differentiable at $v=v_2^{\ast}$.
    }
    \label{fig:gammasandthirdren}
\end{figure*}
This would then be added to the contribution from the states \eqref{psik_renyi}, and putting this into the expression for $S_3$, we would get a sum of two terms: 
\be 
e^{-2S_3(x,t)} \sim e^{-2 \, s_{\rm eq}\,  \bar \Gamma_3(s) t} + e^{- 2\,  s_{\rm eq} \, \Gamma_2(s) \, t},
\ee
where we have assumed that the overall constants in front are of similar magnitude.
From the competition between the two terms, the true growth rate for the third Renyi entropy at late times is given by
\be 
\Gamma_3(s) = \text{min}\le(\, \bar \Gamma_3(s)\, , \, \Gamma_2(s)\, \ri). \label{g3min}
\ee
At the critical value $s=s^{\ast}$  where $\bar \Gamma_3(s)$ and $\Gamma_2(s)$  cross (see left panel of Fig.~\ref{fig:gammasandthirdren}), \eqref{g3min} implies a first-order phase transition in $\Gamma_3(s)$.
For $s>s^{\ast}$, $\Gamma_3(s) = \Gamma_2(s)$, so that in particular the constraint \eqref{gamma_condition} is satisfied. We show the particular case $q=8$, but we find that the curves cross at all values of $q$ that we checked, from $q=3$ to $q=10$. The ratio $s^{\ast}/s_{\rm eq}$ appears to increase monotonically with $q$.

Using the general formula \eqref{leg}, we can find the membrane tension $\sE_3(v)$ as the Legendre transform of $-\Gamma_3(s)$. 
Assuming \eqref{g3min}, there are three possible sources of the maximum value in the Legendre transform: it could come either from the region $0\leq s< s^{\ast}$, where $\Gamma_3(s) = \bar\Gamma_3(s)$, or from the endpoint at $s = s^{\ast}$, or from the region $s>s^{\ast}$.
We find three distinct regimes for the behaviour of $\sE_3(v)$ depending on which of  these options dominates: 
\be \label{mem78}
\sE_3(v) = \begin{cases} 
\bar \sE_3(v) & v \leq v^{\ast}_1 \\
\bar \Gamma_3(s_{\ast}) + \frac{s^{\ast}}{s_{\rm eq}}~v & v^{\ast}_{1} \leq v \leq v^{\ast}_{2} \\
\sE_2(v) & v\geq v^{\ast}_{2} 
\end{cases} 
\ee
Here $\bar\sE_3(v)$ is the Legendre transform of $-\bar \Gamma_3(s)$ restricted to the regime $s< s_c$ where it is convex~\footnote{Note that the Legendre transform of the full function $-\bar \Gamma_3(s)$ is not well-defined as it is not convex.}, and $s_c$ is the value of $s$ at which the second derivative of $\bar \Gamma_3$ changes from negative to positive. $\sE_2(v)$ is the membrane tension for the second Renyi entropy found previously. 
The two critical velocities are
\be 
v_1^{\ast}= -s_{\rm eq} \Gamma_3'(s_c), \quad
 v_2^{\ast}= - s_{\rm eq} \Gamma_2'(s^{\ast}) \, .
 \ee

In the center and right panels of Fig.~\ref{fig:gammasandthirdren}, we show the membrane tension of \eqref{mem78} obtained from $\bar\Gamma_3(s)$ and $\Gamma_2(s)$ shown in the left panel there, and its first derivative.
The membrane tension has a first-order phase transition at $v=v_1^{\ast}$, and a second-order phase transition at $v=v_2^{\ast}$. 
We now make a brief technical note on the difference of the analysis we used above from the $n=2$ case.
In this discussion of this section, we did not directly use the domain wall propagator \eqref{n_prop} to derive the membrane tension, but instead found it indirectly as the Legendre transform of $-\Gamma_3$.
The form of the dispersion relation $E_3(k)$ is such that the  domain wall propagator from the states \eqref{psik_renyi}, 
\be 
\bar{G}_3(x, y, t) = \int_{-\pi}^{\pi} \frac{dk}{2\pi}~  e^{-(E_3(k)+ikv ) t} \, , 
\ee
can be used to obtain the membrane tension $\bar\sE_3(v)$ only up to $v_1^{\ast}$ by similar steps to  \eqref{mem_0}.
For $v>v_1^{\ast}$, the propagator has an oscillating behaviour with time due to a change in the structure of the solutions to the saddle-point equation $E_3'(k)=-iv$.
The propagator in this regime $v>v_1^{\ast}$ does not contribute to  the evolution of the entropy in the combined  saddle-point analysis in the integral over $v$ and $k$ for $e^{-2S_3}$.
From the saddle-point equations \eqref{saddle_point}, we get a well-defined result for $\bar \Gamma_3$ for any $s$.
The Legendre transform of $-\bar\Gamma_3$ by itself is again not well-defined beyond $v_1^{\ast}$, but the Legendre transform of the negative of \eqref{g3min} is well-defined, as discussed above, and this should be seen as the definition of the membrane tension for this case.~\footnote{We thank Raghu Mahajan and Douglas Stanford for helpful discussions on the saddle point analysis.}

\begin{figure*}[!ht]
\includegraphics[scale=0.95]{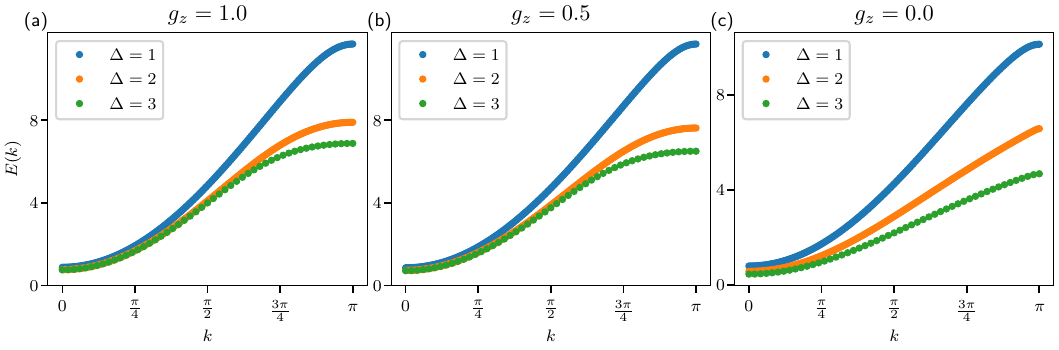}
    \caption{Dispersion relations obtained for various values of $g_Z$ from  minimizing $\braket{\psi_k|P_4|\psi_k}$ for $P_4$ corresponding to the couplings \eqref{halpha_ising} within the subspace spanned by the ansatz $\ket{\psi_k}$ of \eqref{phikdef}, with $\ket{\phi_{x+1, ..., x+\Delta}}$ now interpreted as an arbitrary state in the Hilbert space corresponding to four copies of the spin-1/2 Hilbert space of the original system on $\Delta$ sites. We find rapid convergence of $E(k)$ with $\Delta$ for (a) $g_Z = 1.0$, and (b) $0.5$ where we should expect the model to be chaotic, and slower convergence for (c) $g_Z=0$, where it is integrable.
    To check the convergence of the gaps $E_{\Delta}(k=0)$  for different $g_Z$, we can consider the ratio   $r(g_z) = \frac{E_{\Delta = 2}(k = 0)-E_{\Delta = 3}(k = 0)}{(E_{\Delta = 1}(k = 0)-E_{\Delta = 2}(k = 0)}$.
    We find  $r(g_z = 1) = 0.078$ , $r(g_Z = 0.5) = 0.101$, and $r(g_Z = 0) = 0.515$, giving better indication of convergence for $g_Z \neq 0$.}
\label{fig:general_brownian}
\end{figure*}
\subsubsection{Discussion}
In \cite{zhou_nahum_statmech}, the membrane tension for the third Renyi entropy was studied for random unitary circuits.
The calculations in that context, which were done in expansions for large $q$ and large $v$, also gave  indication of a phase transition in $\sE_3(v)$. The physical mechanism suggested for the ``unbinding'' phase transition there appears similar to the one discussed here. 
From the calculations in the present model at finite $q$ and general $v$, we see that there are two phase transitions in $\sE_3(v)$. Unlike in the large $q$ random circuit calculation of \cite{zhou_nahum_statmech}, where the critical velocity appeared to be $v_B$, we find that both critical velocities are smaller than $v_B$ at finite $q$. 
It is also important to note that another logical consequence of having scattering states contribute to $e^{-S_3(x, t)}$ is that the entanglement domain wall can evolve into two elementary domain walls that need not be close to each other. 
This would then mean that $e^{-S_3(x, t)}$ can also depend on more complicated measures of the initial states such as $\braket{e \cdots e_{x_1} \sigma_{x_1+1} \cdots \sigma_{x_2} \eta_{x_2+1} \cdots \eta|\rho_0, e}$, where $\sigma$ is a permutation with Cayley distance $1$ to both $e$ and $\eta$. 
This object in general cannot be expressed in terms of the entanglement entropies. 
This possibility is not consistent with the membrane picture for entanglement, which posits a simple formula of the form of (\ref{membrane_form}), where $S_3(x, t)$ depends \textit{only} on $S_3(y, 0)$. 
To establish the membrane picture more firmly for higher Renyi entropies, it would be important to rule out such possibilities and rigorously test the conjectures we have made here, which we leave for future work.

\section{Brownian models with fixed coupling operators}
\label{sec:fixed_op}

Let us now consider the family of models in \eqref{htdef}, where the coupling coefficients $\{J_{\alpha}(t)\}$ are random and uncorrelated, but the operators $\{B_{\alpha}\}$ are fixed and act on a system with a $q$-dimensional on-site Hilbert space.
This is a less random and hence more realistic example of a chaotic system than the one considered in Sec. \ref{sec:gue}.

The superhamiltonian \eqref{p2ndef} for this case still has the property that  any of its terms $P_{2n, \alpha}$ that has support on $m$ sites annihilates configurations of the form  $\ket{\sigma}^{\otimes m}$ on  those sites for any $\sigma \in \sS_n$.
This implies that the states \eqref{nth_gs_mt} are still ground states in this case.
In cases where the time-evolution has no symmetries, we expect that these are generically a complete basis for the ground state subspace, so that the saturation value of $S_{n,A}(t)$ is given by the Page value~\cite{page1993}, and $s_{\rm eq} = \log q$.
Unlike in the GUE case, the action of a general $P_{2n, \alpha}$ can now take a general initial state with different permutation states at different sites into the subspace orthogonal to all the permutation states, hence the effective Hilbert space on each site is now $q^{2n}$-dimensional, rather than $n!$ dimensional in the GUE case.
However, since the ground states are still of the ferromagnetic form of (\ref{nth_gs_mt}), it is natural to once again conjecture that the low-energy excitations are well-approximated by the structure in \eqref{eta_e_modes_mt} for $O(1)$ $\Delta$, with the state $\ket{\phi}_{x+1, ..., x+\Delta}$ now living in the full $(q^{2n})^{\Delta}$-dimensional Hilbert space.

Numerically, it is feasible to test the above conjecture using the variational technique of  Appendix \ref{app:var} up to $\Delta =3$ for $q=2$ and $n=2$, where the maximum Hilbert space dimension of the effective Hamiltonian for the variational problem (see Appendix \ref{app:var}) is $\lesssim 16^3 $.
For concreteness, let us take the set of coupling operators to be the following spin-$1/2$ operators in a one-dimensional system of $L$ sites:
\be \label{halpha_ising} 
\{B_{\alpha}\}= \{X_i,~ Z_i, ~ Z_{i}Z_{i+1}\}.
\ee
Taking the $g_{\alpha}$ in the variance of the couplings \eqref{jab_av}  to be some positive numbers $g_X, g_Z, g_{ZZ}$ respectively.
In the discussion below, we will fix $g_X = g_{ZZ}=1$, and consider a variety of different values of $g_Z$.
We expect this time-evolution to be chaotic for generic values of $g_Z$, except at the point $g_Z = 0$, where the time-dependent Hamiltonian is only a linear superposition of $\{X_i\}$ and $\{Z_i Z_{i+1}\}$ operators and hence has a  quadratic (non-interacting) Majorana fermion representation using the Jordan-Wigner transformation.

We show the results of the variational method for this family of models in 
Fig. \ref{fig:general_brownian}, which confirms the expectation that the eigenstates have the structure~\eqref{eta_e_modes_mt} for generic $g_Z$. 
The dispersion relation in the cases $g_Z = 1.0$ and $g_Z = 0.5$  converges rapidly. For $g_Z=0.0$, corresponding to the free fermion case, the dispersion relation
shows slower convergence.

The Brownian free Majorana fermion evolution corresponding to $g_Z=0$ was previously studied in \cite{nahum_freefermion}, where it was found that $P_4$ can be mapped to the ferromagnetic Heisenberg model.
The ground state subspace for this case is much larger than the one spanned by \eqref{eq:P4gs} (which can be attributed to so-called superoperator~\cite{lastres2024nonuniversality} or quadratic symmetries~\cite{zeier2011symmetry}), and it has  gapless low-energy excitations which lead to a growth of $S_2$ proportional to $\sqrt{t}$ rather than $t$.  Hence, the variational ansatz~\eqref{eta_e_modes_mt} is likely to not be a good approximation to the true low energy eigenstates for $g_z=0.0$, consistent with our observations.

The closing of the gap of $P_4$ (or equivalently vanishing of the entanglement velocity $v_E$) can be seen as a precise information-theoretic signature of a transition from chaotic to free-fermion integrable behaviour in this family of models.
Any non-zero $g_Z$ causes the model to lose its free fermion character and recover the  general features of chaotic many-body systems, hence opening up a gap in the thermodynamic limit.
At finite system size, we expect a crossover from chaotic to integrable behaviour at small $g_Z$, similar to the discussion in \cite{nahum_freefermion}.

\begin{figure}[t]
    \centering
    \includegraphics[width=0.5\textwidth]{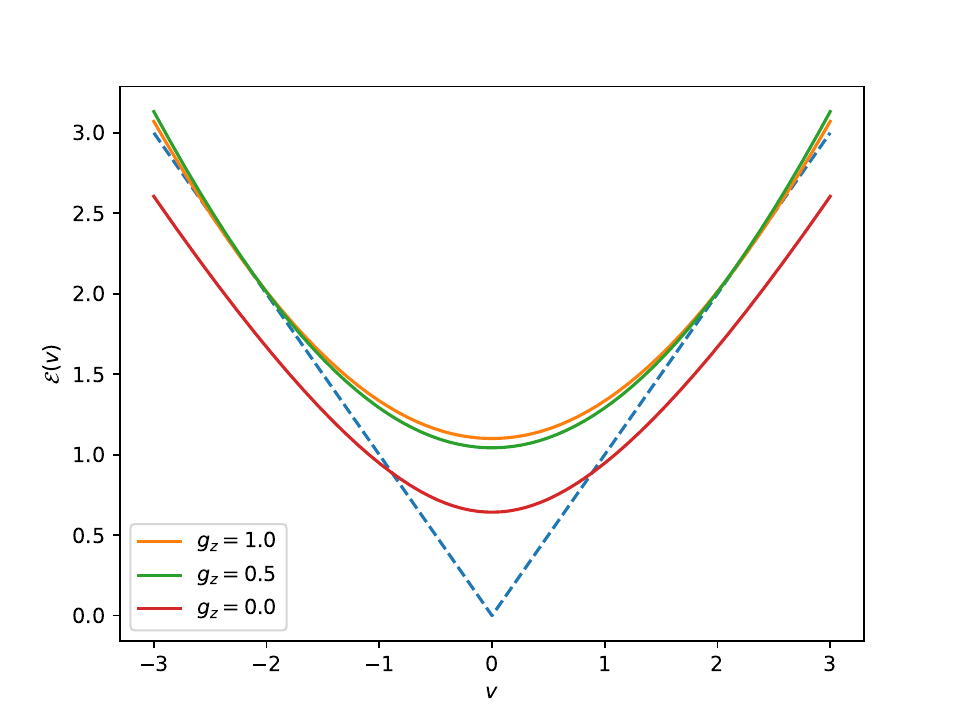}
    \caption{Membrane tensions for the second Renyi entropy for the model of \eqref{halpha_ising}, obtained from the  $\Delta=3$ dispersion relations of Fig. \ref{fig:general_brownian} using the fitting to \eqref{disp_fit}  with $N_{\rm max}=5$. The $g_Z =1, 0.5$ cases satisfy the constraints to a good approximation up to small numerical errors.  We do not expect the $g_Z = 0$ case to satisfy the constraints since the variational computation of the dispersion relation does not converge for the $\Delta$ we can access.}
    \label{fig:fixed_membrane}
\end{figure}
While we do not have an independent check of the dispersion relations from exact diagonalization of $P_4$ in the above  family of models (due to the large on-site Hilbert space dimension $q^4=16$), we  have performed the following two consistency checks, which verify that the dispersion relation is close to convergence for $g_Z=1.0, 0.5$ and far from convergence for $g_z=0.0$:
\begin{enumerate}
\item We directly evaluated the amplitude $e^{-S_2(x,t)} = \braket{D_x| e^{-P_4t}|\rho_0, e}$ using the  TEBD method for imaginary time evolution~\cite{schollwock2013matrix} for an initial pure product state $\rho_0$.
For generic values of $g_Z$, we find that this quantity has an exponential decay regime for a large range of times, corresponding to linear growth of the entropy.
This gives an independent calculation of the entanglement velocity $v_{E,2}$ defined in the first equation of \eqref{ve_disp}, which can be compared to the one found from the gap of the dispersion relation.
For $g_Z=1.0, 0.5$, we find good agreement up to about $10\%$ of the values, which are expected from the level of precision on both sides of the calculation. 
For $g_Z=0$, we do not see a linear growth regime from TEBD.
\item We find the membrane tensions from the $\Delta=3$ dispersion relations in Fig.~\ref{fig:fixed_membrane}.
$\sE(v)$ for $g_Z=0.5, 0.1$ approximately obeys the constraints \eqref{const}, while the $g_Z=0$ case does not. 
\end{enumerate}

\begin{figure*}[!ht]
\centering
\includegraphics[width=0.8\textwidth]{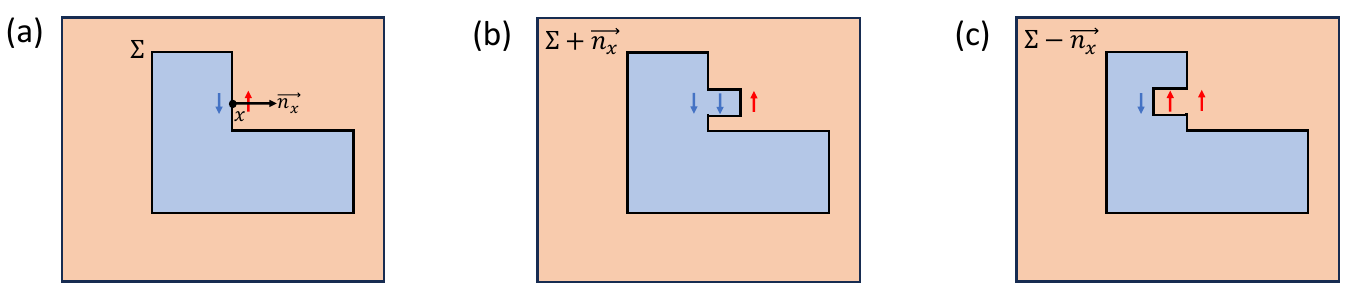}
\caption{In (a), we show we label an arbitrary point $x$ on a surface $\Sigma$ and show the definition of $\vec{n}_x$. In (b) and (c), we show the corresponding states $\bra{D_{\Sigma + \vec{n}_x}}$ and $\bra{D_{\Sigma - \vec{n}_x}}$ appearing on the RHS of \eqref{dsigma}.}
\label{fig:sigma_nx}
\end{figure*}

\begin{figure*}[!ht]
\includegraphics[width=0.8\textwidth]{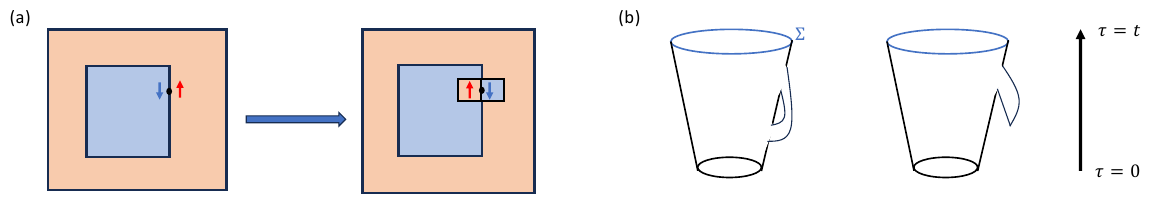}
\caption{In (a), we show an example of how a single domain wall splits under the acting of $A_1$ in $d=2$. In (b), we show some of the resulting  spacetime diagrams which should renormalize the membrane tension.}
\label{fig:splitting}
\end{figure*}
\section{Membrane picture in general spatial dimensions} \label{sec:eom}
Let us briefly discuss how the above picture generalizes to higher spatial dimensions $d\geq 2$.
It is useful to return to the simplest case of the second Renyi entropy in the Brownian GUE model, where the superhamiltonian $P_4$ in arbitrary dimensions is given by \eqref{aleft_def_mt} or (\ref{eq:PEspin_mt}).
To find the second Renyi entropy $S_{2}(\Sigma,t)$ of a region $R$ with boundary $\Sigma$, the final state in \eqref{eq:renyi_avg}  is a domain wall between $\bra{\down}$ and $\bra{\up}$ along $\Sigma$, as shown in Fig. \ref{fig:sigma_nx}(a), which we call $\bra{D_{\Sigma}}$.
As a first approximation in the large $q$ limit, let us again ignore the $A_1$ term of \eqref{aleft_def_mt}.
Even with this simplification, it is not  straightforward in  to diagonalize $A_0$ in the subspace relevant for the evolution of $\bra{D_{\Sigma}}$ for $d\geq 2$. However, we can obtain an expression for the evolution of $S_{2, \Sigma}$ for an infinitesimal time-step of length $\delta t$ by noting that 
\be 
\bra{D_{\Sigma}} A_0 = \sum_{x \in \Sigma} \left(  \bra{D_{\Sigma}} - \frac{1}{q} \left(  \bra{D_{\Sigma + \vec{n}_x}} + \bra{D_{\Sigma - \vec{n}_x}} \ri) \ri) \label{dsigma}
\ee 
where $\Sigma \pm \vec{n_x}$ correspond to inward or outward deformations of $\Sigma$ by a single lattice site in the  direction normal to the surface at $x$.\footnote{We assume that the  shape of the initial surface $\Sigma$ is such that it does not pinch off and split into two domain walls in time $\delta t$.}
See Fig. \ref{fig:sigma_nx} for an illustration in $d=2$.
Hence, close to a small segment of the domain wall, the dynamics in higher dimensions  resemble the one-dimensional dynamics in \eqref{dshift} in the direction normal to the surface.

Now by linearizing the evolution of $\sZ_{2, \Sigma}(t) = e^{-S_{2} (\Sigma, t)} = \braket{D_x|e^{-A_0  t} |\rho_0, e}$ for a short time $\delta t$, we find  
\be 
\frac{\partial  \sZ_{2, \Sigma}(t)}{\partial t}  = \sum_{x \in \Sigma} \left( \sZ_{2, \Sigma}(t) - \frac{1}{q} \left(\sZ_{2, \Sigma+\vec n_x}(t) + \sZ_{2, \Sigma- \vec n_x}(t)\ri) \ri) \label{discrete}
\ee
By taking continuum approximations for the differences and sums in the above expression, we obtain
\be 
\frac{\partial \sZ_{2, \Sigma}(t)}{\partial t} = -\int_{\Sigma} d^{d-1}x  \le[\le(1- \frac{2}{q}\ri) - \frac{1}{q} \partial_{n_x}^2  \ri]\sZ_{2, \Sigma}(t)   \label{z2_diff} 
\ee
For the second Renyi entropy, on ignoring a second derivative term $\partial^2_{n_x} S_2(\Sigma, t)$ which is negligible in the scaling limit of large system size and late time, \eqref{z2_diff} is equivalent to 
\be 
\frac{\partial S_2(\Sigma, t)}{ \partial t} = s_{\rm eq}\int_{\Sigma}  d^{d-1}x  \, \Gamma(s(x)) \label{gamma_int}
\ee
where
\be 
s(x) = \frac{\partial S_2 (\Sigma, t)}{\partial n_{x}} , \quad \Gamma(s) =\frac{1}{s_{\rm eq}}\le( 1 - \frac{2}{q} - \frac{1}{q} s^2\ri) 
\ee
As discussed in \cite{huse}, a differential equation of the form \eqref{gamma_int} is equivalent to the membrane formula in arbitrary dimensions  with $\sE(v)$ given by \eqref{leg}, which for this case is
\be \label{simp_tension}
\sE(v) = \frac{1}{s_{\rm eq}} \le( 1- \frac{2}{q} + \frac{q v^2}{4} \ri) \, . 
\ee 
independent of the spatial dimension $d$.
In particular, the above series of steps are valid in the case $d=1$, where on comparing to \eqref{mem_gue}, we see that we have obtained  the membrane tension we would get from taking first the large $q$ and then the small $v$  limit in $d=1$.
Recall that in order to satisfy the constraints \eqref{const}, the higher order terms in both $v$ and $1/q$ were important.
In order to go beyond the quadratic approximation in $v$, we would need to find an alternative to  taking the continuum limit in \eqref{discrete}.
The higher order corrections in $1/q$ would come from incorporating corrections from the term in the last line of \eqref{aleft_def_mt}, which causes domain walls to split as shown in Fig. \ref{fig:splitting}~(a), and gives rise to spacetime diagrams like those in Fig. \ref{fig:splitting}~(b) which should renormalize the membrane tension.
These are analogous to the diagrams for $d=1$ discussed in Appendix \ref{sec:int_picture}, which provided an alternative way of understanding corrections to the dispersion relation and membrane tension away from the large $q$.
It would be interesting to quantitatively incorporate the effects of such diagrams and see whether they lead to a dimension-dependent membrane tension, as observed in holographic CFTs \cite{mark_membrane}.  

Finally, we note that the superhamiltonian $P_4$ for the GUE model on any lattice in one or higher dimensions is the Peschel-Emery Hamiltonian of Eq.~(\ref{eq:PEspin_mt}) on that lattice. 
For large $q$, the low-energy physics of this Hamiltonian is expected to resemble that of the Transverse-Field Ising Model (TFIM) in the presence of a ``weak" field of $O(1/q)$, which is the Hamiltonian obtained by ignoring the $O(1/q^2)$ term in Eq.~(\ref{eq:PEspin_mt}). 
Indeed, as discussed in Sec.~\ref{subsec:secondrenyi}, many of the properties of $P_4$ in one dimension are similar to those of the one dimensional TFIM, e.g., the symmetry broken degenerate ground states and the structure of the low-energy domain wall excitations.
It is natural to expect similar connections in higher dimensions too at least for large $q$, i.e., the entanglement membrane picture in higher dimensions should be related to the low-energy physics of higher dimensional transverse field Ising models in a weak field. It would be interesting to explore this connection in future work.
\section{Conclusions and Discussion} \label{sec:discussion}
In this work, we have identified the microscopic mechanism which is responsible for the emergence of the membrane picture of entanglement dynamics in ``Brownian'' time-evolutions of the form \eqref{halpha_gen}.
The Lorentzian time-evolution of the Renyi entropies in such models can be mapped to a Euclidean evolution by a ``superhamiltonian'' living on multiple copies of the system.
The ground state of this superhamiltonian is determined by the symmetries of the system, and in evolutions without symmetries, which is what we have restricted to, this ground state manifold is two-fold degenerate, resembling an Ising ferromagnet.
The low-energy excitations on top of this ferromagnet are naturally of the form of dressed domain walls, and we have demonstrated that this structure leads to the membrane picture for the growth of second Renyi entropy in this class of evolutions. 
We demonstrated this in maximally random Brownian GUE models in one spatial dimension using analytical perturbative techniques and exact diagonalization, as well as variational numerical techniques for probing the low-energy eigenstates.
We also provided evidence that the same structure of low-energy modes appears in more generic examples  of Brownian models, independent of details of the interactions, as long as they have no symmetries.
The membrane tensions in all cases are determined by the dispersion relations of the low-energy modes, and we showed that they satisfy the expected physical constraints.
We further used the variational techniques to study the third Renyi entropy, and argued that in this case, the membrane tension exhibits a first order as well as a second order transition as a function of velocity. 
Collectively, these examples provide an understanding of how   universality emerges in the late-time dynamics of entanglement entropies in this class of quantum many-body systems. 
While the above derivation of the membrane formula from the microscopic dynamics is specific to this particular class of Brownian models, we find that it provides a number of useful, generalizable lessons.
First, while conventionally-used Haar random circuits allow for accurate computations of similar objects~\cite{vijay_randomcircuit, zhou_nahum_statmech}, a simple physical picture that should apply more broadly can be hard to extract from them.
On the other hand, these Brownian evolutions encompass a much larger class of systems that hinders exact solvability, but that very feature forces us to think in terms of objects that are \textit{universal} to these class of systems, which in this case are simply the low-energy modes of effective superhamiltonians.
These low-energy modes could also motivate a natural path towards building an {\it effective} theory  for more general chaotic systems, including  fixed Hamiltonians without any randomness.
Some motivation that the structure found here should extend to more general systems comes from considering the $n=1$ superhamiltonians (living on two copies of the system) in $U(1)$-conserving versions of these models, previously studied in  \cite{moudgalya2023symmetries, ogunnaike2023unifying}.
These papers found that the diffusive behaviour of the two-point functions of the charge density comes from certain gapless low-energy eigenstates of the $n=1$ superhamiltonian.
These gapless excitations can be seen as a precise realization of the hydrodynamic modes which are expected to govern diffusive behaviour of two-point functions much more generally.
Hence, we can also expect that the entanglement modes found in this work are likely to also exist in more general systems, consistent with the expectation from~\cite{zhou_nahum}. 
Alternative approaches which start from the assumption that such gapless modes exist have previously been used to construct effective field theories of hydrodynamics~\cite{crossley, rangamani}. 
It is then natural to ask whether one can formulate an effective field theory for entanglement dynamics based on the set of low-lying modes living on four or more copies that we found in this  work.
Note that any such effective field theory would need to contain several new elements relative to those developed in earlier works~\cite{rangamani,crossley}.
The modes leading to the membrane picture differ from hydrodynamic modes in that they  are gapped~\footnote{The fact that the modes are gapped is needed to ensure exponential decay of $e^{-(n-1)S_n}$, or equivalently linear growth of the entropies.}, and also exist without any symmetries of the system, as we have demonstrated in this work. 
Moreover, even above the gap, the dispersion relation at $O(1)$ values of the momentum $k$ is physically relevant for obtaining the correct dynamics of highly entangled initial states.
In particular, since $s_{\rm eq}=O(1)$, a small $k$ expansion of the dispersion relation will not satisfy the condition \eqref{econd};  this UV sensitivity suggests that the conventional approach of derivative expansions might not be the right way to formulate the relevant EFT if we want it to capture the correct physics for all kinds of initial states and for all times.
In holographic theories, the modes that lead to diffusion in the boundary conformal field theory correspond to quasinormal modes in the bulk gravity dual~\cite{horowitz}.
Based on the results of this work, it is natural to ask what the analog of these quasinormal modes is for observables like the $n$-th Renyi entropy and the von Neumann entropy.
It would be interesting to identify such modes in the replicated geometries for the $n$-th Renyi entropy in the prescription of~\cite{rangamani1, rangamani2}, as well as their analytic continuation to $n \to 1$.
In the context of higher Renyi entropies, it would be interesting to better understand the transition that we found in $\sE_3(v)$ as a function of $v$.
At a technical level, it would be good use the methods of~\cite{scattering_particles} for probing multi-quasiparticle excitations to check the existence of the excited states \eqref{free_state}, which is assumed in obtaining our expression \eqref{g3min}.
At a conceptual level, there are interesting questions about what this transition means information-theoretically.
In particular, beyond $v_{2}^{\ast}$, we find that $\sE_3(v)$ becomes equal to $\sE_2(v)$, which in particular implies that $v_B$ defined through \eqref{const} is the same for $n=2$ and $n=3$.
Is there a simple physical reason why the butterfly velocity according to this definition should be the same for all the Renyi entropies?
It would also be good to collect more data on such phase transitions in more general Brownian models and for $n\geq 4$. This would  require technical improvements to the variational methods used in this work, for instance by combining them with a matrix product state (MPS) ansatz (see \cite{mps_review} for a review). 
The physical existence of these transitions in generic non-Brownian models or time-independent systems should also be verified more carefully, for example using direct computations of the entropies.

The methods developed in~\cite{scattering_particles} and their extensions could also be useful for a number of related questions.
Recall that in the large $q$ limit, to show the membrane picture for multiple intervals even for the second Renyi entropy, we used the fact that multiple domain walls were essentially non-interacting in this limit.
For the membrane picture for multiple intervals to be robust at finite $q$ and in more general cases, the quasiparticles consisting of dressed domain walls should be weakly interacting, which can be checked using such numerical methods.
Higher dimensional generalizations of these methods could also be useful in understanding the physical picture for entanglement growth in higher-dimensional systems, which were discussed in Section \ref{sec:eom}, although the fact that domain-walls are no longer particle-like would change the physics significantly in higher dimensions.
It is also important to understand how the physical picture of this work is modified in cases where the system has conservation laws.
The simplest such case one can consider is to take Brownian models like the ones studied in this work with an additional $U(1)$ conservation law.
Renyi entropies in systems with $U(1)$ conservation are said to show a $\sim \sqrt{t}$ growth~\cite{rakovszky2018diffusive, huang_2020, zhou2020diffusive}. However, it is not yet clear  whether these results hold universally, and what the underlying mechanism is~\cite{znidaric2020entanglement, rakovszky2021entanglement}.
In the superhamiltonian for this case,  there is an interesting and rather intricate coupling between the domain wall degrees of freedom discussed in this work, and the hydrodynamic degrees of freedom associated with the conserved charge density.
We plan to report these on results in upcoming work.
This interplay between the entanglement domain walls and hydrodynamic modes can also be explored for a variety of other symmetries, such as dipole or multipole conservation~\cite{guardado2020subdiffusion, feldmeier2020anomalous,  moudgalya2021spectral}, or other unconventional symmetries such as quantum many-body scars~\cite{moudgalya2022exhaustive, gotta2023asymptotic} and Hilbert space fragmentation~\cite{moudgalya2022from, feldmeier2022emergent}.
Hydrodynamic modes for these unconventional symmetries  were derived using the $n = 1$ superhamiltonian in \cite{ogunnaike2023unifying, moudgalya2023symmetries}.
The superhamiltonians found in this work govern the evolution not only of the Renyi entropies, but also of other interesting dynamical quantities that probe chaos and thermalization.
In particular, the out-of-time-ordered correlator (OTOC) can also be written as a transition amplitude under $P_4$:  
\begin{align}
&{\rm OTOC}(x, y, t) \equiv \Tr[V_x(t) W_y^{\dagger} V_x(t) W_y^{\dagger}] \nn 
&= \braket{W_y,  \eta |e^{-\,P_4\,t}| V_x, e }  \, . 
\end{align}
where we use the notation of Sec.~\ref{sec:setup}, and $x$ and $y$ are the spatial locations of the two operators. 
By putting the  eigenstates of $P_4$ that we found in \eqref{eta_e_modes_mt} into this expression, we obtain the  following expression for the OTOC, previously noted  in \cite{zhou_nahum, velocity_dependent} (assuming that $y>x$, and after an average over unitary rotations of the operators $V_x, W_y$, which allows us to write the boundary conditions for this quantity in terms of $\ket{\up}, \ket{\down}$~\cite{zhou_nahum}):
\be
{\rm OTOC}(x, y, t) = \sum_{0\leq v\leq \frac{y-x}{t}} e^{-s_{\rm eq}(\sE_2(v) -v)t}  \label{vsum}
\ee

Using the constraints $\sE_2(v_B)=v_B, \sE_2(v) \geq v$, we see that if $t<(y-x)/v_B$, then $v_B$ is included in the sum over $v$, so that ${\rm OTOC}(x, y, t)\approx 1$.
If $t>(y-x)/v_B$, then the OTOC starts to decay, and is dominated by the endpoint contribution from $v=(y-x)/t$ in \eqref{vsum} due to the convexity of $\sE_2(v)$. 
By expanding close to $v=v_B$ and using \eqref{const}, we find the diffusive broadening of the operator front first noted in \cite{rak_randomcircuit, nahum_randomcircuit}: 
\be 
{\rm OTOC}(t) \approx \exp\le({-\frac{1}{2}\, \sE_2''(v_B)\,\frac{(y-x -v_Bt)^2}{t}}\ri) \label{diff}
\ee

A particular case where this structure should apply is in the  Brownian SYK model~\cite{brownian_syk, sunderhauf2019quantum}.
In this model, there is an alternative effective description in a certain large $N$ limit in terms of collective degrees of freedom known as ``scramblons.'' The diffusive broadening \eqref{diff} is very non-trivial to derive from the Feynman diagrams of the scramblon field theory~\cite{scramblon_loops}.
It would be interesting to understand how the effective modes found in the present work emerge from the scramblons in the appropriate late-time limit. 
Finally, it would be interesting to understand if the modes discussed in this work provide insights into  entanglement dynamics in a variety of other situations, such as systems with long-range interactions~\cite{schuckert2020nonlocal, ogunnaike2023unifying, morningstar2023hydrodynamics, gliozzi2023hierarchical}, Clifford circuits~\cite{sierant2023entanglement, sommers2024zero}, dual unitary circuits~\cite{, rampp2023entanglement}, or in the presence of measurements~\cite{potter2022entanglement}.
\section*{Acknowledgements}
We thank Henry Lin and  Tibor Rakovszky for early collaboration on this work, and Alessio Lerose for a related ongoing collaboration.
We are grateful to Hosho Katsura, Michael Knap, Shota Komatsu, Marco Lastres, Hong Liu,  Raghu Mahajan, Mark Mezei, Lesik Motrunich, Frank Pollmann, Xiaoliang Qi, Douglas Stanford, and Caterina Zerba for useful discussions, Raghu Mahajan for helpful comments on the draft, and  Frank Pollmann for helpful numerical suggestions.
SM thanks Lesik Motrunich for collaboration on a previous work~\cite{moudgalya2023symmetries}, and Marco Lastres and Frank Pollmann for collaboration on \cite{lastres2024nonuniversality}.
SV is supported by Google, and SM from the Munich Center for Quantum Science and Technology (MCQST) and the Deutsche Forschungsgemeinschaft (DFG, German Research Foundation) under Germany’s Excellence Strategy--EXC--2111--390814868. 
We acknowledge the hospitality of the Yukawa Institute of Theoretical Physics (YITP), Kyoto during the workshop “Quantum Information, Quantum
Matter and Quantum Gravity” (YITP-T-23-01), where this work was initiated.
\bibliography{newrefs}

\begin{thebibliography}{92}%
\makeatletter
\providecommand \@ifxundefined [1]{%
 \@ifx{#1\undefined}
}%
\providecommand \@ifnum [1]{%
 \ifnum #1\expandafter \@firstoftwo
 \else \expandafter \@secondoftwo
 \fi
}%
\providecommand \@ifx [1]{%
 \ifx #1\expandafter \@firstoftwo
 \else \expandafter \@secondoftwo
 \fi
}%
\providecommand \natexlab [1]{#1}%
\providecommand \enquote  [1]{``#1''}%
\providecommand \bibnamefont  [1]{#1}%
\providecommand \bibfnamefont [1]{#1}%
\providecommand \citenamefont [1]{#1}%
\providecommand \href@noop [0]{\@secondoftwo}%
\providecommand \href [0]{\begingroup \@sanitize@url \@href}%
\providecommand \@href[1]{\@@startlink{#1}\@@href}%
\providecommand \@@href[1]{\endgroup#1\@@endlink}%
\providecommand \@sanitize@url [0]{\catcode `\\12\catcode `\$12\catcode
  `\&12\catcode `\#12\catcode `\^12\catcode `\_12\catcode `\%12\relax}%
\providecommand \@@startlink[1]{}%
\providecommand \@@endlink[0]{}%
\providecommand \url  [0]{\begingroup\@sanitize@url \@url }%
\providecommand \@url [1]{\endgroup\@href {#1}{\urlprefix }}%
\providecommand \urlprefix  [0]{URL }%
\providecommand \Eprint [0]{\href }%
\providecommand \doibase [0]{https://doi.org/}%
\providecommand \selectlanguage [0]{\@gobble}%
\providecommand \bibinfo  [0]{\@secondoftwo}%
\providecommand \bibfield  [0]{\@secondoftwo}%
\providecommand \translation [1]{[#1]}%
\providecommand \BibitemOpen [0]{}%
\providecommand \bibitemStop [0]{}%
\providecommand \bibitemNoStop [0]{.\EOS\space}%
\providecommand \EOS [0]{\spacefactor3000\relax}%
\providecommand \BibitemShut  [1]{\csname bibitem#1\endcsname}%
\let\auto@bib@innerbib\@empty
\bibitem [{\citenamefont {{Banks}}\ \emph {et~al.}(1998)\citenamefont
  {{Banks}}, \citenamefont {{Douglas}}, \citenamefont {{Horowitz}},\ and\
  \citenamefont {{Martinec}}}]{banks}%
  \BibitemOpen
  \bibfield  {author} {\bibinfo {author} {\bibfnamefont {T.}~\bibnamefont
  {{Banks}}}, \bibinfo {author} {\bibfnamefont {M.~R.}\ \bibnamefont
  {{Douglas}}}, \bibinfo {author} {\bibfnamefont {G.~T.}\ \bibnamefont
  {{Horowitz}}},\ and\ \bibinfo {author} {\bibfnamefont {E.}~\bibnamefont
  {{Martinec}}},\ }\bibfield  {title} {\bibinfo {title} {{AdS Dynamics from
  Conformal Field Theory}},\ }\href@noop {} {\bibfield  {journal} {\bibinfo
  {journal} {arXiv e-prints}\ } (\bibinfo {year} {1998})},\ \Eprint
  {https://arxiv.org/abs/hep-th/9808016} {arXiv:hep-th/9808016 [hep-th]}
  \BibitemShut {NoStop}%
\bibitem [{\citenamefont {{Balasubramanian}}\ \emph {et~al.}(1999)\citenamefont
  {{Balasubramanian}}, \citenamefont {{Kraus}}, \citenamefont {{Lawrence}},\
  and\ \citenamefont {{Trivedi}}}]{trivedi}%
  \BibitemOpen
  \bibfield  {author} {\bibinfo {author} {\bibfnamefont {V.}~\bibnamefont
  {{Balasubramanian}}}, \bibinfo {author} {\bibfnamefont {P.}~\bibnamefont
  {{Kraus}}}, \bibinfo {author} {\bibfnamefont {A.}~\bibnamefont
  {{Lawrence}}},\ and\ \bibinfo {author} {\bibfnamefont {S.~P.}\ \bibnamefont
  {{Trivedi}}},\ }\bibfield  {title} {\bibinfo {title} {{Holographic probes of
  anti-de Sitter spacetimes}},\ }\href
  {https://doi.org/10.1103/PhysRevD.59.104021} {\bibfield  {journal} {\bibinfo
  {journal} {\prd}\ }\textbf {\bibinfo {volume} {59}},\ \bibinfo {eid} {104021}
  (\bibinfo {year} {1999})},\ \Eprint {https://arxiv.org/abs/hep-th/9808017}
  {arXiv:hep-th/9808017 [hep-th]} \BibitemShut {NoStop}%
\bibitem [{\citenamefont {{Page}}(1993)}]{page}%
  \BibitemOpen
  \bibfield  {author} {\bibinfo {author} {\bibfnamefont {D.~N.}\ \bibnamefont
  {{Page}}},\ }\bibfield  {title} {\bibinfo {title} {{Average entropy of a
  subsystem}},\ }\href {https://doi.org/10.1103/PhysRevLett.71.1291} {\bibfield
   {journal} {\bibinfo  {journal} {\prl}\ }\textbf {\bibinfo {volume} {71}},\
  \bibinfo {pages} {1291} (\bibinfo {year} {1993})},\ \Eprint
  {https://arxiv.org/abs/gr-qc/9305007} {arXiv:gr-qc/9305007 [gr-qc]}
  \BibitemShut {NoStop}%
\bibitem [{\citenamefont {{Lubkin}}(1978)}]{lubkin}%
  \BibitemOpen
  \bibfield  {author} {\bibinfo {author} {\bibfnamefont {E.}~\bibnamefont
  {{Lubkin}}},\ }\bibfield  {title} {\bibinfo {title} {{Entropy of an n-system
  from its correlation with a k-reservoir}},\ }\href
  {https://doi.org/10.1063/1.523763} {\bibfield  {journal} {\bibinfo  {journal}
  {Journal of Mathematical Physics}\ }\textbf {\bibinfo {volume} {19}},\
  \bibinfo {pages} {1028} (\bibinfo {year} {1978})}\BibitemShut {NoStop}%
\bibitem [{\citenamefont {{Lloyd}}\ and\ \citenamefont
  {{Pagels}}(1988)}]{pagels}%
  \BibitemOpen
  \bibfield  {author} {\bibinfo {author} {\bibfnamefont {S.}~\bibnamefont
  {{Lloyd}}}\ and\ \bibinfo {author} {\bibfnamefont {H.}~\bibnamefont
  {{Pagels}}},\ }\bibfield  {title} {\bibinfo {title} {{Complexity as
  thermodynamic depth}},\ }\href {https://doi.org/10.1016/0003-4916(88)90094-2}
  {\bibfield  {journal} {\bibinfo  {journal} {Annals of Physics}\ }\textbf
  {\bibinfo {volume} {188}},\ \bibinfo {pages} {186} (\bibinfo {year}
  {1988})}\BibitemShut {NoStop}%
\bibitem [{\citenamefont {{Nadal}}\ \emph {et~al.}(2011)\citenamefont
  {{Nadal}}, \citenamefont {{Majumdar}},\ and\ \citenamefont
  {{Vergassola}}}]{nadal}%
  \BibitemOpen
  \bibfield  {author} {\bibinfo {author} {\bibfnamefont {C.}~\bibnamefont
  {{Nadal}}}, \bibinfo {author} {\bibfnamefont {S.~N.}\ \bibnamefont
  {{Majumdar}}},\ and\ \bibinfo {author} {\bibfnamefont {M.}~\bibnamefont
  {{Vergassola}}},\ }\bibfield  {title} {\bibinfo {title} {{Statistical
  Distribution of Quantum Entanglement for a Random Bipartite State}},\ }\href
  {https://doi.org/10.1007/s10955-010-0108-4} {\bibfield  {journal} {\bibinfo
  {journal} {Journal of Statistical Physics}\ }\textbf {\bibinfo {volume}
  {142}},\ \bibinfo {pages} {403} (\bibinfo {year} {2011})},\ \Eprint
  {https://arxiv.org/abs/1006.4091} {arXiv:1006.4091 [cond-mat.stat-mech]}
  \BibitemShut {NoStop}%
\bibitem [{\citenamefont {{Leutheusser}}\ and\ \citenamefont {{Van
  Raamsdonk}}(2017)}]{leut}%
  \BibitemOpen
  \bibfield  {author} {\bibinfo {author} {\bibfnamefont {S.}~\bibnamefont
  {{Leutheusser}}}\ and\ \bibinfo {author} {\bibfnamefont {M.}~\bibnamefont
  {{Van Raamsdonk}}},\ }\bibfield  {title} {\bibinfo {title} {{Tensor network
  models of unitary black hole evaporation}},\ }\href
  {https://doi.org/10.1007/JHEP08(2017)141} {\bibfield  {journal} {\bibinfo
  {journal} {Journal of High Energy Physics}\ }\textbf {\bibinfo {volume}
  {2017}},\ \bibinfo {eid} {141} (\bibinfo {year} {2017})},\ \Eprint
  {https://arxiv.org/abs/1611.08613} {arXiv:1611.08613 [hep-th]} \BibitemShut
  {NoStop}%
\bibitem [{\citenamefont {Liu}\ and\ \citenamefont
  {Vardhan}(2021)}]{eq_approx}%
  \BibitemOpen
  \bibfield  {author} {\bibinfo {author} {\bibfnamefont {H.}~\bibnamefont
  {Liu}}\ and\ \bibinfo {author} {\bibfnamefont {S.}~\bibnamefont {Vardhan}},\
  }\bibfield  {title} {\bibinfo {title} {Entanglement entropies of equilibrated
  pure states in quantum many-body systems and gravity},\ }\href
  {https://doi.org/10.1103/PRXQuantum.2.010344} {\bibfield  {journal} {\bibinfo
   {journal} {PRX Quantum}\ }\textbf {\bibinfo {volume} {2}},\ \bibinfo {pages}
  {010344} (\bibinfo {year} {2021})}\BibitemShut {NoStop}%
\bibitem [{\citenamefont {{Kim}}\ and\ \citenamefont
  {{Huse}}(2013)}]{kim_huse}%
  \BibitemOpen
  \bibfield  {author} {\bibinfo {author} {\bibfnamefont {H.}~\bibnamefont
  {{Kim}}}\ and\ \bibinfo {author} {\bibfnamefont {D.~A.}\ \bibnamefont
  {{Huse}}},\ }\bibfield  {title} {\bibinfo {title} {{Ballistic Spreading of
  Entanglement in a Diffusive Nonintegrable System}},\ }\href
  {https://doi.org/10.1103/PhysRevLett.111.127205} {\bibfield  {journal}
  {\bibinfo  {journal} {\prl}\ }\textbf {\bibinfo {volume} {111}},\ \bibinfo
  {eid} {127205} (\bibinfo {year} {2013})},\ \Eprint
  {https://arxiv.org/abs/1306.4306} {arXiv:1306.4306 [quant-ph]} \BibitemShut
  {NoStop}%
\bibitem [{\citenamefont {{Liu}}\ and\ \citenamefont {{Suh}}(2014)}]{liu_suh}%
  \BibitemOpen
  \bibfield  {author} {\bibinfo {author} {\bibfnamefont {H.}~\bibnamefont
  {{Liu}}}\ and\ \bibinfo {author} {\bibfnamefont {S.~J.}\ \bibnamefont
  {{Suh}}},\ }\bibfield  {title} {\bibinfo {title} {{Entanglement Tsunami:
  Universal Scaling in Holographic Thermalization}},\ }\href
  {https://doi.org/10.1103/PhysRevLett.112.011601} {\bibfield  {journal}
  {\bibinfo  {journal} {\prl}\ }\textbf {\bibinfo {volume} {112}},\ \bibinfo
  {eid} {011601} (\bibinfo {year} {2014})},\ \Eprint
  {https://arxiv.org/abs/1305.7244} {arXiv:1305.7244 [hep-th]} \BibitemShut
  {NoStop}%
\bibitem [{\citenamefont {{Hartman}}\ and\ \citenamefont
  {{Maldacena}}(2013)}]{hartman_maldacena}%
  \BibitemOpen
  \bibfield  {author} {\bibinfo {author} {\bibfnamefont {T.}~\bibnamefont
  {{Hartman}}}\ and\ \bibinfo {author} {\bibfnamefont {J.}~\bibnamefont
  {{Maldacena}}},\ }\bibfield  {title} {\bibinfo {title} {{Time evolution of
  entanglement entropy from black hole interiors}},\ }\href
  {https://doi.org/10.1007/JHEP05(2013)014} {\bibfield  {journal} {\bibinfo
  {journal} {Journal of High Energy Physics}\ }\textbf {\bibinfo {volume}
  {2013}},\ \bibinfo {eid} {14} (\bibinfo {year} {2013})},\ \Eprint
  {https://arxiv.org/abs/1303.1080} {arXiv:1303.1080 [hep-th]} \BibitemShut
  {NoStop}%
\bibitem [{\citenamefont {{Jonay}}\ \emph {et~al.}(2018)\citenamefont
  {{Jonay}}, \citenamefont {{Huse}},\ and\ \citenamefont {{Nahum}}}]{huse}%
  \BibitemOpen
  \bibfield  {author} {\bibinfo {author} {\bibfnamefont {C.}~\bibnamefont
  {{Jonay}}}, \bibinfo {author} {\bibfnamefont {D.~A.}\ \bibnamefont
  {{Huse}}},\ and\ \bibinfo {author} {\bibfnamefont {A.}~\bibnamefont
  {{Nahum}}},\ }\bibfield  {title} {\bibinfo {title} {{Coarse-grained dynamics
  of operator and state entanglement}},\ }\href@noop {} {\bibfield  {journal}
  {\bibinfo  {journal} {arXiv e-prints}\ } (\bibinfo {year} {2018})},\ \Eprint
  {https://arxiv.org/abs/1803.00089} {arXiv:1803.00089 [cond-mat.stat-mech]}
  \BibitemShut {NoStop}%
\bibitem [{\citenamefont {{Nahum}}\ \emph {et~al.}(2017)\citenamefont
  {{Nahum}}, \citenamefont {{Ruhman}}, \citenamefont {{Vijay}},\ and\
  \citenamefont {{Haah}}}]{vijay_randomcircuit}%
  \BibitemOpen
  \bibfield  {author} {\bibinfo {author} {\bibfnamefont {A.}~\bibnamefont
  {{Nahum}}}, \bibinfo {author} {\bibfnamefont {J.}~\bibnamefont {{Ruhman}}},
  \bibinfo {author} {\bibfnamefont {S.}~\bibnamefont {{Vijay}}},\ and\ \bibinfo
  {author} {\bibfnamefont {J.}~\bibnamefont {{Haah}}},\ }\bibfield  {title}
  {\bibinfo {title} {{Quantum Entanglement Growth under Random Unitary
  Dynamics}},\ }\href {https://doi.org/10.1103/PhysRevX.7.031016} {\bibfield
  {journal} {\bibinfo  {journal} {Physical Review X}\ }\textbf {\bibinfo
  {volume} {7}},\ \bibinfo {eid} {031016} (\bibinfo {year} {2017})},\ \Eprint
  {https://arxiv.org/abs/1608.06950} {arXiv:1608.06950 [cond-mat.stat-mech]}
  \BibitemShut {NoStop}%
\bibitem [{\citenamefont {{Nahum}}\ \emph {et~al.}(2018)\citenamefont
  {{Nahum}}, \citenamefont {{Vijay}},\ and\ \citenamefont
  {{Haah}}}]{nahum_randomcircuit}%
  \BibitemOpen
  \bibfield  {author} {\bibinfo {author} {\bibfnamefont {A.}~\bibnamefont
  {{Nahum}}}, \bibinfo {author} {\bibfnamefont {S.}~\bibnamefont {{Vijay}}},\
  and\ \bibinfo {author} {\bibfnamefont {J.}~\bibnamefont {{Haah}}},\
  }\bibfield  {title} {\bibinfo {title} {{Operator Spreading in Random Unitary
  Circuits}},\ }\href {https://doi.org/10.1103/PhysRevX.8.021014} {\bibfield
  {journal} {\bibinfo  {journal} {Physical Review X}\ }\textbf {\bibinfo
  {volume} {8}},\ \bibinfo {eid} {021014} (\bibinfo {year} {2018})},\ \Eprint
  {https://arxiv.org/abs/1705.08975} {arXiv:1705.08975 [cond-mat.str-el]}
  \BibitemShut {NoStop}%
\bibitem [{\citenamefont {{von Keyserlingk}}\ \emph {et~al.}(2018)\citenamefont
  {{von Keyserlingk}}, \citenamefont {{Rakovszky}}, \citenamefont
  {{Pollmann}},\ and\ \citenamefont {{Sondhi}}}]{rak_randomcircuit}%
  \BibitemOpen
  \bibfield  {author} {\bibinfo {author} {\bibfnamefont {C.~W.}\ \bibnamefont
  {{von Keyserlingk}}}, \bibinfo {author} {\bibfnamefont {T.}~\bibnamefont
  {{Rakovszky}}}, \bibinfo {author} {\bibfnamefont {F.}~\bibnamefont
  {{Pollmann}}},\ and\ \bibinfo {author} {\bibfnamefont {S.~L.}\ \bibnamefont
  {{Sondhi}}},\ }\bibfield  {title} {\bibinfo {title} {{Operator Hydrodynamics,
  OTOCs, and Entanglement Growth in Systems without Conservation Laws}},\
  }\href {https://doi.org/10.1103/PhysRevX.8.021013} {\bibfield  {journal}
  {\bibinfo  {journal} {Physical Review X}\ }\textbf {\bibinfo {volume} {8}},\
  \bibinfo {eid} {021013} (\bibinfo {year} {2018})},\ \Eprint
  {https://arxiv.org/abs/1705.08910} {arXiv:1705.08910 [cond-mat.str-el]}
  \BibitemShut {NoStop}%
\bibitem [{\citenamefont {{Zhou}}\ and\ \citenamefont
  {{Nahum}}(2019)}]{zhou_nahum_statmech}%
  \BibitemOpen
  \bibfield  {author} {\bibinfo {author} {\bibfnamefont {T.}~\bibnamefont
  {{Zhou}}}\ and\ \bibinfo {author} {\bibfnamefont {A.}~\bibnamefont
  {{Nahum}}},\ }\bibfield  {title} {\bibinfo {title} {{Emergent statistical
  mechanics of entanglement in random unitary circuits}},\ }\href
  {https://doi.org/10.1103/PhysRevB.99.174205} {\bibfield  {journal} {\bibinfo
  {journal} {\prb}\ }\textbf {\bibinfo {volume} {99}},\ \bibinfo {eid} {174205}
  (\bibinfo {year} {2019})},\ \Eprint {https://arxiv.org/abs/1804.09737}
  {arXiv:1804.09737 [cond-mat.stat-mech]} \BibitemShut {NoStop}%
\bibitem [{\citenamefont {{Mezei}}(2018)}]{mark_membrane}%
  \BibitemOpen
  \bibfield  {author} {\bibinfo {author} {\bibfnamefont {M.}~\bibnamefont
  {{Mezei}}},\ }\bibfield  {title} {\bibinfo {title} {{Membrane theory of
  entanglement dynamics from holography}},\ }\href
  {https://doi.org/10.1103/PhysRevD.98.106025} {\bibfield  {journal} {\bibinfo
  {journal} {\prd}\ }\textbf {\bibinfo {volume} {98}},\ \bibinfo {eid} {106025}
  (\bibinfo {year} {2018})},\ \Eprint {https://arxiv.org/abs/1803.10244}
  {arXiv:1803.10244 [hep-th]} \BibitemShut {NoStop}%
\bibitem [{\citenamefont {{Zhou}}\ and\ \citenamefont
  {{Nahum}}(2020)}]{zhou_nahum}%
  \BibitemOpen
  \bibfield  {author} {\bibinfo {author} {\bibfnamefont {T.}~\bibnamefont
  {{Zhou}}}\ and\ \bibinfo {author} {\bibfnamefont {A.}~\bibnamefont
  {{Nahum}}},\ }\bibfield  {title} {\bibinfo {title} {{Entanglement Membrane in
  Chaotic Many-Body Systems}},\ }\href
  {https://doi.org/10.1103/PhysRevX.10.031066} {\bibfield  {journal} {\bibinfo
  {journal} {Physical Review X}\ }\textbf {\bibinfo {volume} {10}},\ \bibinfo
  {eid} {031066} (\bibinfo {year} {2020})},\ \Eprint
  {https://arxiv.org/abs/1912.12311} {arXiv:1912.12311 [cond-mat.str-el]}
  \BibitemShut {NoStop}%
\bibitem [{\citenamefont {McCulloch}\ \emph {et~al.}(2023)\citenamefont
  {McCulloch}, \citenamefont {De~Nardis}, \citenamefont {Gopalakrishnan},\ and\
  \citenamefont {Vasseur}}]{mcculloch2023full}%
  \BibitemOpen
  \bibfield  {author} {\bibinfo {author} {\bibfnamefont {E.}~\bibnamefont
  {McCulloch}}, \bibinfo {author} {\bibfnamefont {J.}~\bibnamefont
  {De~Nardis}}, \bibinfo {author} {\bibfnamefont {S.}~\bibnamefont
  {Gopalakrishnan}},\ and\ \bibinfo {author} {\bibfnamefont {R.}~\bibnamefont
  {Vasseur}},\ }\bibfield  {title} {\bibinfo {title} {Full counting statistics
  of charge in chaotic many-body quantum systems},\ }\href
  {https://doi.org/10.1103/PhysRevLett.131.210402} {\bibfield  {journal}
  {\bibinfo  {journal} {Phys. Rev. Lett.}\ }\textbf {\bibinfo {volume} {131}},\
  \bibinfo {pages} {210402} (\bibinfo {year} {2023})}\BibitemShut {NoStop}%
\bibitem [{\citenamefont {Ogunnaike}\ \emph {et~al.}(2023)\citenamefont
  {Ogunnaike}, \citenamefont {Feldmeier},\ and\ \citenamefont
  {Lee}}]{ogunnaike2023unifying}%
  \BibitemOpen
  \bibfield  {author} {\bibinfo {author} {\bibfnamefont {O.}~\bibnamefont
  {Ogunnaike}}, \bibinfo {author} {\bibfnamefont {J.}~\bibnamefont
  {Feldmeier}},\ and\ \bibinfo {author} {\bibfnamefont {J.~Y.}\ \bibnamefont
  {Lee}},\ }\bibfield  {title} {\bibinfo {title} {{Unifying Emergent
  Hydrodynamics and Lindbladian Low-Energy Spectra across Symmetries,
  Constraints, and Long-Range Interactions}},\ }\href
  {https://doi.org/10.1103/PhysRevLett.131.220403} {\bibfield  {journal}
  {\bibinfo  {journal} {Phys. Rev. Lett.}\ }\textbf {\bibinfo {volume} {131}},\
  \bibinfo {pages} {220403} (\bibinfo {year} {2023})}\BibitemShut {NoStop}%
\bibitem [{\citenamefont {Moudgalya}\ and\ \citenamefont
  {Motrunich}(2024{\natexlab{a}})}]{moudgalya2023symmetries}%
  \BibitemOpen
  \bibfield  {author} {\bibinfo {author} {\bibfnamefont {S.}~\bibnamefont
  {Moudgalya}}\ and\ \bibinfo {author} {\bibfnamefont {O.~I.}\ \bibnamefont
  {Motrunich}},\ }\bibfield  {title} {\bibinfo {title} {Symmetries as ground
  states of local superoperators: Hydrodynamic implications},\ }\href
  {https://doi.org/10.1103/PRXQuantum.5.040330} {\bibfield  {journal} {\bibinfo
   {journal} {PRX Quantum}\ }\textbf {\bibinfo {volume} {5}},\ \bibinfo {pages}
  {040330} (\bibinfo {year} {2024}{\natexlab{a}})}\BibitemShut {NoStop}%
\bibitem [{\citenamefont {{Lashkari}}\ \emph {et~al.}(2013)\citenamefont
  {{Lashkari}}, \citenamefont {{Stanford}}, \citenamefont {{Hastings}},
  \citenamefont {{Osborne}},\ and\ \citenamefont {{Hayden}}}]{lashkari}%
  \BibitemOpen
  \bibfield  {author} {\bibinfo {author} {\bibfnamefont {N.}~\bibnamefont
  {{Lashkari}}}, \bibinfo {author} {\bibfnamefont {D.}~\bibnamefont
  {{Stanford}}}, \bibinfo {author} {\bibfnamefont {M.}~\bibnamefont
  {{Hastings}}}, \bibinfo {author} {\bibfnamefont {T.}~\bibnamefont
  {{Osborne}}},\ and\ \bibinfo {author} {\bibfnamefont {P.}~\bibnamefont
  {{Hayden}}},\ }\bibfield  {title} {\bibinfo {title} {{Towards the fast
  scrambling conjecture}},\ }\href {https://doi.org/10.1007/JHEP04(2013)022}
  {\bibfield  {journal} {\bibinfo  {journal} {Journal of High Energy Physics}\
  }\textbf {\bibinfo {volume} {2013}},\ \bibinfo {eid} {22} (\bibinfo {year}
  {2013})},\ \Eprint {https://arxiv.org/abs/1111.6580} {arXiv:1111.6580
  [hep-th]} \BibitemShut {NoStop}%
\bibitem [{\citenamefont {{Xu}}\ and\ \citenamefont
  {{Swingle}}(2019)}]{swingle}%
  \BibitemOpen
  \bibfield  {author} {\bibinfo {author} {\bibfnamefont {S.}~\bibnamefont
  {{Xu}}}\ and\ \bibinfo {author} {\bibfnamefont {B.}~\bibnamefont
  {{Swingle}}},\ }\bibfield  {title} {\bibinfo {title} {{Locality, Quantum
  Fluctuations, and Scrambling}},\ }\href
  {https://doi.org/10.1103/PhysRevX.9.031048} {\bibfield  {journal} {\bibinfo
  {journal} {Physical Review X}\ }\textbf {\bibinfo {volume} {9}},\ \bibinfo
  {eid} {031048} (\bibinfo {year} {2019})},\ \Eprint
  {https://arxiv.org/abs/1805.05376} {arXiv:1805.05376 [cond-mat.str-el]}
  \BibitemShut {NoStop}%
\bibitem [{\citenamefont {{Saad}}\ \emph {et~al.}(2018)\citenamefont {{Saad}},
  \citenamefont {{Shenker}},\ and\ \citenamefont {{Stanford}}}]{brownian_syk}%
  \BibitemOpen
  \bibfield  {author} {\bibinfo {author} {\bibfnamefont {P.}~\bibnamefont
  {{Saad}}}, \bibinfo {author} {\bibfnamefont {S.~H.}\ \bibnamefont
  {{Shenker}}},\ and\ \bibinfo {author} {\bibfnamefont {D.}~\bibnamefont
  {{Stanford}}},\ }\bibfield  {title} {\bibinfo {title} {{A semiclassical ramp
  in SYK and in gravity}},\ }\href@noop {} {\bibfield  {journal} {\bibinfo
  {journal} {arXiv e-prints}\ } (\bibinfo {year} {2018})},\ \Eprint
  {https://arxiv.org/abs/1806.06840} {arXiv:1806.06840 [hep-th]} \BibitemShut
  {NoStop}%
\bibitem [{\citenamefont {S{\"u}nderhauf}\ \emph {et~al.}(2019)\citenamefont
  {S{\"u}nderhauf}, \citenamefont {Piroli}, \citenamefont {Qi}, \citenamefont
  {Schuch},\ and\ \citenamefont {Cirac}}]{sunderhauf2019quantum}%
  \BibitemOpen
  \bibfield  {author} {\bibinfo {author} {\bibfnamefont {C.}~\bibnamefont
  {S{\"u}nderhauf}}, \bibinfo {author} {\bibfnamefont {L.}~\bibnamefont
  {Piroli}}, \bibinfo {author} {\bibfnamefont {X.-L.}\ \bibnamefont {Qi}},
  \bibinfo {author} {\bibfnamefont {N.}~\bibnamefont {Schuch}},\ and\ \bibinfo
  {author} {\bibfnamefont {J.~I.}\ \bibnamefont {Cirac}},\ }\bibfield  {title}
  {\bibinfo {title} {{Quantum chaos in the Brownian SYK model with large finite
  N : OTOCs and tripartite information}},\ }\href
  {https://doi.org/10.1007/JHEP11(2019)038} {\bibfield  {journal} {\bibinfo
  {journal} {Journal of High Energy Physics}\ }\textbf {\bibinfo {volume}
  {2019}},\ \bibinfo {pages} {38} (\bibinfo {year} {2019})}\BibitemShut
  {NoStop}%
\bibitem [{\citenamefont {Bauer}\ \emph {et~al.}(2017)\citenamefont {Bauer},
  \citenamefont {Bernard},\ and\ \citenamefont {Jin}}]{bauer2017stochastic}%
  \BibitemOpen
  \bibfield  {author} {\bibinfo {author} {\bibfnamefont {M.}~\bibnamefont
  {Bauer}}, \bibinfo {author} {\bibfnamefont {D.}~\bibnamefont {Bernard}},\
  and\ \bibinfo {author} {\bibfnamefont {T.}~\bibnamefont {Jin}},\ }\bibfield
  {title} {\bibinfo {title} {{Stochastic dissipative quantum spin chains (I) :
  Quantum fluctuating discrete hydrodynamics}},\ }\href
  {https://doi.org/10.21468/SciPostPhys.3.5.033} {\bibfield  {journal}
  {\bibinfo  {journal} {SciPost Phys.}\ }\textbf {\bibinfo {volume} {3}},\
  \bibinfo {pages} {033} (\bibinfo {year} {2017})}\BibitemShut {NoStop}%
\bibitem [{\citenamefont {Bauer}\ \emph {et~al.}(2019)\citenamefont {Bauer},
  \citenamefont {Bernard},\ and\ \citenamefont {Jin}}]{bauer2019equilibrium}%
  \BibitemOpen
  \bibfield  {author} {\bibinfo {author} {\bibfnamefont {M.}~\bibnamefont
  {Bauer}}, \bibinfo {author} {\bibfnamefont {D.}~\bibnamefont {Bernard}},\
  and\ \bibinfo {author} {\bibfnamefont {T.}~\bibnamefont {Jin}},\ }\bibfield
  {title} {\bibinfo {title} {{Equilibrium fluctuations in maximally noisy
  extended quantum systems}},\ }\href
  {https://doi.org/10.21468/SciPostPhys.6.4.045} {\bibfield  {journal}
  {\bibinfo  {journal} {SciPost Phys.}\ }\textbf {\bibinfo {volume} {6}},\
  \bibinfo {pages} {045} (\bibinfo {year} {2019})}\BibitemShut {NoStop}%
\bibitem [{\citenamefont {Bernard}\ and\ \citenamefont
  {Piroli}(2021)}]{bernard2021entanglement}%
  \BibitemOpen
  \bibfield  {author} {\bibinfo {author} {\bibfnamefont {D.}~\bibnamefont
  {Bernard}}\ and\ \bibinfo {author} {\bibfnamefont {L.}~\bibnamefont
  {Piroli}},\ }\bibfield  {title} {\bibinfo {title} {Entanglement distribution
  in the quantum symmetric simple exclusion process},\ }\href
  {https://doi.org/10.1103/PhysRevE.104.014146} {\bibfield  {journal} {\bibinfo
   {journal} {Phys. Rev. E}\ }\textbf {\bibinfo {volume} {104}},\ \bibinfo
  {pages} {014146} (\bibinfo {year} {2021})}\BibitemShut {NoStop}%
\bibitem [{\citenamefont {Bernard}\ \emph {et~al.}(2022)\citenamefont
  {Bernard}, \citenamefont {Essler}, \citenamefont {Hruza},\ and\ \citenamefont
  {Medenjak}}]{bernard2022dynamics}%
  \BibitemOpen
  \bibfield  {author} {\bibinfo {author} {\bibfnamefont {D.}~\bibnamefont
  {Bernard}}, \bibinfo {author} {\bibfnamefont {F.~H.~L.}\ \bibnamefont
  {Essler}}, \bibinfo {author} {\bibfnamefont {L.}~\bibnamefont {Hruza}},\ and\
  \bibinfo {author} {\bibfnamefont {M.}~\bibnamefont {Medenjak}},\ }\bibfield
  {title} {\bibinfo {title} {{Dynamics of fluctuations in quantum simple
  exclusion processes}},\ }\href
  {https://doi.org/10.21468/SciPostPhys.12.1.042} {\bibfield  {journal}
  {\bibinfo  {journal} {SciPost Phys.}\ }\textbf {\bibinfo {volume} {12}},\
  \bibinfo {pages} {042} (\bibinfo {year} {2022})}\BibitemShut {NoStop}%
\bibitem [{\citenamefont {{Swann}}\ \emph {et~al.}(2023)\citenamefont
  {{Swann}}, \citenamefont {{Bernard}},\ and\ \citenamefont
  {{Nahum}}}]{nahum_freefermion}%
  \BibitemOpen
  \bibfield  {author} {\bibinfo {author} {\bibfnamefont {T.}~\bibnamefont
  {{Swann}}}, \bibinfo {author} {\bibfnamefont {D.}~\bibnamefont {{Bernard}}},\
  and\ \bibinfo {author} {\bibfnamefont {A.}~\bibnamefont {{Nahum}}},\
  }\bibfield  {title} {\bibinfo {title} {{Spacetime picture for entanglement
  generation in noisy fermion chains}},\ }\href@noop {} {\bibfield  {journal}
  {\bibinfo  {journal} {arXiv e-prints}\ } (\bibinfo {year} {2023})},\ \Eprint
  {https://arxiv.org/abs/2302.12212} {arXiv:2302.12212 [cond-mat.stat-mech]}
  \BibitemShut {NoStop}%
\bibitem [{\citenamefont {Zhou}\ and\ \citenamefont
  {Chen}(2019)}]{zhou2019operator}%
  \BibitemOpen
  \bibfield  {author} {\bibinfo {author} {\bibfnamefont {T.}~\bibnamefont
  {Zhou}}\ and\ \bibinfo {author} {\bibfnamefont {X.}~\bibnamefont {Chen}},\
  }\bibfield  {title} {\bibinfo {title} {{Operator dynamics in a Brownian
  quantum circuit}},\ }\href {https://doi.org/10.1103/PhysRevE.99.052212}
  {\bibfield  {journal} {\bibinfo  {journal} {Phys. Rev. E}\ }\textbf {\bibinfo
  {volume} {99}},\ \bibinfo {pages} {052212} (\bibinfo {year}
  {2019})}\BibitemShut {NoStop}%
\bibitem [{\citenamefont {Jian}\ and\ \citenamefont
  {Swingle}(2021)}]{jian2021note}%
  \BibitemOpen
  \bibfield  {author} {\bibinfo {author} {\bibfnamefont {S.-K.}\ \bibnamefont
  {Jian}}\ and\ \bibinfo {author} {\bibfnamefont {B.}~\bibnamefont {Swingle}},\
  }\bibfield  {title} {\bibinfo {title} {{Note on entropy dynamics in the
  Brownian SYK model}},\ }\href {https://doi.org/10.1007/JHEP03(2021)042}
  {\bibfield  {journal} {\bibinfo  {journal} {Journal of High Energy Physics}\
  }\textbf {\bibinfo {volume} {2021}},\ \bibinfo {pages} {42} (\bibinfo {year}
  {2021})}\BibitemShut {NoStop}%
\bibitem [{\citenamefont {Bao}\ \emph {et~al.}(2021)\citenamefont {Bao},
  \citenamefont {Choi},\ and\ \citenamefont {Altman}}]{bao2021symmetry}%
  \BibitemOpen
  \bibfield  {author} {\bibinfo {author} {\bibfnamefont {Y.}~\bibnamefont
  {Bao}}, \bibinfo {author} {\bibfnamefont {S.}~\bibnamefont {Choi}},\ and\
  \bibinfo {author} {\bibfnamefont {E.}~\bibnamefont {Altman}},\ }\bibfield
  {title} {\bibinfo {title} {Symmetry enriched phases of quantum circuits},\
  }\href {https://doi.org/https://doi.org/10.1016/j.aop.2021.168618} {\bibfield
   {journal} {\bibinfo  {journal} {Annals of Physics}\ }\textbf {\bibinfo
  {volume} {435}},\ \bibinfo {pages} {168618} (\bibinfo {year} {2021})},\
  \bibinfo {note} {special issue on Philip W. Anderson}\BibitemShut {NoStop}%
\bibitem [{\citenamefont {{Jian}}\ \emph {et~al.}(2022)\citenamefont {{Jian}},
  \citenamefont {{Bentsen}},\ and\ \citenamefont {{Swingle}}}]{jian2022linear}%
  \BibitemOpen
  \bibfield  {author} {\bibinfo {author} {\bibfnamefont {S.-K.}\ \bibnamefont
  {{Jian}}}, \bibinfo {author} {\bibfnamefont {G.}~\bibnamefont {{Bentsen}}},\
  and\ \bibinfo {author} {\bibfnamefont {B.}~\bibnamefont {{Swingle}}},\
  }\bibfield  {title} {\bibinfo {title} {{Linear Growth of Circuit Complexity
  from Brownian Dynamics}},\ }\href@noop {} {\bibfield  {journal} {\bibinfo
  {journal} {arXiv e-prints}\ } (\bibinfo {year} {2022})},\ \Eprint
  {https://arxiv.org/abs/2206.14205} {arXiv:2206.14205 [quant-ph]} \BibitemShut
  {NoStop}%
\bibitem [{\citenamefont {Leontica}\ and\ \citenamefont
  {McGinley}(2023)}]{leontica2023purification}%
  \BibitemOpen
  \bibfield  {author} {\bibinfo {author} {\bibfnamefont {S.}~\bibnamefont
  {Leontica}}\ and\ \bibinfo {author} {\bibfnamefont {M.}~\bibnamefont
  {McGinley}},\ }\bibfield  {title} {\bibinfo {title} {Purification dynamics in
  a continuous-time hybrid quantum circuit model},\ }\href
  {https://doi.org/10.1103/PhysRevB.108.174308} {\bibfield  {journal} {\bibinfo
   {journal} {Phys. Rev. B}\ }\textbf {\bibinfo {volume} {108}},\ \bibinfo
  {pages} {174308} (\bibinfo {year} {2023})}\BibitemShut {NoStop}%
\bibitem [{\citenamefont {Moudgalya}\ and\ \citenamefont
  {Motrunich}(2023)}]{moudgalya2022from}%
  \BibitemOpen
  \bibfield  {author} {\bibinfo {author} {\bibfnamefont {S.}~\bibnamefont
  {Moudgalya}}\ and\ \bibinfo {author} {\bibfnamefont {O.~I.}\ \bibnamefont
  {Motrunich}},\ }\bibfield  {title} {\bibinfo {title} {{From symmetries to
  commutant algebras in standard Hamiltonians}},\ }\href
  {https://doi.org/https://doi.org/10.1016/j.aop.2023.169384} {\bibfield
  {journal} {\bibinfo  {journal} {Annals of Physics}\ }\textbf {\bibinfo
  {volume} {455}},\ \bibinfo {pages} {169384} (\bibinfo {year}
  {2023})}\BibitemShut {NoStop}%
\bibitem [{\citenamefont {{Calabrese}}\ and\ \citenamefont
  {{Cardy}}(2007)}]{calabrese}%
  \BibitemOpen
  \bibfield  {author} {\bibinfo {author} {\bibfnamefont {P.}~\bibnamefont
  {{Calabrese}}}\ and\ \bibinfo {author} {\bibfnamefont {J.}~\bibnamefont
  {{Cardy}}},\ }\bibfield  {title} {\bibinfo {title} {{Entanglement and
  correlation functions following a local quench: a conformal field theory
  approach}},\ }\href {https://doi.org/10.1088/1742-5468/2007/10/P10004}
  {\bibfield  {journal} {\bibinfo  {journal} {Journal of Statistical Mechanics:
  Theory and Experiment}\ }\textbf {\bibinfo {volume} {2007}},\ \bibinfo
  {pages} {10004} (\bibinfo {year} {2007})},\ \Eprint
  {https://arxiv.org/abs/0708.3750} {arXiv:0708.3750 [cond-mat.stat-mech]}
  \BibitemShut {NoStop}%
\bibitem [{\citenamefont {{Leichenauer}}\ and\ \citenamefont
  {{Moosa}}(2015)}]{leichenauer}%
  \BibitemOpen
  \bibfield  {author} {\bibinfo {author} {\bibfnamefont {S.}~\bibnamefont
  {{Leichenauer}}}\ and\ \bibinfo {author} {\bibfnamefont {M.}~\bibnamefont
  {{Moosa}}},\ }\bibfield  {title} {\bibinfo {title} {{Entanglement tsunami in
  (1 +1 )-dimensions}},\ }\href {https://doi.org/10.1103/PhysRevD.92.126004}
  {\bibfield  {journal} {\bibinfo  {journal} {\prd}\ }\textbf {\bibinfo
  {volume} {92}},\ \bibinfo {eid} {126004} (\bibinfo {year} {2015})},\ \Eprint
  {https://arxiv.org/abs/1505.04225} {arXiv:1505.04225 [hep-th]} \BibitemShut
  {NoStop}%
\bibitem [{\citenamefont {{Casini}}\ \emph {et~al.}(2016)\citenamefont
  {{Casini}}, \citenamefont {{Liu}},\ and\ \citenamefont {{Mezei}}}]{spread}%
  \BibitemOpen
  \bibfield  {author} {\bibinfo {author} {\bibfnamefont {H.}~\bibnamefont
  {{Casini}}}, \bibinfo {author} {\bibfnamefont {H.}~\bibnamefont {{Liu}}},\
  and\ \bibinfo {author} {\bibfnamefont {M.}~\bibnamefont {{Mezei}}},\
  }\bibfield  {title} {\bibinfo {title} {{Spread of entanglement and
  causality}},\ }\href {https://doi.org/10.1007/JHEP07(2016)077} {\bibfield
  {journal} {\bibinfo  {journal} {Journal of High Energy Physics}\ }\textbf
  {\bibinfo {volume} {2016}},\ \bibinfo {eid} {77} (\bibinfo {year} {2016})},\
  \Eprint {https://arxiv.org/abs/1509.05044} {arXiv:1509.05044 [hep-th]}
  \BibitemShut {NoStop}%
\bibitem [{\citenamefont {{Haegeman}}\ \emph {et~al.}(2013)\citenamefont
  {{Haegeman}}, \citenamefont {{Michalakis}}, \citenamefont {{Nachtergaele}},
  \citenamefont {{Osborne}}, \citenamefont {{Schuch}},\ and\ \citenamefont
  {{Verstraete}}}]{elementary_excitations}%
  \BibitemOpen
  \bibfield  {author} {\bibinfo {author} {\bibfnamefont {J.}~\bibnamefont
  {{Haegeman}}}, \bibinfo {author} {\bibfnamefont {S.}~\bibnamefont
  {{Michalakis}}}, \bibinfo {author} {\bibfnamefont {B.}~\bibnamefont
  {{Nachtergaele}}}, \bibinfo {author} {\bibfnamefont {T.~J.}\ \bibnamefont
  {{Osborne}}}, \bibinfo {author} {\bibfnamefont {N.}~\bibnamefont
  {{Schuch}}},\ and\ \bibinfo {author} {\bibfnamefont {F.}~\bibnamefont
  {{Verstraete}}},\ }\bibfield  {title} {\bibinfo {title} {{Elementary
  Excitations in Gapped Quantum Spin Systems}},\ }\href
  {https://doi.org/10.1103/PhysRevLett.111.080401} {\bibfield  {journal}
  {\bibinfo  {journal} {\prl}\ }\textbf {\bibinfo {volume} {111}},\ \bibinfo
  {eid} {080401} (\bibinfo {year} {2013})},\ \Eprint
  {https://arxiv.org/abs/1305.2176} {arXiv:1305.2176 [quant-ph]} \BibitemShut
  {NoStop}%
\bibitem [{\citenamefont {{Haegeman}}\ \emph {et~al.}(2012)\citenamefont
  {{Haegeman}}, \citenamefont {{Pirvu}}, \citenamefont {{Weir}}, \citenamefont
  {{Cirac}}, \citenamefont {{Osborne}}, \citenamefont {{Verschelde}},\ and\
  \citenamefont {{Verstraete}}}]{variational_ansatz}%
  \BibitemOpen
  \bibfield  {author} {\bibinfo {author} {\bibfnamefont {J.}~\bibnamefont
  {{Haegeman}}}, \bibinfo {author} {\bibfnamefont {B.}~\bibnamefont {{Pirvu}}},
  \bibinfo {author} {\bibfnamefont {D.~J.}\ \bibnamefont {{Weir}}}, \bibinfo
  {author} {\bibfnamefont {J.~I.}\ \bibnamefont {{Cirac}}}, \bibinfo {author}
  {\bibfnamefont {T.~J.}\ \bibnamefont {{Osborne}}}, \bibinfo {author}
  {\bibfnamefont {H.}~\bibnamefont {{Verschelde}}},\ and\ \bibinfo {author}
  {\bibfnamefont {F.}~\bibnamefont {{Verstraete}}},\ }\bibfield  {title}
  {\bibinfo {title} {{Variational matrix product ansatz for dispersion
  relations}},\ }\href {https://doi.org/10.1103/PhysRevB.85.100408} {\bibfield
  {journal} {\bibinfo  {journal} {\prb}\ }\textbf {\bibinfo {volume} {85}},\
  \bibinfo {eid} {100408} (\bibinfo {year} {2012})},\ \Eprint
  {https://arxiv.org/abs/1103.2286} {arXiv:1103.2286 [quant-ph]} \BibitemShut
  {NoStop}%
\bibitem [{\citenamefont {{Vanderstraeten}}\ \emph {et~al.}(2015)\citenamefont
  {{Vanderstraeten}}, \citenamefont {{Verstraete}},\ and\ \citenamefont
  {{Haegeman}}}]{scattering_particles}%
  \BibitemOpen
  \bibfield  {author} {\bibinfo {author} {\bibfnamefont {L.}~\bibnamefont
  {{Vanderstraeten}}}, \bibinfo {author} {\bibfnamefont {F.}~\bibnamefont
  {{Verstraete}}},\ and\ \bibinfo {author} {\bibfnamefont {J.}~\bibnamefont
  {{Haegeman}}},\ }\bibfield  {title} {\bibinfo {title} {{Scattering particles
  in quantum spin chains}},\ }\href
  {https://doi.org/10.1103/PhysRevB.92.125136} {\bibfield  {journal} {\bibinfo
  {journal} {\prb}\ }\textbf {\bibinfo {volume} {92}},\ \bibinfo {eid} {125136}
  (\bibinfo {year} {2015})},\ \Eprint {https://arxiv.org/abs/1506.01008}
  {arXiv:1506.01008 [cond-mat.str-el]} \BibitemShut {NoStop}%
\bibitem [{\citenamefont {Fisher}\ \emph {et~al.}(2023)\citenamefont {Fisher},
  \citenamefont {Khemani}, \citenamefont {Nahum},\ and\ \citenamefont
  {Vijay}}]{fisher2023random}%
  \BibitemOpen
  \bibfield  {author} {\bibinfo {author} {\bibfnamefont {M.~P.}\ \bibnamefont
  {Fisher}}, \bibinfo {author} {\bibfnamefont {V.}~\bibnamefont {Khemani}},
  \bibinfo {author} {\bibfnamefont {A.}~\bibnamefont {Nahum}},\ and\ \bibinfo
  {author} {\bibfnamefont {S.}~\bibnamefont {Vijay}},\ }\bibfield  {title}
  {\bibinfo {title} {Random quantum circuits},\ }\href
  {https://doi.org/https://doi.org/10.1146/annurev-conmatphys-031720-030658}
  {\bibfield  {journal} {\bibinfo  {journal} {Annual Review of Condensed Matter
  Physics}\ }\textbf {\bibinfo {volume} {14}},\ \bibinfo {pages} {335}
  (\bibinfo {year} {2023})}\BibitemShut {NoStop}%
\bibitem [{\citenamefont {Harrow}\ and\ \citenamefont
  {Low}(2009)}]{harrow2009random}%
  \BibitemOpen
  \bibfield  {author} {\bibinfo {author} {\bibfnamefont {A.~W.}\ \bibnamefont
  {Harrow}}\ and\ \bibinfo {author} {\bibfnamefont {R.~A.}\ \bibnamefont
  {Low}},\ }\bibfield  {title} {\bibinfo {title} {Random quantum circuits are
  approximate 2-designs},\ }\href {https://doi.org/10.1007/s00220-009-0873-6}
  {\bibfield  {journal} {\bibinfo  {journal} {Communications in Mathematical
  Physics}\ }\textbf {\bibinfo {volume} {291}},\ \bibinfo {pages} {257}
  (\bibinfo {year} {2009})}\BibitemShut {NoStop}%
\bibitem [{\citenamefont {Brand{\~a}o}\ \emph {et~al.}(2016)\citenamefont
  {Brand{\~a}o}, \citenamefont {Harrow},\ and\ \citenamefont
  {Horodecki}}]{brandao2016local}%
  \BibitemOpen
  \bibfield  {author} {\bibinfo {author} {\bibfnamefont {F.~G. S.~L.}\
  \bibnamefont {Brand{\~a}o}}, \bibinfo {author} {\bibfnamefont {A.~W.}\
  \bibnamefont {Harrow}},\ and\ \bibinfo {author} {\bibfnamefont
  {M.}~\bibnamefont {Horodecki}},\ }\bibfield  {title} {\bibinfo {title} {Local
  random quantum circuits are approximate polynomial-designs},\ }\href
  {https://doi.org/10.1007/s00220-016-2706-8} {\bibfield  {journal} {\bibinfo
  {journal} {Communications in Mathematical Physics}\ }\textbf {\bibinfo
  {volume} {346}},\ \bibinfo {pages} {397} (\bibinfo {year}
  {2016})}\BibitemShut {NoStop}%
\bibitem [{\citenamefont {{Hunter-Jones}}(2019)}]{hunterjones2019unitary}%
  \BibitemOpen
  \bibfield  {author} {\bibinfo {author} {\bibfnamefont {N.}~\bibnamefont
  {{Hunter-Jones}}},\ }\bibfield  {title} {\bibinfo {title} {{Unitary designs
  from statistical mechanics in random quantum circuits}},\ }\href@noop {}
  {\bibfield  {journal} {\bibinfo  {journal} {arXiv e-prints}\ } (\bibinfo
  {year} {2019})},\ \Eprint {https://arxiv.org/abs/1905.12053}
  {arXiv:1905.12053 [quant-ph]} \BibitemShut {NoStop}%
\bibitem [{\citenamefont {{Lastres}}\ \emph {et~al.}(2024)\citenamefont
  {{Lastres}}, \citenamefont {{Pollmann}},\ and\ \citenamefont
  {{Moudgalya}}}]{lastres2024nonuniversality}%
  \BibitemOpen
  \bibfield  {author} {\bibinfo {author} {\bibfnamefont {M.}~\bibnamefont
  {{Lastres}}}, \bibinfo {author} {\bibfnamefont {F.}~\bibnamefont
  {{Pollmann}}},\ and\ \bibinfo {author} {\bibfnamefont {S.}~\bibnamefont
  {{Moudgalya}}},\ }\bibfield  {title} {\bibinfo {title} {{Non-Universality
  from Conserved Superoperators in Unitary Circuits}},\ }\bibfield  {journal}
  {\bibinfo  {journal} {arXiv e-prints}\ }\href
  {https://doi.org/10.48550/arXiv.2409.11407} {10.48550/arXiv.2409.11407}
  (\bibinfo {year} {2024}),\ \Eprint {https://arxiv.org/abs/2409.11407}
  {arXiv:2409.11407 [quant-ph]} \BibitemShut {NoStop}%
\bibitem [{\citenamefont {Hearth}\ \emph {et~al.}(2025)\citenamefont {Hearth},
  \citenamefont {Flynn}, \citenamefont {Chandran},\ and\ \citenamefont
  {Laumann}}]{hearth2023unitary}%
  \BibitemOpen
  \bibfield  {author} {\bibinfo {author} {\bibfnamefont {S.~N.}\ \bibnamefont
  {Hearth}}, \bibinfo {author} {\bibfnamefont {M.~O.}\ \bibnamefont {Flynn}},
  \bibinfo {author} {\bibfnamefont {A.}~\bibnamefont {Chandran}},\ and\
  \bibinfo {author} {\bibfnamefont {C.~R.}\ \bibnamefont {Laumann}},\
  }\bibfield  {title} {\bibinfo {title} {Unitary $k$-designs from random
  number-conserving quantum circuits},\ }\href
  {https://doi.org/10.1103/PhysRevX.15.021022} {\bibfield  {journal} {\bibinfo
  {journal} {Phys. Rev. X}\ }\textbf {\bibinfo {volume} {15}},\ \bibinfo
  {pages} {021022} (\bibinfo {year} {2025})}\BibitemShut {NoStop}%
\bibitem [{\citenamefont {Guo}\ \emph {et~al.}(2024)\citenamefont {Guo},
  \citenamefont {Sasieta},\ and\ \citenamefont {Swingle}}]{guo2024complexity}%
  \BibitemOpen
  \bibfield  {author} {\bibinfo {author} {\bibfnamefont {S.}~\bibnamefont
  {Guo}}, \bibinfo {author} {\bibfnamefont {M.}~\bibnamefont {Sasieta}},\ and\
  \bibinfo {author} {\bibfnamefont {B.}~\bibnamefont {Swingle}},\ }\bibfield
  {title} {\bibinfo {title} {{Complexity is not enough for randomness}},\
  }\href {https://doi.org/10.21468/SciPostPhys.17.6.151} {\bibfield  {journal}
  {\bibinfo  {journal} {SciPost Phys.}\ }\textbf {\bibinfo {volume} {17}},\
  \bibinfo {pages} {151} (\bibinfo {year} {2024})}\BibitemShut {NoStop}%
\bibitem [{\citenamefont {{Tang}}(2024)}]{haifeng}%
  \BibitemOpen
  \bibfield  {author} {\bibinfo {author} {\bibfnamefont {H.}~\bibnamefont
  {{Tang}}},\ }\bibfield  {title} {\bibinfo {title} {{Brownian Gaussian Unitary
  Ensemble: non-equilibrium dynamics, efficient $k$-design and application in
  classical shadow tomography}},\ }\href@noop {} {\bibfield  {journal}
  {\bibinfo  {journal} {arXiv e-prints}\ } (\bibinfo {year} {2024})},\ \Eprint
  {https://arxiv.org/abs/2406.11320} {arXiv:2406.11320 [hep-th]} \BibitemShut
  {NoStop}%
\bibitem [{\citenamefont {You}\ and\ \citenamefont
  {Gu}(2018)}]{you2018entanglement}%
  \BibitemOpen
  \bibfield  {author} {\bibinfo {author} {\bibfnamefont {Y.-Z.}\ \bibnamefont
  {You}}\ and\ \bibinfo {author} {\bibfnamefont {Y.}~\bibnamefont {Gu}},\
  }\bibfield  {title} {\bibinfo {title} {Entanglement features of random
  hamiltonian dynamics},\ }\href {https://doi.org/10.1103/PhysRevB.98.014309}
  {\bibfield  {journal} {\bibinfo  {journal} {Phys. Rev. B}\ }\textbf {\bibinfo
  {volume} {98}},\ \bibinfo {pages} {014309} (\bibinfo {year}
  {2018})}\BibitemShut {NoStop}%
\bibitem [{\citenamefont {Kuo}\ \emph {et~al.}(2020)\citenamefont {Kuo},
  \citenamefont {Akhtar}, \citenamefont {Arovas},\ and\ \citenamefont
  {You}}]{kuo2020markovian}%
  \BibitemOpen
  \bibfield  {author} {\bibinfo {author} {\bibfnamefont {W.-T.}\ \bibnamefont
  {Kuo}}, \bibinfo {author} {\bibfnamefont {A.~A.}\ \bibnamefont {Akhtar}},
  \bibinfo {author} {\bibfnamefont {D.~P.}\ \bibnamefont {Arovas}},\ and\
  \bibinfo {author} {\bibfnamefont {Y.-Z.}\ \bibnamefont {You}},\ }\bibfield
  {title} {\bibinfo {title} {Markovian entanglement dynamics under locally
  scrambled quantum evolution},\ }\href
  {https://doi.org/10.1103/PhysRevB.101.224202} {\bibfield  {journal} {\bibinfo
   {journal} {Phys. Rev. B}\ }\textbf {\bibinfo {volume} {101}},\ \bibinfo
  {pages} {224202} (\bibinfo {year} {2020})}\BibitemShut {NoStop}%
\bibitem [{\citenamefont {Peschel}\ and\ \citenamefont
  {Emery}(1981)}]{peschel1981calculation}%
  \BibitemOpen
  \bibfield  {author} {\bibinfo {author} {\bibfnamefont {I.}~\bibnamefont
  {Peschel}}\ and\ \bibinfo {author} {\bibfnamefont {V.~J.}\ \bibnamefont
  {Emery}},\ }\bibfield  {title} {\bibinfo {title} {{Calculation of spin
  correlations in two-dimensional Ising systems from one-dimensional kinetic
  models}},\ }\href {https://doi.org/10.1007/BF01297524} {\bibfield  {journal}
  {\bibinfo  {journal} {Zeitschrift f{\"u}r Physik B Condensed Matter}\
  }\textbf {\bibinfo {volume} {43}},\ \bibinfo {pages} {241} (\bibinfo {year}
  {1981})}\BibitemShut {NoStop}%
\bibitem [{\citenamefont {Katsura}\ \emph {et~al.}(2015)\citenamefont
  {Katsura}, \citenamefont {Schuricht},\ and\ \citenamefont
  {Takahashi}}]{katsura2015exact}%
  \BibitemOpen
  \bibfield  {author} {\bibinfo {author} {\bibfnamefont {H.}~\bibnamefont
  {Katsura}}, \bibinfo {author} {\bibfnamefont {D.}~\bibnamefont {Schuricht}},\
  and\ \bibinfo {author} {\bibfnamefont {M.}~\bibnamefont {Takahashi}},\
  }\bibfield  {title} {\bibinfo {title} {{Exact ground states and topological
  order in interacting Kitaev/Majorana chains}},\ }\href
  {https://doi.org/10.1103/PhysRevB.92.115137} {\bibfield  {journal} {\bibinfo
  {journal} {Phys. Rev. B}\ }\textbf {\bibinfo {volume} {92}},\ \bibinfo
  {pages} {115137} (\bibinfo {year} {2015})}\BibitemShut {NoStop}%
\bibitem [{\citenamefont {\ifmmode \check{Z}\else
  \v{Z}\fi{}nidari\ifmmode~\check{c}\else
  \v{c}\fi{}}(2008)}]{znidaric2008exact}%
  \BibitemOpen
  \bibfield  {author} {\bibinfo {author} {\bibfnamefont {M.}~\bibnamefont
  {\ifmmode \check{Z}\else \v{Z}\fi{}nidari\ifmmode~\check{c}\else
  \v{c}\fi{}}},\ }\bibfield  {title} {\bibinfo {title} {Exact convergence times
  for generation of random bipartite entanglement},\ }\href
  {https://doi.org/10.1103/PhysRevA.78.032324} {\bibfield  {journal} {\bibinfo
  {journal} {Phys. Rev. A}\ }\textbf {\bibinfo {volume} {78}},\ \bibinfo
  {pages} {032324} (\bibinfo {year} {2008})}\BibitemShut {NoStop}%
\bibitem [{\citenamefont {{Suzuki}}\ \emph {et~al.}(2024)\citenamefont
  {{Suzuki}}, \citenamefont {{Katsura}}, \citenamefont {{Mitsuhashi}},
  \citenamefont {{Soejima}}, \citenamefont {{Eisert}},\ and\ \citenamefont
  {{Yoshioka}}}]{suzuki2024more}%
  \BibitemOpen
  \bibfield  {author} {\bibinfo {author} {\bibfnamefont {R.}~\bibnamefont
  {{Suzuki}}}, \bibinfo {author} {\bibfnamefont {H.}~\bibnamefont {{Katsura}}},
  \bibinfo {author} {\bibfnamefont {Y.}~\bibnamefont {{Mitsuhashi}}}, \bibinfo
  {author} {\bibfnamefont {T.}~\bibnamefont {{Soejima}}}, \bibinfo {author}
  {\bibfnamefont {J.}~\bibnamefont {{Eisert}}},\ and\ \bibinfo {author}
  {\bibfnamefont {N.}~\bibnamefont {{Yoshioka}}},\ }\bibfield  {title}
  {\bibinfo {title} {{More global randomness from less random local gates}},\
  }\bibfield  {journal} {\bibinfo  {journal} {arXiv e-prints}\ }\href
  {https://doi.org/10.48550/arXiv.2410.24127} {10.48550/arXiv.2410.24127}
  (\bibinfo {year} {2024}),\ \Eprint {https://arxiv.org/abs/2410.24127}
  {arXiv:2410.24127 [quant-ph]} \BibitemShut {NoStop}%
\bibitem [{\citenamefont {{Gu}}\ \emph {et~al.}(2017)\citenamefont {{Gu}},
  \citenamefont {{Lucas}},\ and\ \citenamefont {{Qi}}}]{syk_chain}%
  \BibitemOpen
  \bibfield  {author} {\bibinfo {author} {\bibfnamefont {Y.}~\bibnamefont
  {{Gu}}}, \bibinfo {author} {\bibfnamefont {A.}~\bibnamefont {{Lucas}}},\ and\
  \bibinfo {author} {\bibfnamefont {X.-L.}\ \bibnamefont {{Qi}}},\ }\bibfield
  {title} {\bibinfo {title} {{Spread of entanglement in a Sachdev-Ye-Kitaev
  chain}},\ }\href {https://doi.org/10.1007/JHEP09(2017)120} {\bibfield
  {journal} {\bibinfo  {journal} {Journal of High Energy Physics}\ }\textbf
  {\bibinfo {volume} {2017}},\ \bibinfo {eid} {120} (\bibinfo {year} {2017})},\
  \Eprint {https://arxiv.org/abs/1708.00871} {arXiv:1708.00871 [hep-th]}
  \BibitemShut {NoStop}%
\bibitem [{\citenamefont {Page}(1993)}]{page1993}%
  \BibitemOpen
  \bibfield  {author} {\bibinfo {author} {\bibfnamefont {D.~N.}\ \bibnamefont
  {Page}},\ }\bibfield  {title} {\bibinfo {title} {Average entropy of a
  subsystem},\ }\href {https://doi.org/10.1103/PhysRevLett.71.1291} {\bibfield
  {journal} {\bibinfo  {journal} {Physical Review Letters}\ }\textbf {\bibinfo
  {volume} {71}},\ \bibinfo {pages} {1291} (\bibinfo {year}
  {1993})}\BibitemShut {NoStop}%
\bibitem [{\citenamefont {Zeier}\ and\ \citenamefont
  {Schulte-Herbrüggen}(2011)}]{zeier2011symmetry}%
  \BibitemOpen
  \bibfield  {author} {\bibinfo {author} {\bibfnamefont {R.}~\bibnamefont
  {Zeier}}\ and\ \bibinfo {author} {\bibfnamefont {T.}~\bibnamefont
  {Schulte-Herbrüggen}},\ }\bibfield  {title} {\bibinfo {title} {Symmetry
  principles in quantum systems theory},\ }\href
  {https://doi.org/10.1063/1.3657939} {\bibfield  {journal} {\bibinfo
  {journal} {Journal of Mathematical Physics}\ }\textbf {\bibinfo {volume}
  {52}},\ \bibinfo {pages} {113510} (\bibinfo {year} {2011})},\ \Eprint
  {https://arxiv.org/abs/https://doi.org/10.1063/1.3657939}
  {https://doi.org/10.1063/1.3657939} \BibitemShut {NoStop}%
\bibitem [{\citenamefont {Schollw{\"o}ck}(2013)}]{schollwock2013matrix}%
  \BibitemOpen
  \bibfield  {author} {\bibinfo {author} {\bibfnamefont {U.}~\bibnamefont
  {Schollw{\"o}ck}},\ }\bibinfo {title} {Matrix product state algorithms: Dmrg,
  tebd and relatives},\ in\ \href {https://doi.org/10.1007/978-3-642-35106-8_3}
  {\emph {\bibinfo {booktitle} {Strongly Correlated Systems: Numerical
  Methods}}},\ \bibinfo {editor} {edited by\ \bibinfo {editor} {\bibfnamefont
  {A.}~\bibnamefont {Avella}}\ and\ \bibinfo {editor} {\bibfnamefont
  {F.}~\bibnamefont {Mancini}}}\ (\bibinfo  {publisher} {Springer Berlin
  Heidelberg},\ \bibinfo {address} {Berlin, Heidelberg},\ \bibinfo {year}
  {2013})\ pp.\ \bibinfo {pages} {67--98}\BibitemShut {NoStop}%
\bibitem [{\citenamefont {{Crossley}}\ \emph {et~al.}(2017)\citenamefont
  {{Crossley}}, \citenamefont {{Glorioso}},\ and\ \citenamefont
  {{Liu}}}]{crossley}%
  \BibitemOpen
  \bibfield  {author} {\bibinfo {author} {\bibfnamefont {M.}~\bibnamefont
  {{Crossley}}}, \bibinfo {author} {\bibfnamefont {P.}~\bibnamefont
  {{Glorioso}}},\ and\ \bibinfo {author} {\bibfnamefont {H.}~\bibnamefont
  {{Liu}}},\ }\bibfield  {title} {\bibinfo {title} {{Effective field theory of
  dissipative fluids}},\ }\href {https://doi.org/10.1007/JHEP09(2017)095}
  {\bibfield  {journal} {\bibinfo  {journal} {Journal of High Energy Physics}\
  }\textbf {\bibinfo {volume} {2017}},\ \bibinfo {eid} {95} (\bibinfo {year}
  {2017})}\BibitemShut {NoStop}%
\bibitem [{\citenamefont {{Haehl}}\ \emph {et~al.}(2016)\citenamefont
  {{Haehl}}, \citenamefont {{Loganayagam}},\ and\ \citenamefont
  {{Rangamani}}}]{rangamani}%
  \BibitemOpen
  \bibfield  {author} {\bibinfo {author} {\bibfnamefont {F.~M.}\ \bibnamefont
  {{Haehl}}}, \bibinfo {author} {\bibfnamefont {R.}~\bibnamefont
  {{Loganayagam}}},\ and\ \bibinfo {author} {\bibfnamefont {M.}~\bibnamefont
  {{Rangamani}}},\ }\bibfield  {title} {\bibinfo {title} {{Topological sigma
  models \& dissipative hydrodynamics}},\ }\href
  {https://doi.org/10.1007/JHEP04(2016)039} {\bibfield  {journal} {\bibinfo
  {journal} {Journal of High Energy Physics}\ }\textbf {\bibinfo {volume}
  {2016}},\ \bibinfo {eid} {39} (\bibinfo {year} {2016})},\ \Eprint
  {https://arxiv.org/abs/1511.07809} {arXiv:1511.07809 [hep-th]} \BibitemShut
  {NoStop}%
\bibitem [{\citenamefont {{Horowitz}}\ and\ \citenamefont
  {{Hubeny}}(2000)}]{horowitz}%
  \BibitemOpen
  \bibfield  {author} {\bibinfo {author} {\bibfnamefont {G.~T.}\ \bibnamefont
  {{Horowitz}}}\ and\ \bibinfo {author} {\bibfnamefont {V.~E.}\ \bibnamefont
  {{Hubeny}}},\ }\bibfield  {title} {\bibinfo {title} {{Quasinormal modes of
  AdS black holes and the approach to thermal equilibrium}},\ }\href
  {https://doi.org/10.1103/PhysRevD.62.024027} {\bibfield  {journal} {\bibinfo
  {journal} {\prd}\ }\textbf {\bibinfo {volume} {62}},\ \bibinfo {eid} {024027}
  (\bibinfo {year} {2000})},\ \Eprint {https://arxiv.org/abs/hep-th/9909056}
  {arXiv:hep-th/9909056 [hep-th]} \BibitemShut {NoStop}%
\bibitem [{\citenamefont {{Colin-Ellerin}}\ \emph
  {et~al.}(2021{\natexlab{a}})\citenamefont {{Colin-Ellerin}}, \citenamefont
  {{Dong}}, \citenamefont {{Marolf}}, \citenamefont {{Rangamani}},\ and\
  \citenamefont {{Wang}}}]{rangamani1}%
  \BibitemOpen
  \bibfield  {author} {\bibinfo {author} {\bibfnamefont {S.}~\bibnamefont
  {{Colin-Ellerin}}}, \bibinfo {author} {\bibfnamefont {X.}~\bibnamefont
  {{Dong}}}, \bibinfo {author} {\bibfnamefont {D.}~\bibnamefont {{Marolf}}},
  \bibinfo {author} {\bibfnamefont {M.}~\bibnamefont {{Rangamani}}},\ and\
  \bibinfo {author} {\bibfnamefont {Z.}~\bibnamefont {{Wang}}},\ }\bibfield
  {title} {\bibinfo {title} {{Real-time gravitational replicas: formalism and a
  variational principle}},\ }\href {https://doi.org/10.1007/JHEP05(2021)117}
  {\bibfield  {journal} {\bibinfo  {journal} {Journal of High Energy Physics}\
  }\textbf {\bibinfo {volume} {2021}},\ \bibinfo {eid} {117} (\bibinfo {year}
  {2021}{\natexlab{a}})},\ \Eprint {https://arxiv.org/abs/2012.00828}
  {arXiv:2012.00828 [hep-th]} \BibitemShut {NoStop}%
\bibitem [{\citenamefont {{Colin-Ellerin}}\ \emph
  {et~al.}(2021{\natexlab{b}})\citenamefont {{Colin-Ellerin}}, \citenamefont
  {{Dong}}, \citenamefont {{Marolf}}, \citenamefont {{Rangamani}},\ and\
  \citenamefont {{Wang}}}]{rangamani2}%
  \BibitemOpen
  \bibfield  {author} {\bibinfo {author} {\bibfnamefont {S.}~\bibnamefont
  {{Colin-Ellerin}}}, \bibinfo {author} {\bibfnamefont {X.}~\bibnamefont
  {{Dong}}}, \bibinfo {author} {\bibfnamefont {D.}~\bibnamefont {{Marolf}}},
  \bibinfo {author} {\bibfnamefont {M.}~\bibnamefont {{Rangamani}}},\ and\
  \bibinfo {author} {\bibfnamefont {Z.}~\bibnamefont {{Wang}}},\ }\bibfield
  {title} {\bibinfo {title} {{Real-time gravitational replicas: Low dimensional
  examples}},\ }\href@noop {} {\bibfield  {journal} {\bibinfo  {journal} {arXiv
  e-prints}\ } (\bibinfo {year} {2021}{\natexlab{b}})},\ \Eprint
  {https://arxiv.org/abs/2105.07002} {arXiv:2105.07002 [hep-th]} \BibitemShut
  {NoStop}%
\bibitem [{\citenamefont {{Cirac}}\ \emph {et~al.}(2021)\citenamefont
  {{Cirac}}, \citenamefont {{P{\'e}rez-Garc{\'\i}a}}, \citenamefont
  {{Schuch}},\ and\ \citenamefont {{Verstraete}}}]{mps_review}%
  \BibitemOpen
  \bibfield  {author} {\bibinfo {author} {\bibfnamefont {J.~I.}\ \bibnamefont
  {{Cirac}}}, \bibinfo {author} {\bibfnamefont {D.}~\bibnamefont
  {{P{\'e}rez-Garc{\'\i}a}}}, \bibinfo {author} {\bibfnamefont
  {N.}~\bibnamefont {{Schuch}}},\ and\ \bibinfo {author} {\bibfnamefont
  {F.}~\bibnamefont {{Verstraete}}},\ }\bibfield  {title} {\bibinfo {title}
  {{Matrix product states and projected entangled pair states: Concepts,
  symmetries, theorems}},\ }\href
  {https://doi.org/10.1103/RevModPhys.93.045003} {\bibfield  {journal}
  {\bibinfo  {journal} {Reviews of Modern Physics}\ }\textbf {\bibinfo {volume}
  {93}},\ \bibinfo {eid} {045003} (\bibinfo {year} {2021})}\BibitemShut
  {NoStop}%
\bibitem [{\citenamefont {Rakovszky}\ \emph {et~al.}(2018)\citenamefont
  {Rakovszky}, \citenamefont {Pollmann},\ and\ \citenamefont {von
  Keyserlingk}}]{rakovszky2018diffusive}%
  \BibitemOpen
  \bibfield  {author} {\bibinfo {author} {\bibfnamefont {T.}~\bibnamefont
  {Rakovszky}}, \bibinfo {author} {\bibfnamefont {F.}~\bibnamefont
  {Pollmann}},\ and\ \bibinfo {author} {\bibfnamefont {C.~W.}\ \bibnamefont
  {von Keyserlingk}},\ }\bibfield  {title} {\bibinfo {title} {Diffusive
  hydrodynamics of out-of-time-ordered correlators with charge conservation},\
  }\href {https://doi.org/10.1103/PhysRevX.8.031058} {\bibfield  {journal}
  {\bibinfo  {journal} {Phys. Rev. X}\ }\textbf {\bibinfo {volume} {8}},\
  \bibinfo {pages} {031058} (\bibinfo {year} {2018})}\BibitemShut {NoStop}%
\bibitem [{\citenamefont {Huang}(2020)}]{huang_2020}%
  \BibitemOpen
  \bibfield  {author} {\bibinfo {author} {\bibfnamefont {Y.}~\bibnamefont
  {Huang}},\ }\bibfield  {title} {\bibinfo {title} {Dynamics of rényi
  entanglement entropy in diffusive qudit systems},\ }\href
  {https://doi.org/10.1088/2633-1357/abd1e2} {\bibfield  {journal} {\bibinfo
  {journal} {IOP SciNotes}\ }\textbf {\bibinfo {volume} {1}},\ \bibinfo {pages}
  {035205} (\bibinfo {year} {2020})}\BibitemShut {NoStop}%
\bibitem [{\citenamefont {Zhou}\ and\ \citenamefont
  {Ludwig}(2020)}]{zhou2020diffusive}%
  \BibitemOpen
  \bibfield  {author} {\bibinfo {author} {\bibfnamefont {T.}~\bibnamefont
  {Zhou}}\ and\ \bibinfo {author} {\bibfnamefont {A.~W.~W.}\ \bibnamefont
  {Ludwig}},\ }\bibfield  {title} {\bibinfo {title} {Diffusive scaling of
  r\'enyi entanglement entropy},\ }\href
  {https://doi.org/10.1103/PhysRevResearch.2.033020} {\bibfield  {journal}
  {\bibinfo  {journal} {Phys. Rev. Res.}\ }\textbf {\bibinfo {volume} {2}},\
  \bibinfo {pages} {033020} (\bibinfo {year} {2020})}\BibitemShut {NoStop}%
\bibitem [{\citenamefont
  {{\v{Z}}nidari{\v{c}}}(2020)}]{znidaric2020entanglement}%
  \BibitemOpen
  \bibfield  {author} {\bibinfo {author} {\bibfnamefont {M.}~\bibnamefont
  {{\v{Z}}nidari{\v{c}}}},\ }\bibfield  {title} {\bibinfo {title} {Entanglement
  growth in diffusive systems},\ }\href
  {https://doi.org/10.1038/s42005-020-0366-7} {\bibfield  {journal} {\bibinfo
  {journal} {Communications Physics}\ }\textbf {\bibinfo {volume} {3}},\
  \bibinfo {pages} {100} (\bibinfo {year} {2020})}\BibitemShut {NoStop}%
\bibitem [{\citenamefont {Rakovszky}\ \emph {et~al.}(2021)\citenamefont
  {Rakovszky}, \citenamefont {Pollmann},\ and\ \citenamefont {von
  Keyserlingk}}]{rakovszky2021entanglement}%
  \BibitemOpen
  \bibfield  {author} {\bibinfo {author} {\bibfnamefont {T.}~\bibnamefont
  {Rakovszky}}, \bibinfo {author} {\bibfnamefont {F.}~\bibnamefont
  {Pollmann}},\ and\ \bibinfo {author} {\bibfnamefont {C.}~\bibnamefont {von
  Keyserlingk}},\ }\bibfield  {title} {\bibinfo {title} {Entanglement growth in
  diffusive systems with large spin},\ }\href
  {https://doi.org/10.1038/s42005-021-00594-4} {\bibfield  {journal} {\bibinfo
  {journal} {Communications Physics}\ }\textbf {\bibinfo {volume} {4}},\
  \bibinfo {pages} {91} (\bibinfo {year} {2021})}\BibitemShut {NoStop}%
\bibitem [{\citenamefont {Guardado-Sanchez}\ \emph {et~al.}(2020)\citenamefont
  {Guardado-Sanchez}, \citenamefont {Morningstar}, \citenamefont {Spar},
  \citenamefont {Brown}, \citenamefont {Huse},\ and\ \citenamefont
  {Bakr}}]{guardado2020subdiffusion}%
  \BibitemOpen
  \bibfield  {author} {\bibinfo {author} {\bibfnamefont {E.}~\bibnamefont
  {Guardado-Sanchez}}, \bibinfo {author} {\bibfnamefont {A.}~\bibnamefont
  {Morningstar}}, \bibinfo {author} {\bibfnamefont {B.~M.}\ \bibnamefont
  {Spar}}, \bibinfo {author} {\bibfnamefont {P.~T.}\ \bibnamefont {Brown}},
  \bibinfo {author} {\bibfnamefont {D.~A.}\ \bibnamefont {Huse}},\ and\
  \bibinfo {author} {\bibfnamefont {W.~S.}\ \bibnamefont {Bakr}},\ }\bibfield
  {title} {\bibinfo {title} {{Subdiffusion and Heat Transport in a Tilted
  Two-Dimensional {Fermi-Hubbard} System}},\ }\href
  {https://doi.org/10.1103/PhysRevX.10.011042} {\bibfield  {journal} {\bibinfo
  {journal} {Phys. Rev. X}\ }\textbf {\bibinfo {volume} {10}},\ \bibinfo
  {pages} {011042} (\bibinfo {year} {2020})}\BibitemShut {NoStop}%
\bibitem [{\citenamefont {Feldmeier}\ \emph {et~al.}(2020)\citenamefont
  {Feldmeier}, \citenamefont {Sala}, \citenamefont {De~Tomasi}, \citenamefont
  {Pollmann},\ and\ \citenamefont {Knap}}]{feldmeier2020anomalous}%
  \BibitemOpen
  \bibfield  {author} {\bibinfo {author} {\bibfnamefont {J.}~\bibnamefont
  {Feldmeier}}, \bibinfo {author} {\bibfnamefont {P.}~\bibnamefont {Sala}},
  \bibinfo {author} {\bibfnamefont {G.}~\bibnamefont {De~Tomasi}}, \bibinfo
  {author} {\bibfnamefont {F.}~\bibnamefont {Pollmann}},\ and\ \bibinfo
  {author} {\bibfnamefont {M.}~\bibnamefont {Knap}},\ }\bibfield  {title}
  {\bibinfo {title} {{Anomalous Diffusion in Dipole- and
  Higher-Moment-Conserving Systems}},\ }\href
  {https://doi.org/10.1103/PhysRevLett.125.245303} {\bibfield  {journal}
  {\bibinfo  {journal} {Phys. Rev. Lett.}\ }\textbf {\bibinfo {volume} {125}},\
  \bibinfo {pages} {245303} (\bibinfo {year} {2020})}\BibitemShut {NoStop}%
\bibitem [{\citenamefont {Moudgalya}\ \emph {et~al.}(2021)\citenamefont
  {Moudgalya}, \citenamefont {Prem}, \citenamefont {Huse},\ and\ \citenamefont
  {Chan}}]{moudgalya2021spectral}%
  \BibitemOpen
  \bibfield  {author} {\bibinfo {author} {\bibfnamefont {S.}~\bibnamefont
  {Moudgalya}}, \bibinfo {author} {\bibfnamefont {A.}~\bibnamefont {Prem}},
  \bibinfo {author} {\bibfnamefont {D.~A.}\ \bibnamefont {Huse}},\ and\
  \bibinfo {author} {\bibfnamefont {A.}~\bibnamefont {Chan}},\ }\bibfield
  {title} {\bibinfo {title} {Spectral statistics in constrained many-body
  quantum chaotic systems},\ }\href
  {https://doi.org/10.1103/PhysRevResearch.3.023176} {\bibfield  {journal}
  {\bibinfo  {journal} {Phys. Rev. Res.}\ }\textbf {\bibinfo {volume} {3}},\
  \bibinfo {pages} {023176} (\bibinfo {year} {2021})}\BibitemShut {NoStop}%
\bibitem [{\citenamefont {Moudgalya}\ and\ \citenamefont
  {Motrunich}(2024{\natexlab{b}})}]{moudgalya2022exhaustive}%
  \BibitemOpen
  \bibfield  {author} {\bibinfo {author} {\bibfnamefont {S.}~\bibnamefont
  {Moudgalya}}\ and\ \bibinfo {author} {\bibfnamefont {O.~I.}\ \bibnamefont
  {Motrunich}},\ }\bibfield  {title} {\bibinfo {title} {Exhaustive
  characterization of quantum many-body scars using commutant algebras},\
  }\href {https://doi.org/10.1103/PhysRevX.14.041069} {\bibfield  {journal}
  {\bibinfo  {journal} {Phys. Rev. X}\ }\textbf {\bibinfo {volume} {14}},\
  \bibinfo {pages} {041069} (\bibinfo {year} {2024}{\natexlab{b}})}\BibitemShut
  {NoStop}%
\bibitem [{\citenamefont {Gotta}\ \emph {et~al.}(2023)\citenamefont {Gotta},
  \citenamefont {Moudgalya},\ and\ \citenamefont
  {Mazza}}]{gotta2023asymptotic}%
  \BibitemOpen
  \bibfield  {author} {\bibinfo {author} {\bibfnamefont {L.}~\bibnamefont
  {Gotta}}, \bibinfo {author} {\bibfnamefont {S.}~\bibnamefont {Moudgalya}},\
  and\ \bibinfo {author} {\bibfnamefont {L.}~\bibnamefont {Mazza}},\ }\bibfield
   {title} {\bibinfo {title} {{Asymptotic Quantum Many-Body Scars}},\ }\href
  {https://doi.org/10.1103/PhysRevLett.131.190401} {\bibfield  {journal}
  {\bibinfo  {journal} {Phys. Rev. Lett.}\ }\textbf {\bibinfo {volume} {131}},\
  \bibinfo {pages} {190401} (\bibinfo {year} {2023})}\BibitemShut {NoStop}%
\bibitem [{\citenamefont {Feldmeier}\ \emph {et~al.}(2022)\citenamefont
  {Feldmeier}, \citenamefont {Witczak-Krempa},\ and\ \citenamefont
  {Knap}}]{feldmeier2022emergent}%
  \BibitemOpen
  \bibfield  {author} {\bibinfo {author} {\bibfnamefont {J.}~\bibnamefont
  {Feldmeier}}, \bibinfo {author} {\bibfnamefont {W.}~\bibnamefont
  {Witczak-Krempa}},\ and\ \bibinfo {author} {\bibfnamefont {M.}~\bibnamefont
  {Knap}},\ }\bibfield  {title} {\bibinfo {title} {Emergent tracer dynamics in
  constrained quantum systems},\ }\href
  {https://doi.org/10.1103/PhysRevB.106.094303} {\bibfield  {journal} {\bibinfo
   {journal} {Phys. Rev. B}\ }\textbf {\bibinfo {volume} {106}},\ \bibinfo
  {pages} {094303} (\bibinfo {year} {2022})}\BibitemShut {NoStop}%
\bibitem [{\citenamefont {{Khemani}}\ \emph {et~al.}(2018)\citenamefont
  {{Khemani}}, \citenamefont {{Huse}},\ and\ \citenamefont
  {{Nahum}}}]{velocity_dependent}%
  \BibitemOpen
  \bibfield  {author} {\bibinfo {author} {\bibfnamefont {V.}~\bibnamefont
  {{Khemani}}}, \bibinfo {author} {\bibfnamefont {D.~A.}\ \bibnamefont
  {{Huse}}},\ and\ \bibinfo {author} {\bibfnamefont {A.}~\bibnamefont
  {{Nahum}}},\ }\bibfield  {title} {\bibinfo {title} {{Velocity-dependent
  Lyapunov exponents in many-body quantum, semiclassical, and classical
  chaos}},\ }\href {https://doi.org/10.1103/PhysRevB.98.144304} {\bibfield
  {journal} {\bibinfo  {journal} {\prb}\ }\textbf {\bibinfo {volume} {98}},\
  \bibinfo {eid} {144304} (\bibinfo {year} {2018})},\ \Eprint
  {https://arxiv.org/abs/1803.05902} {arXiv:1803.05902 [cond-mat.stat-mech]}
  \BibitemShut {NoStop}%
\bibitem [{\citenamefont {{Stanford}}\ \emph {et~al.}(2023)\citenamefont
  {{Stanford}}, \citenamefont {{Vardhan}},\ and\ \citenamefont
  {{Yao}}}]{scramblon_loops}%
  \BibitemOpen
  \bibfield  {author} {\bibinfo {author} {\bibfnamefont {D.}~\bibnamefont
  {{Stanford}}}, \bibinfo {author} {\bibfnamefont {S.}~\bibnamefont
  {{Vardhan}}},\ and\ \bibinfo {author} {\bibfnamefont {S.}~\bibnamefont
  {{Yao}}},\ }\bibfield  {title} {\bibinfo {title} {{Scramblon loops}},\
  }\href@noop {} {\bibfield  {journal} {\bibinfo  {journal} {arXiv e-prints}\ }
  (\bibinfo {year} {2023})},\ \Eprint {https://arxiv.org/abs/2311.12121}
  {arXiv:2311.12121 [hep-th]} \BibitemShut {NoStop}%
\bibitem [{\citenamefont {Schuckert}\ \emph {et~al.}(2020)\citenamefont
  {Schuckert}, \citenamefont {Lovas},\ and\ \citenamefont
  {Knap}}]{schuckert2020nonlocal}%
  \BibitemOpen
  \bibfield  {author} {\bibinfo {author} {\bibfnamefont {A.}~\bibnamefont
  {Schuckert}}, \bibinfo {author} {\bibfnamefont {I.}~\bibnamefont {Lovas}},\
  and\ \bibinfo {author} {\bibfnamefont {M.}~\bibnamefont {Knap}},\ }\bibfield
  {title} {\bibinfo {title} {Nonlocal emergent hydrodynamics in a long-range
  quantum spin system},\ }\href {https://doi.org/10.1103/PhysRevB.101.020416}
  {\bibfield  {journal} {\bibinfo  {journal} {Phys. Rev. B}\ }\textbf {\bibinfo
  {volume} {101}},\ \bibinfo {pages} {020416} (\bibinfo {year}
  {2020})}\BibitemShut {NoStop}%
\bibitem [{\citenamefont {Morningstar}\ \emph {et~al.}(2023)\citenamefont
  {Morningstar}, \citenamefont {O'Dea},\ and\ \citenamefont
  {Richter}}]{morningstar2023hydrodynamics}%
  \BibitemOpen
  \bibfield  {author} {\bibinfo {author} {\bibfnamefont {A.}~\bibnamefont
  {Morningstar}}, \bibinfo {author} {\bibfnamefont {N.}~\bibnamefont {O'Dea}},\
  and\ \bibinfo {author} {\bibfnamefont {J.}~\bibnamefont {Richter}},\
  }\bibfield  {title} {\bibinfo {title} {Hydrodynamics in long-range
  interacting systems with center-of-mass conservation},\ }\href
  {https://doi.org/10.1103/PhysRevB.108.L020304} {\bibfield  {journal}
  {\bibinfo  {journal} {Phys. Rev. B}\ }\textbf {\bibinfo {volume} {108}},\
  \bibinfo {pages} {L020304} (\bibinfo {year} {2023})}\BibitemShut {NoStop}%
\bibitem [{\citenamefont {Gliozzi}\ \emph {et~al.}(2023)\citenamefont
  {Gliozzi}, \citenamefont {May-Mann}, \citenamefont {Hughes},\ and\
  \citenamefont {De~Tomasi}}]{gliozzi2023hierarchical}%
  \BibitemOpen
  \bibfield  {author} {\bibinfo {author} {\bibfnamefont {J.}~\bibnamefont
  {Gliozzi}}, \bibinfo {author} {\bibfnamefont {J.}~\bibnamefont {May-Mann}},
  \bibinfo {author} {\bibfnamefont {T.~L.}\ \bibnamefont {Hughes}},\ and\
  \bibinfo {author} {\bibfnamefont {G.}~\bibnamefont {De~Tomasi}},\ }\bibfield
  {title} {\bibinfo {title} {Hierarchical hydrodynamics in long-range
  multipole-conserving systems},\ }\href
  {https://doi.org/10.1103/PhysRevB.108.195106} {\bibfield  {journal} {\bibinfo
   {journal} {Phys. Rev. B}\ }\textbf {\bibinfo {volume} {108}},\ \bibinfo
  {pages} {195106} (\bibinfo {year} {2023})}\BibitemShut {NoStop}%
\bibitem [{\citenamefont {Sierant}\ \emph {et~al.}(2023)\citenamefont
  {Sierant}, \citenamefont {Schir\`o}, \citenamefont {Lewenstein},\ and\
  \citenamefont {Turkeshi}}]{sierant2023entanglement}%
  \BibitemOpen
  \bibfield  {author} {\bibinfo {author} {\bibfnamefont {P.}~\bibnamefont
  {Sierant}}, \bibinfo {author} {\bibfnamefont {M.}~\bibnamefont {Schir\`o}},
  \bibinfo {author} {\bibfnamefont {M.}~\bibnamefont {Lewenstein}},\ and\
  \bibinfo {author} {\bibfnamefont {X.}~\bibnamefont {Turkeshi}},\ }\bibfield
  {title} {\bibinfo {title} {Entanglement growth and minimal membranes in
  ($d+1$) random unitary circuits},\ }\href
  {https://doi.org/10.1103/PhysRevLett.131.230403} {\bibfield  {journal}
  {\bibinfo  {journal} {Phys. Rev. Lett.}\ }\textbf {\bibinfo {volume} {131}},\
  \bibinfo {pages} {230403} (\bibinfo {year} {2023})}\BibitemShut {NoStop}%
\bibitem [{\citenamefont {Sommers}\ \emph {et~al.}(2024)\citenamefont
  {Sommers}, \citenamefont {Gopalakrishnan}, \citenamefont {Gullans},\ and\
  \citenamefont {Huse}}]{sommers2024zero}%
  \BibitemOpen
  \bibfield  {author} {\bibinfo {author} {\bibfnamefont {G.~M.}\ \bibnamefont
  {Sommers}}, \bibinfo {author} {\bibfnamefont {S.}~\bibnamefont
  {Gopalakrishnan}}, \bibinfo {author} {\bibfnamefont {M.~J.}\ \bibnamefont
  {Gullans}},\ and\ \bibinfo {author} {\bibfnamefont {D.~A.}\ \bibnamefont
  {Huse}},\ }\bibfield  {title} {\bibinfo {title} {Zero-temperature
  entanglement membranes in quantum circuits},\ }\href
  {https://doi.org/10.1103/PhysRevB.110.064311} {\bibfield  {journal} {\bibinfo
   {journal} {Phys. Rev. B}\ }\textbf {\bibinfo {volume} {110}},\ \bibinfo
  {pages} {064311} (\bibinfo {year} {2024})}\BibitemShut {NoStop}%
\bibitem [{\citenamefont {Rampp}\ \emph {et~al.}(2024)\citenamefont {Rampp},
  \citenamefont {Rather},\ and\ \citenamefont
  {Claeys}}]{rampp2023entanglement}%
  \BibitemOpen
  \bibfield  {author} {\bibinfo {author} {\bibfnamefont {M.~A.}\ \bibnamefont
  {Rampp}}, \bibinfo {author} {\bibfnamefont {S.~A.}\ \bibnamefont {Rather}},\
  and\ \bibinfo {author} {\bibfnamefont {P.~W.}\ \bibnamefont {Claeys}},\
  }\bibfield  {title} {\bibinfo {title} {Entanglement membrane in exactly
  solvable lattice models},\ }\href
  {https://doi.org/10.1103/PhysRevResearch.6.033271} {\bibfield  {journal}
  {\bibinfo  {journal} {Phys. Rev. Res.}\ }\textbf {\bibinfo {volume} {6}},\
  \bibinfo {pages} {033271} (\bibinfo {year} {2024})}\BibitemShut {NoStop}%
\bibitem [{\citenamefont {Potter}\ and\ \citenamefont
  {Vasseur}(2022)}]{potter2022entanglement}%
  \BibitemOpen
  \bibfield  {author} {\bibinfo {author} {\bibfnamefont {A.~C.}\ \bibnamefont
  {Potter}}\ and\ \bibinfo {author} {\bibfnamefont {R.}~\bibnamefont
  {Vasseur}},\ }\bibinfo {title} {Entanglement dynamics in hybrid quantum
  circuits},\ in\ \href {https://doi.org/10.1007/978-3-031-03998-0_9} {\emph
  {\bibinfo {booktitle} {Entanglement in Spin Chains: From Theory to Quantum
  Technology Applications}}},\ \bibinfo {editor} {edited by\ \bibinfo {editor}
  {\bibfnamefont {A.}~\bibnamefont {Bayat}}, \bibinfo {editor} {\bibfnamefont
  {S.}~\bibnamefont {Bose}},\ and\ \bibinfo {editor} {\bibfnamefont
  {H.}~\bibnamefont {Johannesson}}}\ (\bibinfo  {publisher} {Springer
  International Publishing},\ \bibinfo {address} {Cham},\ \bibinfo {year}
  {2022})\ pp.\ \bibinfo {pages} {211--249}\BibitemShut {NoStop}%
\bibitem [{\citenamefont {Zaletel}\ \emph {et~al.}(2014)\citenamefont
  {Zaletel}, \citenamefont {Mong},\ and\ \citenamefont
  {Pollmann}}]{zaletel2014flux}%
  \BibitemOpen
  \bibfield  {author} {\bibinfo {author} {\bibfnamefont {M.~P.}\ \bibnamefont
  {Zaletel}}, \bibinfo {author} {\bibfnamefont {R.~S.~K.}\ \bibnamefont
  {Mong}},\ and\ \bibinfo {author} {\bibfnamefont {F.}~\bibnamefont
  {Pollmann}},\ }\bibfield  {title} {\bibinfo {title} {Flux insertion,
  entanglement, and quantized responses},\ }\href
  {https://doi.org/10.1088/1742-5468/2014/10/P10007} {\bibfield  {journal}
  {\bibinfo  {journal} {Journal of Statistical Mechanics: Theory and
  Experiment}\ }\textbf {\bibinfo {volume} {2014}},\ \bibinfo {pages} {P10007}
  (\bibinfo {year} {2014})}\BibitemShut {NoStop}%
\bibitem [{\citenamefont {Olver}(1997)}]{olver1997asymptotics}%
  \BibitemOpen
  \bibfield  {author} {\bibinfo {author} {\bibfnamefont {F.}~\bibnamefont
  {Olver}},\ }\href@noop {} {\emph {\bibinfo {title} {Asymptotics and special
  functions}}}\ (\bibinfo  {publisher} {AK Peters/CRC Press},\ \bibinfo {year}
  {1997})\BibitemShut {NoStop}%
\bibitem [{\citenamefont {Bhardwaj}\ \emph {et~al.}(2025)\citenamefont
  {Bhardwaj}, \citenamefont {Bottini}, \citenamefont {Sch\"afer-Nameki},\ and\
  \citenamefont {Tiwari}}]{bhardwaj2025illustrating}%
  \BibitemOpen
  \bibfield  {author} {\bibinfo {author} {\bibfnamefont {L.}~\bibnamefont
  {Bhardwaj}}, \bibinfo {author} {\bibfnamefont {L.~E.}\ \bibnamefont
  {Bottini}}, \bibinfo {author} {\bibfnamefont {S.}~\bibnamefont
  {Sch\"afer-Nameki}},\ and\ \bibinfo {author} {\bibfnamefont {A.}~\bibnamefont
  {Tiwari}},\ }\bibfield  {title} {\bibinfo {title} {Illustrating the
  categorical landau paradigm in lattice models},\ }\href
  {https://doi.org/10.1103/PhysRevB.111.054432} {\bibfield  {journal} {\bibinfo
   {journal} {Phys. Rev. B}\ }\textbf {\bibinfo {volume} {111}},\ \bibinfo
  {pages} {054432} (\bibinfo {year} {2025})}\BibitemShut {NoStop}%
\bibitem [{\citenamefont {Moudgalya}\ \emph {et~al.}(2018)\citenamefont
  {Moudgalya}, \citenamefont {Rachel}, \citenamefont {Bernevig},\ and\
  \citenamefont {Regnault}}]{moudgalya2018exact}%
  \BibitemOpen
  \bibfield  {author} {\bibinfo {author} {\bibfnamefont {S.}~\bibnamefont
  {Moudgalya}}, \bibinfo {author} {\bibfnamefont {S.}~\bibnamefont {Rachel}},
  \bibinfo {author} {\bibfnamefont {B.~A.}\ \bibnamefont {Bernevig}},\ and\
  \bibinfo {author} {\bibfnamefont {N.}~\bibnamefont {Regnault}},\ }\bibfield
  {title} {\bibinfo {title} {Exact excited states of nonintegrable models},\
  }\href {https://doi.org/10.1103/PhysRevB.98.235155} {\bibfield  {journal}
  {\bibinfo  {journal} {Physical Review B}\ }\textbf {\bibinfo {volume} {98}},\
  \bibinfo {pages} {235155} (\bibinfo {year} {2018})}\BibitemShut {NoStop}%
\bibitem [{\citenamefont {{Roberts}}\ \emph {et~al.}(2015)\citenamefont
  {{Roberts}}, \citenamefont {{Stanford}},\ and\ \citenamefont
  {{Susskind}}}]{roberts}%
  \BibitemOpen
  \bibfield  {author} {\bibinfo {author} {\bibfnamefont {D.~A.}\ \bibnamefont
  {{Roberts}}}, \bibinfo {author} {\bibfnamefont {D.}~\bibnamefont
  {{Stanford}}},\ and\ \bibinfo {author} {\bibfnamefont {L.}~\bibnamefont
  {{Susskind}}},\ }\bibfield  {title} {\bibinfo {title} {{Localized shocks}},\
  }\href {https://doi.org/10.1007/JHEP03(2015)051} {\bibfield  {journal}
  {\bibinfo  {journal} {Journal of High Energy Physics}\ }\textbf {\bibinfo
  {volume} {2015}},\ \bibinfo {eid} {51} (\bibinfo {year} {2015})},\ \Eprint
  {https://arxiv.org/abs/1409.8180} {arXiv:1409.8180 [hep-th]} \BibitemShut
  {NoStop}%
\bibitem [{\citenamefont {{Roberts}}\ \emph {et~al.}(2018)\citenamefont
  {{Roberts}}, \citenamefont {{Stanford}},\ and\ \citenamefont
  {{Streicher}}}]{syk}%
  \BibitemOpen
  \bibfield  {author} {\bibinfo {author} {\bibfnamefont {D.~A.}\ \bibnamefont
  {{Roberts}}}, \bibinfo {author} {\bibfnamefont {D.}~\bibnamefont
  {{Stanford}}},\ and\ \bibinfo {author} {\bibfnamefont {A.}~\bibnamefont
  {{Streicher}}},\ }\bibfield  {title} {\bibinfo {title} {{Operator growth in
  the SYK model}},\ }\href {https://doi.org/10.1007/JHEP06(2018)122} {\bibfield
   {journal} {\bibinfo  {journal} {Journal of High Energy Physics}\ }\textbf
  {\bibinfo {volume} {2018}},\ \bibinfo {eid} {122} (\bibinfo {year} {2018})},\
  \Eprint {https://arxiv.org/abs/1802.02633} {arXiv:1802.02633 [hep-th]}
  \BibitemShut {NoStop}%
\end{thebibliography}%

\onecolumngrid
\appendix

\section{Derivation of the equilibrium approximation for Brownian models}
\label{app:eq}

In this Appendix, we provide some details of the derivation of the equilibrium approximation for Brownian models, discussed in Sec.~\ref{sec:eq}, as well as some  explicit examples.  

Let us first justify the assumption that the states $\ket{m_1, ..., m_n;~\sigma}$ defined in \eqref{basis_2n} can be treated as orthonormal.
First note that since $\{Q_m\}$ is an orthonormal basis for $\sC$, we have $\braket{m'_1, ..., m'_n; \sigma | m_1, ..., m_n; \sigma} = \prod_{j}{\delta_{m_j, m'_j}}$.
Next, consider the overlap between  states with $\sigma \neq \tau$.   
For $\sigma, \tau \in \sS_n$, suppose  $\sigma^{-1}\tau$ has $k$ cycles, with lengths $l_1, ..., l_k$ respectively.
More explicitly, say the cycle decomposition  is given by 
\be 
\sigma^{-1}\tau= (a_{1, 1} ... a_{1, l_1}) \, (a_{2, 1} ... a_{2, l_2}) \, ... \, (a_{k, 1} ... a_{k, l_k}) \, \, . \label{permdefs}
\ee
Then we have the following general expression for the overlap:
\begin{equation}
\braket{m'_1, ..., m'_n ;~\sigma | m_1, ..., m_n;~\tau} = \prod_{i=1}^k\Tr[ Q_{m'_{a_{i,1}}}^{\dagger} Q_{m_{a_{i,1}}} Q_{m'_{a_{i,2}}}^{\dagger} Q_{m_{a_{i,2}}}... Q_{m'_{a_{i,l_i}}}^{\dagger} Q_{m_{a_{i,l_i}}} ] \, \label{offd}
\end{equation}
In most examples of interest, the products \eqref{offd} for $\sigma \neq \tau$ are suppressed compared to 1 in powers of the total Hilbert space dimension.
For example, in the case with no symmetries, we only have the option $m=1$ corresponding to the normalized identity operator $I/\sqrt{D}$, and hence the overlap in \eqref{offd} is $D^{k-n}$, where $D$ is the total Hilbert space dimension.
Another simple example we can consider is a spin-1/2 chain  of length $L$ with conserved $U(1)$ charge $\sum_{i=1}^L Z_i$.
In this case, we can take take a basis for $\sC$ consisting of projectors onto sectors with different values of the total  spin.
Let $P_0$ be the projector $\ket{0}\bra{0}$ onto the one-site state with  spin $-1$, and $P_1$ the projector $\ket{1}\bra{1}$ onto the state with  spin 1.
Then we can take the following basis of $L+1$ operators for $\sC$:  
\be 
Q_m~ = ~\frac{1}{\sqrt{{L \choose m} }}\sum_{i_1< ...<i_m} \otimes_{k=1}^m (P_1)_{i_k} \otimes_{p \notin \{i_1, ..., i_m \}} (P_0)_p \, \, , \quad m = 0, ..., L   \, .  \label{qm_u1}
\ee
For this case, we can check that 
$\braket{m'_1, ..., m'_n ;~\sigma | m_1, ..., m_n;~\tau}$ for $\sigma \neq \tau$  is either exactly zero, or suppressed in powers of ${L \choose m_i}$, ${L \choose m_i'}$, which are typically large.

Based on the above argument, the states $\ket{m_1, ..., m_n;~\sigma}$  can be treated as an approximately orthonormal basis for the ground state subspace for $P_{2n}$.
Then by putting \eqref{late_t_proj} into the expression \eqref{eq:renyi_avg} for the Renyi entropy, we find 
\begin{equation} 
\lim_{t\to\infty}\overline{\Tr[\rho_A(t)^n]}  = \sum_{m_1, ..., m_n; \sigma}\sum_{\sigma}\braket{\eta_R \otimes e_{\bar R}| Q_{m_1}, ..., Q_{m_n} \, ; \sigma } \braket{Q_{m_1}, ..., Q_{m_n} \, ; \sigma | \rho_0, e} \label{renyilate}
\end{equation}
Now suppose that $\sigma$ has $k$ cycles, and its cycle structure is
\be 
\sigma=  (b_{1, 1} ... b_{1, l_1}) \, (b_{2, 1} ... b_{2, l_2}) \, ... \, (b_{k, 1} ... b_{k, l_k})  \, . 
\ee
Then
\be 
\braket{Q_{m_1}, ..., Q_{m_n} \, ; \sigma | \rho_0, e} = \prod_{i=1}^k\Tr[Q_{m_{b_{i,1}}}^{\dagger}\rho_0... Q_{m_{b_{i,l_i}}}^{\dagger}\rho_0 ] 
\ee
In the case where $\rho_0$ is a pure state, the above expression simplifies to 
\be 
\braket{Q_{m_1}, ..., Q_{m_n} \, ; \sigma | \rho_0, e}  = \prod_{i=1}^n \Tr[Q_{m_i}^{\dagger} \rho_0]
\ee
Combining this with the other factor inside the sum in \eqref{renyilate}, we find 
\begin{align} \label{eqa} 
\lim_{t\to\infty}\overline{\Tr[\rho_R(t)^n]} = \sum_{\sigma}\braket{\eta_R \otimes e_{\bar R}| \rho^{\rm (eq)}, \sigma } 
\end{align}
with
\be \label{rho_eq}
\rho^{\rm (eq)} = \sum_m\Tr[Q_m^{\dagger}\rho_0] Q_m \, . 
\ee

In the case with no symmetries, $\rho^{\rm (eq)}$ is always the infinite temperature density matrix $I/D$.
In the spin-1/2 $U(1)$ case discussed above, suppose $\rho_0$ is an initial state with a fixed total spin $q$.
Then using \eqref{qm_u1}, 
\be 
\rho^{\rm (eq)} = ~\frac{1}{{L \choose q }}\sum_{i_1< ...<i_q} \otimes_{k=1}^q (P_1)_{i_k} \otimes_{p \notin \{i_1, ..., i_q \}} (P_0)_p \, 
\ee
which is the maximally mixed density matrix restricted to the spin $q$ quantum number sector. 
\section{ Details on the Peschel-Emery Hamiltonian}\label{app:num}
In this section, we will discuss the details of the GUE superhamiltonian $P_4$ of Eq.~(\ref{aleft_def_mt}), show the mapping to the Peschel-Emery Hamiltonian of Eq.~(\ref{eq:PEspin_mt}), and discuss  numerical and analytical methods  to probe its low-energy spectrum. 
\subsection{Mapping to the Peschel-Emery Hamiltonian}\label{subsec:mapping}
For numerical exact diagonalization, it is convenient to work in the orthonormal basis $\{\ket{+}, \ket{-}\}$ of Eq.~(\ref{eq:orthobasis}), instead of the biorthogonal system $\{\ket{\up}, \ket{\bup}, \ket{\dn}, \ket{\bdn}\}$ of Eqs.~(\ref{spin_def}) and (\ref{bar_spin_def}) that are convenient for analytical calculations discussed in this work.
In this biorthogonal system, operators can be given the following matrix representation:\footnote{Note that one could also work in a different biorthogonal system obtained by interchanging the $\{\ket{\bup}, \ket{\bdn}\}$ and $\{\ket{\up}, \ket{\dn}\}$ in Eq.~(\ref{eq:leftbiorthogonalops}), and the matrix representation of operators in that system would be related as  $\hO^{(\up\dn, \bup\bdn)} = (\hO^{(\bup\bdn, \up\dn)})^T$.}
\begin{equation}
    \hO = a_{\up\bup} \ketbra{\bup}{\up} + a_{\up\bdn} \ketbra{\bup}{\dn} + a_{\dn\bup} \ketbra{\bdn}{\up} + a_{\dn\bdn} \ketbra{\bdn}{\dn} \;\;\;\iff\;\;\;O^{(\bup \bdn, \up\dn)} =
    \begin{pmatrix}
    a_{\up\bup} & a_{\up\bdn} \\
    a_{\dn\bup} & a_{\dn\bdn} 
    \end{pmatrix},\;\;\;a_{s'\bar{s}} \defn \bra{s'} \hO \ket{\bar{s}},
\label{eq:leftbiorthogonalops}
\end{equation}
whereas in the orthonormal basis of (\ref{eq:orthobasis}), the same operator can be represented as:
\begin{equation}
    \hO = a_{++} \ketbra{+}{+} + a_{+-} \ketbra{+}{-} a_{-+} \ketbra{-}{+} + a_{--} \ketbra{-}{-} \;\;\;\iff\;\;\;O^{(+-,+-)} =
    \begin{pmatrix}
    a_{++} & a_{+-} \\
    a_{-+} & a_{--} 
    \end{pmatrix},\;\;\;a_{ss'} = \bra{s}\hO\ket{s'}.
\label{eq:orthonormalops}
\end{equation}
The transformation of the vector (matrix) representations of states (operators) between the biorthogonal system and the orthonormal basis can be compactly described as a similarity transformation.
Denoting the vector in the basis $\{\ket{a}, \ket{b}\}$ with a superscript such as $v^{(ab)}$, we can use Eq.~(\ref{eq:orthobasis}) to directly see that
\begin{equation}
    v^{(+-)} =  S v^{(\bup\bdn)} = (S^{-1})^T v^{(\up\dn)},\;\;\;S = 
    \begin{pmatrix}
        \frac{1}{\sqrt{2(1 + \frac{1}{q})}} & \frac{1}{\sqrt{2(1 + \frac{1}{q})}}\\
        \frac{1}{\sqrt{2(1 - \frac{1}{q})}} & -\frac{1}{\sqrt{2(1 - \frac{1}{q})}}
    \end{pmatrix},\;\;\; S^{-1} = 
    \begin{pmatrix}
        \sqrt{\frac{q+1}{2q}} & \sqrt{\frac{q-1}{2q}} \\
        \sqrt{\frac{q+1}{2q}} & -\sqrt{\frac{q-1}{2q}}
    \end{pmatrix}
\label{eq:basistransforms}
\end{equation}
This  leads to the relation between the matrix representations of an operator $\hO$ in the biorthogonal system and orthonormal basis as
\begin{equation}
    O^{(+-,+-)} = S O^{(\bup\bdn, \up\dn)} S^{-1}.
\label{eq:leftbiorthsimilarity}
\end{equation} 
In the biorthonormal basis, the nearest neighbor term of the superhamiltonian $P_4$ of Eq.~(\ref{aleft_def_mt}) reads
\begin{equation}
    H^{B}_{nn} = 
    \begin{pmatrix}
    0 & 0 & 0 & 0\\
    -\frac{1}{q} & 1 & \frac{1}{q^2} & -\frac{1}{q} \\
    -\frac{1}{q} & \frac{1}{q^2} & 1 & -\frac{1}{q}\\
    0 & 0 & 0 & 0
    \end{pmatrix},
\label{eq:biorthogonalPE}
\end{equation}
with basis elements on the rows (columns) ordered as $\{\up\up, \up\dn, \dn\up, \dn\dn\}$ ($\{\bup\bup, \bup\bdn, \bdn\bup, \bdn\bdn\}$).
This can be converted into the orthonormal basis ordered as $\{++, +-, -+, --\}$ using a similarity transformation with the matrix $S \otimes S$:
\begin{align}
    H^{O}_{nn} &= (S \otimes S) H^{B}_{nn} (S^{-1} \otimes S^{-1}) = \frac{1}{2}[\mathds{1} \otimes \mathds{1} - X \otimes X - \frac{1}{q}(Z \otimes \mathds{1} + \mathds{1} \otimes Z) + \frac{1}{q^2}(Z \otimes Z + X \otimes X)],
\label{eq:orthonormalPE}
\end{align}
where $X, Y, Z$ are the $2 \times 2$ Pauli matrices in the $\{\ket{+}, \ket{-}\}$ basis.
This leads to the expression of (\ref{eq:PEspin_mt}) for $P_4$ on any general lattice.
This Hamiltonian has exact frustration-free ground states $\ket{G_\up}$ and $\ket{G_\dn}$ are of the form of Eq.~(\ref{eq:P4gs}), as evident from the fact that $H^O_{nn}\ket{\up\up} = H^O_{nn}\ket{\dn\dn} = 0$.
The representation Eq.~(\ref{eq:PEspin}) in the orthonormal basis of \eqref{eq:orthobasis} also clearly reveals the $Z_2$ symmetry of the system, and that $\ket{G_\up}$ and $\ket{G_\dn}$ spontaneously break this symmetry,
\begin{equation}
    Q = \prodal{j}{}{Z_j},\;\;\;Q\ket{G_\up} = \ket{G_\dn},\;\;Q\ket{G_\dn} = \ket{G_\up}.
\label{eq:Z2symmetry}
\end{equation}
Restricted to one spatial dimension with $L$ sites labelled from $1$ to $L$, the Hamiltonian of (\ref{eq:PEspin_mt}) reads
\begin{equation}
    H = \frac{1}{2}\sum_{j = 1}^{L_{max}}{[1 - X_i X_{j+1} - \frac{1}{q}(Z_j + Z_{j+1}) + \frac{1}{q^2}(Z_j Z_{j+1} + X_j X_{j+1})]},
\label{eq:PEspin}
\end{equation}
where $L_{max} = L-1$ for OBC and $L_{max} = L$ for PBC. 
As we now show, this exactly lies on the Peschel-Emery line~\cite{peschel1981calculation, katsura2015exact} in the Ising ferromagnetic phase.
This can be explicitly seen in the fermion language using a Jordan-Wigner transformation with the substitutions
\begin{equation}
    X_j = (-1)^{\sum_{k < j}{n_k}}(\cd_j + c_j),\;\;Y_j = -i(-1)^{\sum_{k < j}{n_k}}(\cd_j - c_j),\;\;Z_j = -(-1)^{n_j} = 2 n_j - 1,
\label{eq:JWtransforms}
\end{equation}
where $\cd_j$, $c_j$, and $n_j \defn \cd_j c_j$ are the fermion creation, annihilation, and number operators. 
Up to an overall constant of $(L-1)/2$, the Hamiltonian of Eq.~(\ref{eq:PEspin}) then maps to the standard form of the Kitaev Hamiltonian, given by Eq.~(4) of \cite{katsura2015exact}:
\begin{equation}
    H_{\text{Kitaev}} = - t\sum_{j}{}{(\cd_j c_{j+1} + h.c.)} - \Delta \sum_j{(\cd_{j} \cd_{j+1} + h.c.)} - \frac{\mu}{2}\sum_j{(n_j + n_{j+1} - 1)} + U\sum_j{(2n_j - 1 ) (2 n_{j+1} - 1)}, 
\label{eq:PEfermionstandard}
\end{equation}
with the parameters
\begin{equation}
    \Delta = t = \frac{1}{2}(1 - \frac{1}{q^2}),\;\;\mu = \frac{2}{q},\;\;U = \frac{1}{2 q^2}\;\;\;\implies \mu = 4\sqrt{U^2 + t U}. 
\end{equation}
This satisfies the Peschel-Emery condition for having frustration-free exact ground states~\cite{katsura2015exact}. 
To relate these parameters to the phase diagram of the Kitaev model in Fig.~1 of \cite{katsura2015exact}, we can use the parametrization
\begin{equation}
    \frac{\mu}{t} = \frac{4q}{q^2 - 1},\;\;\;\frac{U}{t} = \frac{1}{q^2 - 1}. 
\end{equation}
It is clear that this entire line lies in the Ising $Z_2$ symmetry broken phase, which is also the topological phase in the Kitaev chain.
Large $q$ corresponds to the weak interaction limit, and smaller $q$ to stronger interactions. 
Moreover, $q = 1$ is a pathological point, since Eq.~(\ref{eq:PEspin}) becomes a commuting projector Hamiltonian with an integer spectrum.
\subsection{Low-energy excitations from twisted boundary conditions}
We now describe a method to numerically study the low-energy excitations of the Peschel-Emery Hamiltonian of Eq.~(\ref{eq:PEspin_mt}) using exact diagonalization.
Since its exact ground states spontaneously break the $Z_2$ symmetry of Eq.~(\ref{eq:Z2symmetry}), we expect the low-energy excitations to be gapped Domain Walls (DW) between the two ground states of the form (\ref{phikdef}).
A technical obstacle to obtaining the dispersion relation of the low-energy mode with OBC is the boundary condition.
While momentum is not well-defined with OBC due to lack of translation invariance in a finite system, PBC forbids single DW excitations just by geometry, i.e., DWs always occur in pairs with PBC.  
However, the single DW dispersion relation can be probed by imposing symmetry-twisted boundary conditions (also called anti-periodic BCs) by inserting a symmetry flux into the system, which pins the position of one of the DWs while allowing the other DW to move freely.
A standard procedure for inserting a symmetry flux to any symmetric Hamiltonian $H$ with an on-site internal symmetry $Q = \prod_{j}{g_j}$ is described in \cite{zaletel2014flux}, which we summarize here.
The chain with PBC is divided into three contiguous parts $A$, $B$, $C$, and the symmetry operations restricted to each part is defined as $Q_\alpha = \prod_{j \in \alpha}{g_j}$, $\alpha \in \{A, B, C\}$.
The Hamiltonian is also divided into three disjoint parts $H = H_{AB} + H_{BC} + H_{CA}$, where $H_{\alpha\beta}$ for $\alpha,\beta \in \{A, B, C\}$ only contains terms that are completely within the region $\alpha \cup \beta$. 
The twisted Hamiltonian is then given by $H_{tw} = Q_A H_{AB} Q_{A}^\dagger + H_{BC} + H_{CA}$. 
Since the terms of $H$ that are completely within or completely outside of the region $A$ commute with $Q_A$ (due to its on-site structure), the addition of a symmetry flux only changes the Hamiltonian terms that straddle $A$ and $B$. 
While the twisting breaks the translation symmetry of the system, it preserves a twisted translation symmetry, given by the operator $T_{tw}$, which satisfies $T_{tw}^L = Q$, where $Q$ is the symmetry operator.
The eigenstates of $H$ can hence be labelled by a momentum w.r.t. the twisted translation operator, in particular the symmetry sector corresponding to symmetry $Q$ eigenvalue $e^{i\theta}$ splits into $L$ sectors labelled by twisted momenta $k \in (2 \pi \mathbb{Z} + \theta)/L$.
For large enough system sizes, we expect the low-energy excitations of the system with twisted BCs, which are single DWs, matches that of OBC, which are also single DWs. 
However, the correspondence is not one to one, for example, the symmetry-broken ground states of the OBCs are themselves not the ground states of the Hamiltonian with twisted BCs.  
Inserting a $Z_2$ symmetry flux in the PBC Peschel-Emery Hamiltonian of Eq.~(\ref{eq:PEspin}), we obtain the twisted Hamiltonian
\begin{equation}
    H_{tw} = \frac{1}{2}\sum_{j = 1}^{L-1}{[1 - X_j X_{j+1} - \frac{1}{q}(Z_j + Z_{j+1}) + \frac{1}{q^2}(Z_j Z_{j+1} + X_j X_{j+1})]} + \frac{1}{2}[1 + X_L X_{1} - \frac{1}{q}(Z_L + Z_{1}) + \frac{1}{q^2}(Z_L Z_{1} - X_L X_{1})],
\label{eq:PEtwist}
\end{equation}
which has a twisted translation symmetry that satisfies $T_{tw}^L = \prod_{j}{Z_j}$. 
Hence we can obtain its eigenstates separately in the $\prod_{j}{Z_j} = + 1$ sector, where the momenta are quantized as $2\pi\mathbb{Z}/L$ and in the $\prod_{j}{Z_j} = - 1$, where the momenta are quantized as $(2\mathbb{Z}+1)\pi/L$.
It is easy to check that the last term in Eq.~(\ref{eq:PEtwist}) has ground states of the form $\{\ket{\up\dn}, \ket{\dn\up}\}$ unlike the terms in the sum, which have ground states $\{\ket{\up\up}, \ket{\dn\dn}\}$. 
Hence in the low-energy spectrum, one DW is pinned to the boundary  whereas the other DW can disperse, revealing a clear band of one DW states shown in Fig.~\ref{fig:PEspectra}. 
We have verified that states of similar energies also appear in the spectrum of the OBC Hamiltonian, which shows the correspondence between systems with OBC and twisted BCs. 
This physics can also be checked in the exactly solvable transverse-field Ising model in the ferromagnetic phase.
\subsection{Analytical estimate of the dispersion relation}\label{subsec:varcalc}
In this section, we give an estimate of the dispersion relation in the Peschel-Emery model at finite $q$ by evaluating its expectation value in the state  \eqref{phikdef} with $\Delta=0$.
This calculation can be done semi-analytically, and gives a reasonable estimate for the exact dispersion relation.
We also show that the corresponding variance is small for large $q$.

The simplest expression for a single DW state is obtained by setting $\Delta = 0$ in Eq.~(\ref{phikdef}):
\begin{equation}
    \ket{\psi_n} \defn \frac{1}{\sqrt{\mN_n}}\sum_{x=1}^{L-1}{c_{n,x}\ket{D_x}},\;\;\;\mN_n \defn \sum_{x,x'}{c^\ast_{n,x} c_{n,x'} \braket{D_x|D_{x'}}} = \sum_{x,x'}{c^\ast_{n,x} c_{n,x'} \frac{1}{q^{|x-x'|}}},\;\;\;c_{n,x} = \exp(-i\frac{n \pi}{L} x)
\label{eq:normexp}
\end{equation}
where $\ket{D_x} = \ket{\dn \cdots \dn_x \up_{x+1}\cdots \up}$, $\mN_n$ is a normalization constant, and we have chosen momentum to be quantized in units of $\frac{n \pi}{L}$, which is the case with OBC.
While $\ket{\psi_n}$ is not orthogonal to the ground states $\ket{G_\up}$ and $\ket{G_\dn}$, we find that orthonormalizing it does not change the computations below.
For these computations, it is convenient to invoke the representation of the Hamiltonian in the non-orthogonal $\{\bup, \bdn\}$ basis, which is different from the biorthogonal and the orthonormal basis discussed in Sec.~\ref{subsec:mapping}. 
Following the discussion there, we can derive the matrix representation of the nearest-neighbor term of Eq.~(\ref{eq:orthonormalPE}) in this basis by a simple transformation (which is not a similarity transformation, since the basis is no longer biorthogonal):
\begin{equation}
    H^{(NO)}_{nn} = (S^{-1})^T H^{O}_{nn} S^{-1}\;\; \implies\;\;\hH_{nn} = \left(1 - \frac{1}{q^2}\right)^2(\ketbra{\bup\bdn}{\bup\bdn} + \ketbra{\bdn\bup}{\bdn\bup}).
\label{eq:Hnnbadbasis}
\end{equation}
With this expression, and denoting the Hamiltonian as $H = \sum_j{\hh_{j,j+1}}$, where $\hh_{j,j+1}$ is the nearest-neighbor term, the expectation value of the energy in the state (\ref{eq:normexp}):
\begin{align}
    E_n &\defn \bra{\psi_n} H \ket{\psi_n} = \frac{1}{\mN_n}\sum_{x, x'}{c^\ast_{n,x} c_{n,x'} \bra{D_x} H \ket{D_{x'}}} = \frac{1}{\mN_n}\sum_{x, x', j}{c^\ast_{n,x} c_{n,x'} \bra{D_x} \hh_{j,j+1} \ket{D_{x'}}},\nonumber \\
    &= \frac{1}{\mN_n}\sum_{x=1}^{L-1}{c^\ast_{n,x} c_{n,x} \bra{\dn\up} \hH_{nn} \ket{\dn\up}} = \frac{1}{\mN_n}\left(1 - \frac{1}{q^2}\right)^2 \sum_{x=1}^{L-1}{c^\ast_{n,x} c_{n,x}} = \frac{L-1}{\mN_n}\left(1 - \frac{1}{q^2}\right)^2.
\label{eq:enerexp}
\end{align}
where in the second line we have used the fact that $\hh_{j,j+1}\ket{D_x} = 0$ unless $x = j$.
We numerically observe that it can be written as
\begin{equation}
    E_n = \frac{L-1}{\mN_n}\left(1 - \frac{1}{q^2}\right)^2 = 1 - \frac{2}{q + q^{-1}}\cos(\frac{n \pi}{L}) + \delta_n,
\label{eq:Evarapprox}
\end{equation}
where $|\delta_n|$ appears to be of $O(1/q^3)$ and decreasing with $L$ for the system sizes we can access.
Setting $\delta_n = 0$, we obtain a ``toy" dispersion relation that gives rise to a membrane tension that satisfied the constraints of (\ref{const}), as discussed in App.~\ref{sec:hermitiantoy}.
Indeed, the curves shown in Fig.~\ref{fig:variational_gue_2renyi} obtained using the methods of Appendix~\ref{app:var} for larger values of $\Delta$ are also well approximated by Eq.~(\ref{eq:Evarapprox}) with $\delta_n = 0$.

To understand why $\{\ket{\psi_n}\}$ are good trial states, we compute their energy variances, which are given by
\begin{align}
    \sigma^2_n &\defn \bra{\psi_n} H^2 \ket{\psi_n} - (E_n)^2 = \frac{1}{\mN_n}\sum_{x, x'}{c^\ast_{n,x} c_{n,x'} \bra{D_x} H^2 \ket{D_{x'}}} - (E_n)^2\nonumber\\
    &= \frac{1}{\mN_n}\sum_{x = 1}^{L-1}{c^\ast_{n,x} c_{n,x} \bra{D_x} \hh_{x, x+1}^2 \ket{D_{x}}} + \frac{1}{\mN_n}\sum_{x=2}^{L-1}{(c^\ast_{n,{x-1}} c_{n,x} + c^\ast_{n,{x}} c_{n,x-1}) \bra{D_{x-1}} \hh_{x-1, x} \hh_{x, x+1} \ket{D_{x}}} - (E_n)^2\nonumber \\
    &= \frac{1}{\mN_n}\left(1 - \frac{1}{q^2}\right)^2\sum_{x=1}^{L-1}{c^\ast_{n,x} c_{n,x}} - \frac{1}{q \mN_n}\left(1 - \frac{1}{q^2}\right)^3 \sum_{x=2}^{L-1}{(c^\ast_{n,x-1} c_{n,x}+c^\ast_{n,x} c_{n,x-1})} - (E^{(var)}_n)^2 \nonumber \\
    &= \frac{L-1}{\mN_n}\left(1 - \frac{1}{q^2}\right)^2 \left[1 - \frac{2(L-2)}{q(L-1)}\left(1 - \frac{1}{q^2}\right)\cos(\frac{n \pi}{L}) - \frac{L-1}{\mN_n}\left(1 - \frac{1}{q^2}\right)^2 \right],
\label{eq:varexp}
\end{align}
where we have evaluated the various matrix elements using the expressions of (\ref{eq:Hnnbadbasis}).
Using (\ref{eq:Evarapprox}), we further obtain
\begin{equation}
\sigma^2_n = E_n \left(\delta_n + \cos(\frac{n \pi}{L})\left[\frac{2}{q + q^{-1}} - \frac{2(L-2)}{q(L-1)}\left(1 -\frac{1}{q^2}\right)\right]\right) \approx E_n \left(\delta_n + \frac{2}{q^3(q^2+1)} \cos(\frac{n \pi}{L})\right),
\end{equation}
where in the last step we have used that $L$ large.
Since $|E_n| \sim O(1)$, we have that $\sigma^2_n \sim O(|\delta_n|)$, which from our numerical observations is $O(1/q^3)$ and appears to be decreasing with system size $L$. 
\section{Exactly solvable superhamiltonians}\label{app:exactDW}
In this section we discuss exactly solvable toy models with bare domain walls as exact excitations.
This will help us understand the qualitative feature of more general GUE models.
We consider models that acts on the $\ket{\up}$ and $\ket{\dn}$ degrees of freedom, defined in Eq.~(\ref{spin_def}), and has an action that resembles that of $A_0$\footnote{Note that due to our unfortunate choice of convention, what we refer to as the left eigenstates of $A_0$ in the main text are the right eigenstates of $H$ in this section.} in Eqs.~(\ref{aleft_def_mt}) and (\ref{dshift}), i.e.,
\begin{equation}
    H\ket{\up\up} = H\ket{\dn\dn} = 0,\;\;\;H\ket{\up\dn} = \ket{\up\dn} - \ell(\ket{\up\up} + \ket{\dn\dn}),\;\;\;H\ket{\dn\up} = \ket{\dn\up} - \ell(\ket{\up\up} + \ket{\dn\dn}),
\end{equation}
where we treat $\ell$ in this section as a free parameter.
While $H$ is a non-Hermitian for general $\ell$, by expressing it in the orthonormal basis of Eq.~(\ref{eq:orthobasis}), we can check that it is Hermitian when $\ell = \frac{1}{q + q^{-1}}$, and we discuss this case in more detail in Sec.~\ref{sec:hermitiantoy}. 
\subsection{Exact solution of single-domain wall excitations}\label{sec:exactsolution}
For completeness, we start by defining the two kinds of domain walls, $\ket{D_x}$ and $\sket{\widetilde{D}_x}$ as
\begin{equation}
    \ket{D_x} \defn \ket{\dn \cdots \dn_x \up_x \cdots \up},\;\;\;\sket{\widetilde{D}_x} = \ket{\up \cdots \up_x \dn_{x+1} \cdots \dn}.
\label{eq:twoDW}
\end{equation}
Note that this is not an orthogonal basis, hence it is convenient to work in a biorthogonal basis by also defining the objects
\begin{equation}
    \ket{\overline{D}_x} \defn \ket{\overline{\dn} \cdots \overline{\dn}_x \overline{\up}_x \cdots \overline{\up}},\;\;\;\sket{\widetilde{D}_x} = \ket{\overline{\up} \cdots \overline{\up}_x \overline{\dn}_{x+1} \cdots \overline{\dn}}.
\label{eq:twoDWbar}
\end{equation}
Analogs of these with higher numbers of domain walls can be similarly defined.
The action of the Hamiltonian on the domain walls of Eq.~(\ref{eq:twoDW}) then reads
\begin{gather}
    H\ket{D_x} = \ket{D_x} - \ell (\ket{D_{x-1}} + \ket{D_{x+1}}),\;\;\text{if}\;\;2 \leq x \leq L-2,\nonumber \\
    H\ket{D_1} = \ket{D_1} - \ell(\ket{D_2} + \ket{G_\up}),\;\;H\ket{D_{L-1}} = \ket{D_{L-1}} - \ell(\ket{D_{L-2}} + \ket{G_\dn}),\nonumber \\
    H\sket{\widetilde{D}_x} = \sket{\widetilde{D}_x} - \ell (\sket{\widetilde{D}_{x-1}} + \sket{\widetilde{D}_{x+1}}),\;\;\text{if}\;\;2 \leq x \leq L-2,\nonumber \\
    H\sket{\widetilde{D}_1} = \sket{\widetilde{D}_1} - \ell(\sket{\widetilde{D}_2} + \sket{G_\dn}),\;\;H\sket{\widetilde{D}_{L-1}} = \sket{\widetilde{D}_{L-1}} - \ell(\sket{\widetilde{D}_{L-2}} + \sket{G_\up}).
\label{eq:Hactionsapp}
\end{gather}  
This shows that domain walls in the bulk are allowed to hop under the action of $H$, whereas on the boundaries they can be annihilated.
It is easy to also show that the action of $H$ on a state with multiple domain walls is similar, i.e., the domain walls in the bulk hop independently, and those on the boundaries can be annihilated.
Hence the action of $H$ can only \textit{decrease} the number of domain walls in the system, since they can be annihilated but not created at the boundary of the system.
Hence, in a biorthogonal basis ordered by the increasing number of domain walls, the matrix of $H$ has a block upper triangular structure, i.e.,
\begin{equation}
    H = 
    \begin{pmatrix}
        M_0 & M_{0,1} & 0 & \cdots & 0\\
        0 & M_1 & M_{1,2} & \ddots & \vdots \\
        \vdots & \ddots & \ddots & \ddots & \vdots \\
        \vdots & \ddots & 0 & M_{L-2} & M_{L-2, L-1} \\
        0 & \cdots & \cdots & 0 & M_{L-1}
    \end{pmatrix},
\label{eq:Hmatrix}
\end{equation}
where $M_n$ is a matrix that denotes the action of $H$ on $n$ domain walls, $M_{n,n+1}$ are rectangular matrices of appropriate sizes, and the matrix elements are numbers of the form $\bra{\overline{D}_{\cdots}}H\ket{D_{\cdots}}$ (or analogs with $D \rightarrow \widetilde{D}$).
Due to this structure, if we are interested in eigenstates that have at most $n$ domain walls, we can truncated $H$ to only the first two blocks, composed of $M_0$, $M_{0,1}$, $M_1$ without changing its right eigenvectors. 
Using the actions of $H$ from Eq.~(\ref{eq:Hactionsapp}), this truncated matrix has dimensions $2 L \times 2L$, and in the basis ordered by $\{\ket{G_\up}, \ket{G_\dn}, \ket{D_1}, \cdots \ket{D_{L-1}}, \sket{\widetilde{D}_1}, \cdots, \sket{\widetilde{D}_{L-1}}\}$ can be written as
\begin{equation}
    M = 
    \begin{pmatrix}
    M_0 & M_{0,1} \\
    0 & M_1
    \end{pmatrix} = 
    \begin{pmatrix}
        0 & 0 & v^T & w^T\\
        0 & 0 & w^T & v^T \\
        0 & 0 & B & 0 \\
        0 & 0 & 0 & B
        \end{pmatrix},
\label{eq:SPmat}
\end{equation}
where we have defined
\begin{equation}
    v^T = 
    \begin{pmatrix}
    -\ell & 0 & \cdots & 0
    \end{pmatrix}_{1 \times L-1},\;\;\;
    w^T = 
    \begin{pmatrix}
    0 & \cdots & 0 & -\ell
    \end{pmatrix}_{1 \times L-1},\;\;\;
    B = 
    \begin{pmatrix}
    1 & -\ell & 0 & \cdots & 0 \\
    -\ell & 1 & -\ell & \ddots & \vdots \\
    0 & \ddots & \ddots & \ddots & 0 \\
    \vdots & \ddots & -\ell & 1 & -\ell \\
    0 & \cdots & 0 & -\ell & 1
    \end{pmatrix}_{L-1 \times L-1}.
\end{equation}
Using the block upper-triangular structure of $M$, it is easy to see that all the eigenvalues of $M$ are two-fold degenerate, and they are just the eigenvalues of the diagonal blocks -- there are two zeros and two each of the eigenvalues of the matrix $B$. 
The structure of the right eigenvectors are a bit more complicated, as we discuss below.
The zero right eigenvectors of $M$ are simply of the form $\begin{pmatrix} 1 & 0 & \cdots & 0\end{pmatrix}^T$ and $\begin{pmatrix} 0 & 1 & \cdots & 0\end{pmatrix}^T$ respectively, and the eigenvectors of $H$ thus are the two ground states
\begin{equation}
    H\ket{G_\up} = H\ket{G_\dn} = 0. 
\end{equation}
The other (unnormalized) right eigenvectors of $M$ are of the form 
\begin{equation}
    M\Psi_\alpha = \varepsilon_\alpha\Psi_\alpha,\;\;\;M\widetilde{\Psi}_\alpha = \varepsilon_\alpha\widetilde{\Psi}_\alpha,\;\;\;\Psi_\alpha = \begin{pmatrix}
    \varepsilon_\alpha^{-1} v^T \psi_\alpha \\
    \varepsilon_\alpha^{-1} w^T \psi_\alpha \\
    \psi_\alpha \\
    0
    \end{pmatrix},\;\;\;
    \widetilde{\Psi}_\alpha = \begin{pmatrix}
    \varepsilon_\alpha^{-1} w^T \psi_\alpha \\
    \varepsilon_\alpha^{-1} v^T \psi_\alpha \\
    0 \\
    \psi_\alpha
    \end{pmatrix},
\end{equation}
where $\varepsilon_\alpha$ and $\psi_{\alpha}$ are the eigenvectors of $B$, which (without normalization) read
\begin{equation}
    B\psi_\alpha = \varepsilon_\alpha \psi_\alpha,\;\;\;\psi_{\alpha, x} = \sin\left(\frac{\pi \alpha x}{L}\right),\;\;\varepsilon_\alpha = 1 - 2 \ell \cos\left(\frac{\pi \alpha}{L}\right),\;\;1\leq x, \alpha \leq L-1.
\label{eq:Beig}
\end{equation}
This means that the corresponding (unnormalized) right eigenvectors of $H$, that satisfy $H\ket{\Psi_\alpha} = \varepsilon_\alpha \ket{\Psi_\alpha}$ and $H\ket{\widetilde{\Psi}_\alpha} = \varepsilon_\alpha \ket{\widetilde{\Psi}_\alpha}$ are
\begin{gather}
    \ket{\Psi_\alpha} = \sum_{x = 1}^{L-1}{\psi_{\alpha, x} \ket{D_x}} - \varepsilon_\alpha^{-1}\ell (\psi_{\alpha, 1} \ket{G_\up} + \psi_{\alpha, L-1}\ket{G_\dn}), \nonumber \\
    \sket{\widetilde{\Psi}_\alpha} = \sum_{x = 1}^{L-1}{\psi_{\alpha, x} \sket{\widetilde{D}_x}} - \varepsilon_\alpha^{-1}\ell (\psi_{\alpha, 1} \ket{G_\dn} + \psi_{\alpha, L-1}\ket{G_\up}).
\label{eq:PsiPsitild}
\end{gather}
The complete set of right eigenvectors of $H$ in the single-particle sector are thus $\ket{G_\up}$, $\ket{G_{\dn}}$, $\{\ket{\Psi_\alpha}\}$ and $\{\sket{\widetilde{\Psi}_\alpha}\}$.

It is natural to wonder what the left eigenvectors for general $\ell$ look like. 
These can be computed exactly in principle by computing the right eigenvectors of the matrix $H^T$ for $H$ of Eq.~(\ref{eq:Hmatrix}), and expressing the resulting vectors in terms of $\ket{\overline{D}_{\cdots}}$ instead of $\ket{D_{\cdots}}$.
While these are guaranteed to coincide with the right eigenvectors in the Hermitian case, that is not true for general $\ell$. 
Here we will not pursue that computation since we are primarily interested in Hermitian superhamiltonians, and setting $\ell = \frac{1}{q + q^{-1}}$ provides us with a simple example of such a case. 
\subsection{Hermitian case}\label{sec:hermitiantoy}
When $\ell = \frac{1}{q + q^{-1}}$, i.e., when $H$ is Hermitian, and this can be verified by expressing it in the orthonormal basis of Eq.~(\ref{eq:orthobasis}) by using the similarity transformations defined in Sec.~\ref{subsec:mapping}.
In this orthogonal basis, the Hamiltonian reads
\begin{equation}
    H(\ell = \frac{1}{q + q^{-1}}) = \sum_{j}{\frac{1}{2} - \frac{1}{2 (q + q^{-1})}[q X_j X_{j+1} + q^{-1} Y_j Y_{j+1} + Z_j + Z_{j+1}]}.
\end{equation}
Note that under the Jordan-Wigner transformation, this maps onto free fermions, which is consistent with the exact solvability illustrated in Sec.~\ref{sec:exactsolution}.
In such a case, the right eigenvectors of $H$ are also the left eigenvectors. 
In that case, we can also verify (with tedious algebra) that the many-body eigenstates corresponding to the different energies are orthogonal.
An orthogonal basis can also be defined as 
\begin{equation}
    \ket{G_\pm} \defn \ket{G_\up} \pm \ket{G_\dn},\;\;\ket{\Psi_{\alpha, \pm}} = \ket{\Psi_\alpha} \pm \sket{\widetilde{\Psi}_\alpha},
\label{eq:orthbasistoy}
\end{equation}
which satisfy
\begin{equation}
    H\ket{G_{\pm}} = 0,\;\;\;H\ket{\Psi_{\alpha, \pm}} = \left[1 - \frac{2}{q + q^{-1}}\cos(\frac{\pi \alpha}{L})\right] \ket{\Psi_{\alpha, \pm}}.
\label{eq:easyreference}
\end{equation}
In the continuum limit, this dispersion reads
\begin{equation}
    E(k) = 1 - \frac{2}{q + q^{-1}}\cos(k),
\label{eq:toydispersion}
\end{equation}
and it is easy to verify that it satisfies the membrane constraint $E(i \log q) = 0$ of Eq.~(\ref{econd}), we discuss some physical aspects of this in App.~\ref{sec:commentsmembrane}. 
Further, $v_B$ can be computed for this dispersion relaton using Eq.~(\ref{vbdef}) as
\begin{equation}
    v_B = \frac{q - q^{-1}}{q + q^{-1}}.
\label{eq:vBtoy}
\end{equation}
Various computations also require the norms of these eigenstates, which can be computed to be
\begin{gather}
\braket{G_{\pm}|G_{\pm}} = r_{\pm}\nonumber \\
N_{\alpha, \pm} \defn \braket{\Psi_{\alpha, \pm}|\Psi_{\alpha, \pm}} =   \sum_{x = 1}^{L-1}{\sum_{x'=1}^{L-1}{\psi_{\alpha, x'}\psi_{\alpha,x} f_{\pm}(|x-x'|)}} - 2\varepsilon_\alpha^{-1}\ell\ p_{\alpha, \pm} \sum_{x = 1}^{L-1}{  \psi_{\alpha,x} f_{\pm}(x)} + \varepsilon_\alpha^{-2}\ell^2 r_{\pm}\ p_{\alpha, \pm}^2,
\label{eq:norms}
\end{gather}
where we have defined
\begin{equation}
p_{\alpha, \pm} \defn \psi_{\alpha, 1} \pm \psi_{\alpha, L-1},\;\;\;r_{\pm} \defn 1 \pm q^{-L},\;\;f_{\pm}(x) \defn q^{-x} \pm q^{-(L-x)}.
\label{eq:convdefns}
\end{equation}
Using Eq.~(\ref{eq:Beig}), we can evaluate the expressions in Eq.~(\ref{eq:norms}) find the identities
\begin{gather}
    \sum_{x = 1}^{L-1}{\psi_{\alpha, x} f_{\pm}(x)} = \varepsilon_{\alpha}^{-1} \ell\ r_{\pm} p_{\alpha, \pm},\nonumber\\
    \sum_{x = 1}^{L-1}\sum_{x' =1}^{L-1}{\psi_{\alpha, x}\psi_{\alpha,x'} f_{\pm}(|x - x'|)} = 2\varepsilon_\alpha^{-1}\ell\left[\sin(\frac{\pi \alpha}{L})\varepsilon_{\alpha}^{-1} \ell r_{\pm} p_{\alpha,\pm} - f_{\pm}(1) \frac{L}{2} + r_{\pm} \frac{L}{2}\cos(\frac{\pi \alpha}{L})\right] + \frac{L}{2}.
\end{gather}
Combining these, the norms can be expressed and approximated in the large $L$ limit as
\begin{equation}
    N_{\alpha, \pm} = \frac{L}{2}  \left[1 +  2 \varepsilon_\alpha^{-1}\ell (r_{\pm}\cos(\frac{\pi \alpha}{L}) - f_{\pm}(1))\right]
    + \varepsilon_\alpha^{-2} \ell^2 r_{\pm} p_{\alpha, \pm}^2 \approx \frac{L}{2}\times \frac{q - q^{-1}}{q + q^{-1}} \times \varepsilon_\alpha^{-1}
\label{eq:normfinal}
\end{equation}
where we have used that $p_{\alpha, +} = 2\sin(\frac{\pi\alpha}{L})\delta_{\alpha, \textrm{odd}}$, $p_{\alpha, -} = 2\sin(\frac{\pi\alpha}{L})\delta_{\alpha, \textrm{even}}$, $r_{\pm} \approx 1$, and neglected the second term in the approximation.

\subsection{Comments on the membrane constraint}\label{sec:commentsmembrane}
We now comment on a few subtle aspects of the membrane constraint relation as derived in the main text in (\ref{econd}), which attributes the origin of this constraint to the structure of the ground state of $P_4$. 
At first sight, this might appear a bit strange since the constraint is on the excited state dispersion relation, which is usually thought to be independent of the structure of the ground state since the excited states and ground states are orthogonal. 
Here we derive this constraint for this toy model using an alternate method, and clarify this subtlety. 
As discussed in the main text, this membrane constraint is equivalent to the fact that entropy growth rate $\Gamma_2(s)$ for a mixed state of entropy density $s$ should vanish for a maximally mixed state, where $s = s_{eq} = \log(q)$. 
For a mixed states $\rho_s$ that satisfies (\ref{initial_s}), we can write down the purity and initial growth rate at $t = 0$ as
\begin{equation}
    e^{-S_2} = \bra{D_x} e^{-H t} \ket{\rho_s} \implies \Gamma_2(s) = \frac{\partial S_2}{\partial t}|_{t = 0} = \frac{\bra{D_x} H \ket{\rho_s}}{\braket{D_x|\rho_s}}.
\label{eq:puritygrowht}
\end{equation}
Using the action of $H$ in Eq.~(\ref{eq:Hactionsapp}) for domain walls in the bulk (with $\ell = \frac{1}{q + q^{-1}}$), and the fact that $\braket{D_x|\rho_s} = e^{-s x}$, this can be simplified to 
\begin{equation}
    \Gamma_2(s) = \frac{-\ell\braket{D_{x+1}|\rho_s} - \ell \braket{D_{x-1}|\rho_s} + \braket{D_x|\rho_s}}{\braket{D_x|\rho_s}} = 1 - \frac{2}{q + q^{-1}}\cosh(s) = E(i s).
\label{eq:growthratesimp}
\end{equation}
It is then clear that the membrane constraint of $E(i\log q) = 0$ follows from the fact that the growth rate $\Gamma_2(s)$ should vanish at $s = \log q$, i.e., for maximally mixed states. 
Note that this computation, unlike the one presented in the main text, does not rely on the structure of the ground state but only on the bulk propagation of domain walls. 
Note that there is a constraint on the dispersion relation of the domain wall excitations in spite of it being orthogonal to the ground state (i.e., $\ket{\rho_{s = \log q}, e}$) because the ``bulk" of the domain walls is not orthogonal to the ground states. 
That is, as discussed in Eq.~(\ref{eq:PsiPsitild}), the domain wall excitations at any finite size are of the form 
\begin{equation}
    \sket{\Psi_\alpha} = c_e \ket{G_\up} + \sum_{x}{c_{\alpha, x} \ket{D_x}} + c_\eta \ket{G_\dn}.  
\label{eq:Dkexpression}
\end{equation}
While the full state is orthogonal to the ground states, the bulk part of the state, i.e., just $\sum_x{c_x \ket{D_x}}$, which is close to the full eigenstate for large $L$, is not orthogonal to the ground state. 
This bulk state is sufficient to determine the dispersion relation, and since it is not orthogonal to the ground state, the ground state imposes a constraint on the dispersion relation of the domain wall state, as discussed around \eqref{459}. 
\section{Methods for extracting the membrane tension}\label{app:membraneextract}
In this appendix, we provide some details on extracting the membrane tension using different methods, and comparison with numerical methods.  
We also demonstrate both methods on the toy model discussed in Sec.~\ref{sec:hermitiantoy}.
\subsection{Dispersion relation method}\label{subsec:dispersion}
To extract the membrane tension numerically from the dispersion relation of the domain wall, we use a simple fitting function of the form of Eq.~(\ref{disp_fit}).
The precise membrane tension can be extracted using the Legendre transform of $E(k)$, given by
\begin{equation}
\mE(v) = \frac{E(i \chi) + \chi v}{\log q} = \frac{\sum_{n = 0}^{N_{\max}} c_n \cosh(n \chi) + \chi v}{\log q} ,\;\;\text{where}\;\;E'(i \chi) = iv \iff \sum_{n = 0}^{N_{\max}}{c_n n \sinh(n \chi)} = -v.
\label{eq:membranetensionlegendre}
\end{equation}
This can be solved numerically to extract $\mE(v)$ from any general $E(k)$.
For the toy superhamiltonian discussed in Sec.~\ref{sec:hermitiantoy}, from the dispersion relation of Eq.~(\ref{eq:easyreference}), we can compute its Legendre transform to be
\begin{equation}
    \sE(v) = \frac{1}{\log q} \le[1 - \sqrt{\left(\frac{2}{q+q^{-1}}\right)^2 + v^2}+ v \, \text{arcsinh}\le(\frac{v (q + q^{-1})}{2}\ri)\ri]. \label{eq:mem_toy} 
\end{equation}
It is easy to verify that this satisfies the membrane constraints $\sE(v_B) = v_B$, where $v_B$ is defined in Eq.~(\ref{eq:vBtoy}).
\subsection{Propagator method}\label{subsec:propagator}
Alternately, we can also extract the membrane tension using the domain wall propagator, which was referred to as the modified partition function in \cite{zhou_nahum}, where it conjectured that
\begin{equation}
    G(x, y, t) \defn \bra{D_x} e^{-H t} \ket{\bD_y} \sim e^{- \log(q) \mE(v) t} + \cdots, 
\label{eq:modifiedZ}
\end{equation}
where the $\cdots$ are subdominant at large times, and the domain wall states are defined in Eqs.~(\ref{eq:Dxdefn}) and (\ref{eq:Dxbardefn}).
Note that this is simply the domain wall propagator discussed in the main text, e.g., in Eq.~(\ref{g0def}), hence Eq.~(\ref{eq:modifiedZ}) is essentially Eq.~(\ref{436}). 
Given a superhamiltonian $H$, we can numerically compute the membrane tension $\mE(v)$ by computing $F(x)_v \defn G(x, y = 0, t = x/v)$ for each $v$ using standard tensor network methods such as TEBD, and extracting the slope of $-\log F(x = v t)_v$ as a function of $t$. 
Further, we can also quantitatively verify Eq.~(\ref{eq:modifiedZ}) for toy superhamiltonians such as those discussed in App.~\ref{app:exactDW} where the bare domain walls are exact excitations.
We will set $\ell = \frac{1}{q + q^{-1}}$ in this computation, since we are ultimately interested in Hermitian superhamiltonians, see App.~\ref{sec:hermitiantoy}. 
We can unpack the expression for $Z(x,y,t)$ in Eq.~(\ref{eq:modifiedZ}) in terms of the eigenstates of $H$ as
\begin{equation}
    G(x, y, t) = \sum_{n}{\braket{D_x|E_n}\braket{E_n|\bD_y} e^{-E_n t}}.
\label{eq:partfuncunpack}
\end{equation}
The \textit{complete} weight of $\ket{D_x}$ lies within the space spanned by domain wall states of the form of $\ket{\Psi_{\alpha}}$, $\ket{\widetilde{\Psi}_\alpha}$ of Eq.~(\ref{plane_waves}) and the ground states $\ket{G_\up}$, $\ket{G_{\dn}}$ [i.e., $\braket{D_x|E_n} = 0$ for all other eigenstates] we can write
\begin{equation}
G(x, y, t) = \frac{1}{2N_{\alpha, +}}\sum_{\alpha}{\braket{D_x|\Psi_{\alpha,+}}\sbraket{\Psi_{\alpha,+}}{\overline{D}_y} e^{-\varepsilon_\alpha t}} + \frac{1}{2 N_{\alpha, -}}\sum_{\alpha}{\braket{D_x|\Psi_{\alpha,-}}\sbraket{\Psi_{\alpha,-}}{\overline{D}_y} e^{-\varepsilon_\alpha t}},
\label{eq:propexpr}
\end{equation}
where we have used the orthogonal basis of Eq.~(\ref{eq:orthbasistoy}), and $N_{\alpha, \pm}$ are defined in Eq.~(\ref{eq:norms}).
The exact expressions for the overlaps in Eq.~(\ref{eq:propexpr}) and their large $L$ approximations read
\begin{align}
\braket{D_x|\Psi_{\alpha, \pm}} &= \sum_{x' = 1}^{L-1}{\psi_{\alpha,x'} (q^{-|x-x'| } \pm q^{-(L - |x -x'|)})} - \varepsilon_\alpha^{-1}\ell(\psi_{\alpha,1} \pm \psi_{\alpha, L-1})(q^{-x} \pm q^{-(L-x)})\nonumber \\
&= \sum_{x' = 1}^{L-1}{\psi_{\alpha,x'} f_{\pm}(|x-x'|)} - \varepsilon_\alpha^{-1}\ell\ p_{\alpha, \pm}\ f_{\pm}(x) = 
\sin\left(\frac{\pi \alpha x}{L}\right)\left[2\varepsilon_\alpha^{-1}\ell (r_{\pm}\cos\left(\frac{\pi\alpha}{L}\right) - f_{\pm}(1)) + r_{\pm}\right] \nonumber \\
&\approx \varepsilon_\alpha^{-1}\frac{q - q^{-1}}{q + q^{-1}}\sin\left(\frac{\pi \alpha x}{L}\right) \nonumber \\
\sbraket{\Psi_{\alpha,\pm}}{\overline{D}_y} &= \psi_{\alpha, y} = \sin\left(\frac{\pi \alpha y}{L}\right),\label{eq:DPsioverlaps}
\end{align}
where have used the notation of Eq.~(\ref{eq:convdefns}) the fact that
\begin{equation}
\braket{G_\pm|\overline{D}_y} = \sbraket{\widetilde{D}_x}{\overline{D}_y} = 0,\;\;\;\sbraket{D_x}{\overline{D}_y} \sim \delta_{x,y}.
\end{equation}
Taking everything together, we can express the propagator as
\begin{equation}
G(x, y, t) = \frac{1}{2}\sum_{\alpha = 1}^L{\psi_{\alpha, y}e^{-\varepsilon_\alpha t}\left(\frac{\braket{D_x|\Psi_{\alpha,+}}}{N_{\alpha, +}} + \frac{\braket{D_x|\Psi_{\alpha,-}}}{N_{\alpha, -}}\right)} \approx \frac{2}{L}\sum_{\alpha = 1}^{L-1}{\psi_{\alpha,x}\psi_{\alpha,y} e^{-\varepsilon_\alpha t}},
\end{equation}
where we have used the overlaps are in Eq.~(\ref{eq:DPsioverlaps}), and the energies $\varepsilon_\alpha$ and the wavefunctions $\psi_{\alpha,x}$ are in Eq.~(\ref{eq:Beig}) (with $\ell = \frac{1}{q + q^{-1}}$).
For large $L$, we can then write $G(x, y, t)$ as an integral as
\begin{equation}
    G(x, y, t) \approx \frac{2}{\pi}\int_{0}^\pi{\mathrm{d}k\ \sin(k x) \sin(k y) e^{-E(k) t}} = \frac{1}{\pi}\int_{0}^\pi{\mathrm{d}k\ \cos(k (x-y))\ e^{-E(k) t}} - \frac{1}{\pi}\int_{0}^\pi{\mathrm{d}k\ \cos(k (x+y))\ e^{-E(k) t}},
\label{eq:propintegral}
\end{equation}
where $E(k)$ is the continuum dispersion relation in Eq.~(\ref{eq:toydispersion}).
We can then evaluate the integrals of the form as in Eq.~(\ref{eq:propintegral}) 
\begin{equation}
    \frac{1}{\pi}\int_{0}^{\pi}{\mathrm{d}k\ \cos(k X)\ e^{- (1 - 2 \ell \cos(k))t}} = e^{-t} I_X(2\ell t)\;\;\;\textrm{if } X \in \mathbb{Z}, 
\label{eq:besselidentity}
\end{equation}
where $I_X(\bullet)$ is the modified Bessel function of the first kind.
Hence the expression of the propagator reads
\begin{equation}
    G(x, y, t) \approx e^{-t} \left[I_{x-y}\left(\frac{2t}{q + q^{-1}}\right) - I_{x+y}\left(\frac{2t}{q + q^{-1}}\right)\right]\approx e^{-t} I_{vt}\left(\frac{2t}{q + q^{-1}}\right)\;\;\;\textrm{if }vt \in \mathbb{Z},
\label{eq:propagatorbessel}
\end{equation}
where the second approximation is justified since we are interested in the regime where $x, y \sim \mathcal{O}(L)$ in the middle of the system, $|x - y| = v t$ with $v \sim \mathcal{O}(1)$,  and times $\mathcal{O}(1) \ll t \ll \mathcal{O}(L)$.
This expression of $G(x,  y, t)$ is completely independent of the full system size $L$ and only depends on the velocity $v$ and time $t$, similar to the expressions in the main text.
For large $t$, the asymptotics of this modified Bessel function can be analyzed using the Debye expansions, which says that~\cite{olver1997asymptotics}
\begin{equation}
    I_{\nu}(\nu \zeta) \sim \frac{1}{\sqrt{2\pi \nu}}(1 + \zeta^2)^{-\frac{1}{4}} e^{\nu \eta(\zeta)} + \mathcal{O}\left(\frac{1}{\nu}\right)\;\;\text{for large $\nu$,}\;\;\;\eta(\zeta) = \sqrt{1 + \zeta^2} + \log \frac{\zeta}{1 + \sqrt{1 + \zeta^2}}.
\label{eq:Debye}
\end{equation}
Using this, we can substitute $\nu = vt$ and $\zeta = \frac{2}{(q + q^{-1}) v}$ to obtain that 
\begin{equation}
   e^{-t} I_{vt}\left(\frac{2t}{q + q^{-1}}\right) \sim e^{- t + v t\ \eta\left(\frac{2}{(q + q^{-1}) v}\right)} \sim e^{-\log(q)\mE(v) t}\;\;\;\text{for large $t$},
\label{eq:Debyeresult}
\end{equation}
where $\mE(v)$ is the membrane tension of the toy model, shown in Eq.~(\ref{eq:mem_toy}), where we have used that $\textrm{arcsinh}(x) = \log(x + \sqrt{1 + x^2})$. 
Hence, combining Eqs.~(\ref{eq:propagatorbessel}) and (\ref{eq:Debyeresult}), we obtain the scaling of the propagator as expected in Eq.~(\ref{eq:modifiedZ}).
As we discussed in the main text, the low-energy excitations of more general superhamiltonians are \textit{dressed} domain walls, hence we expect that the analytical computations above should also approximately hold for the those. 
It would be interesting to check this more carefully in future work.
\section{Superhamiltonians for higher Renyi entropies}\label{app:superhamhigher}
In this section, we write down expressions for the superhamiltonians that govern the behavior of the higher Renyi entropies. 
As discussed in Sec.~\ref{sec:setup}, in general, for $2n$ copies, we can write down a local basis of permutation states as identities (normalized) that couple $n$ copies of $U$ with $n$ copies of $U^\ast$, hence the $2n$-copy super-Hamiltonian has basis states from elements of $S_n$. 
The permutation states $\ket{\sigma}$ that span the local Hilbert space, are defined in Eq.~(\ref{eq:sigmadefn}).
It is convenient to define normalized versions of these permutation states $\ket{\sigma}$, which are defined as 
\begin{equation}
    \sket{\wtsigma} = q^{-\frac{n}{2}}\ket{\sigma},
\end{equation}
which are normalized as $\sbraket{\wtsigma}{\wtsigma} = 1$.
Note that when $n = 2$, we have $\ket{\up} = \sket{\wte}$ and $\ket{\dn} = \sket{\wteta}$, which is the relation in Eq.~(\ref{spin_def}).
Further, similar to Eq.~(\ref{spin_def}), $\sket{\wtsigma}$ can be expressed in terms of $\ket{\textrm{MAX}}$ dimers [defined as in Eq.~(\ref{eq:MAXdefn})] as, 
\begin{equation}
    \sket{\wtsigma} = \bigotimes_{\ell = 1}^n{\ket{\textrm{MAX}}_{f_\ell b_{\sigma(\ell)}}},\;\;\;\braket{\wtsigma|\wttau} = q^{- d(\sigma, \tau)}
\label{eq:sigmatilddefn}
\end{equation}
where $d(\sigma, \tau)$ is the Cayley distance between the permutations $\sigma$ and $\tau$. 
The general superhamiltonian for the GUE model is given in Eq.~(\ref{p2n_gue}).
We now focus on $n = 3$ and illustrate the structure of $P_6$,  which has been used in the discussion in Sec.~\ref{sec:higher_renyi}.
We can denote the 6 on-site permutations as
\begin{equation}
    e = (1\ 2\ 3),\;\;\;\kappa = (2\ 1\ 3),\;\;\;\kappa' = (1\ 3\ 2),\;\;\;\tau = (2\ 3\ 1),\;\;\;\mu = (3\ 2\ 1),\;\;\;\eta = (3\ 1\ 2)
\label{eq:permreps}
\end{equation}
The Cayley graph has the following distance structure:
\begin{gather}
    d(e, \kappa) = d(e, \kappa') = d(e, \mu) = d(\eta, \kappa) = d(\eta, \kappa') = d(\eta, \mu) = 1\nonumber \\
    d(e, \eta) = d(e, \tau) = d(\eta, \tau) = d(\kappa, \kappa') = d(\kappa, \mu) = d(\kappa', \mu) = 2.
\end{gather}
So this Cayley graph is bipartite with two sublattices $A = \{\kappa, \kappa',\rho\}$ and $B = \{e, \eta, \tau\}$, with 
\begin{equation}
    d(A, B) = 1,\;\;\;d(A, A) = d(B, B) = 2. 
\end{equation}
We can further denote the normalized versions of these states as $\{\sket{\wte},  \sket{\wtkappa},  \sket{\wtkappa'},  \sket{\wttau},  \sket{\wtmu},  \sket{\wteta}\}$,  which are defined as in Eq.~(\ref{eq:sigmatilddefn}).
The action of the superhamiltonian $P_6$ of Eq.~(\ref{p2n_gue}) on these states then reads
\begin{gather}
    P_6 \sket{\wtalpha\wtalpha} = 0,\;\;\;\alpha \in S_3\nonumber \\
    P_6 \ket{\wtalpha \wtalpha'} = \frac{1}{2}[3\ket{\alpha \alpha'} - \frac{1}{q}\sum_{\wtbeta}{(\ket{\wtalpha \wtbeta} + \ket{\wtbeta \wtalpha'})} + \frac{1}{q^2}\sum_{\beta, \beta'}{(-1)^{\delta_{\beta, \beta'}}\ket{\wtbeta \wtbeta'}}],\;\;\;\alpha,  \alpha'\in A,\;\;\alpha \neq \alpha',\;\;\;\beta, \beta' \in B \nonumber \\
    P_6 \ket{\wtalpha \wtbeta} = \frac{1}{2}[2 \ket{\wtalpha \wtbeta} - \frac{2}{q}(\ket{\wtalpha \wtalpha} + \ket{\wtbeta\wtbeta}) + \frac{2}{q^2}\ket{\wtbeta \wtalpha}],\;\;\;\alpha \in A,\;\;\;\beta \in B.
    \label{eq:P6action}
\end{gather}
The first line shows the 6 ferromagnetic ground states. 
The second line shows that a domain wall between permutation states of Cayley distance $2$ can dissociate into ``more elementary" domain walls at leading order in $1/q$,  i.e., those whose Cayley distance is closer.
In particular, for the $e-\eta$ domain wall that is of physical interest, we have that 
\begin{equation}
    P_6\ket{e\eta} = \frac{1}{2}{[3\ket{e \eta} - \frac{1}{q}(\ket{e\nu} + \ket{\nu\eta}) + \frac{1}{q^2}( -2( \ket{\kappa\kappa} + \ket{\kappa'\kappa'} + \ket{\mu\mu}) + \ket{\nu\nu})]},
\label{eq:eetaaction}
\end{equation}
where $\ket{\nu} \defn \ket{\sigma} + \ket{\sigma'} + \ket{\mu}$.
The third line in Eq.~(\ref{eq:P6action}) shows that permutation degrees of freedom with Cayley distance $1$ behave similar to $\ket{\up}$ and $\ket{\dn}$ in the case of $n = 2$, i.e., in Eq.~(\ref{aleft_def_mt}). 
Note that the bare energy of the $e-\eta$ domain wall is lower than that of the $e-\nu$ and $\nu-\eta$ elementary domain walls combined, which leads to the rich physics discussed in the main text in Sec.~\ref{sec:higher_renyi}.
For completeness,  we also write down expressions for the biorthogonal basis states $\{\ket{\overline{\wtalpha}}\}$ corresponding to the permutation states $\{\ket{\wtalpha}\}$, which satisfy the property $\braket{\overline{\wtalpha}|\wtbeta} = \delta_{\alpha,\beta}$.
The analogous states for $n = 2$ are shown in Eq.~(\ref{bar_spin_def}).
The expressions for these read
\begin{equation}
\ket{\overline{\wtalpha}} = \frac{q^2}{(q^2 - 4)(q^2 - 1)}[(q^2 - 2)\ket{\wtalpha} + \sum_{\alpha' \neq \alpha}\ket{\wtalpha'} - q\sum_\beta{\ket{\wtbeta}}],
\end{equation}
where $A, A'$ are states from one sublattice and $B$ is a state on the other sublattice of the Cayley graph. 
The expressions for $\ket{\overline{B}}$ are obtained by just switching the sublattices.
For numerical purposes, it is convenient to work with an orthonormal set of  basis vectors analogous to (\ref{eq:orthobasis}) for $n = 3$. 
When $n = 3$, a convenient candidate is of the form
\begin{align}
    \ket{0, \up} &\defn \frac{q}{\sqrt{6(q+1)(q+2)}}(\ket{e} + \ket{\tau} + \ket{\eta} + \ket{\sigma} + \ket{\mu} + \ket{\sigma'}) \nonumber \\
    \ket{0, \dn} &\defn \frac{q}{\sqrt{6(q-1)(q-2)}}(\ket{e} + \ket{\tau} + \ket{\eta} - \ket{\sigma} - \ket{\mu} - \ket{\sigma'})\nonumber \\
    \ket{1, \up} &\defn \frac{q}{\sqrt{6(q^2 - 1)}}(\ket{e} + \omega^2 \ket{\tau} + \omega\ket{\eta} + \omega^2\ket{\sigma} + \omega\ket{\mu} + \ket{\sigma'}) \nonumber \\
    \ket{1, \dn} &\defn \frac{q}{\sqrt{6(q^2 - 1)}}(\ket{e} + \omega^2 \ket{\tau} + \omega\ket{\eta} - \omega^2 \ket{\sigma} - \omega\ket{\mu} - \ket{\sigma'})\nonumber\\
    \ket{2, \up} &\defn \frac{q}{\sqrt{6(q^2 - 1)}}(\ket{e} + \omega \ket{\tau} + \omega^2 \ket{\eta} + \omega\ket{\sigma} + \omega^2\ket{\mu} + \ket{\sigma'})\nonumber \\
    \ket{2, \dn} &\defn \frac{q}{\sqrt{6(q^2 - 1)}}(\ket{e} + \omega \ket{\tau} + \omega^2 \ket{\eta} - \omega\ket{\sigma} - \omega^2\ket{\mu} - \ket{\sigma'}),
\label{eq:orthobasishigher}
\end{align}
where we have labeled them by a qutrit-qubit degrees of freedom, as common for $S_3$ models~\cite{bhardwaj2025illustrating}, and $\omega = \frac{-1+ i \sqrt{3}}{2}$. 
As a sidenote, when $q = 2$, we have $\ket{0, \dn} = 0$,  which means that the permutation states are linearly dependent.
This overcompleteness of the permutation states has been noted in \cite{hearth2023unitary}, and one way to see that is the overlap matrix is not invertible.
But they still seem to have 6 ground states, so perhaps everything is ok, except for the local basis. 
Contrary to the case where the ``dimers" connecting the spins are singlets (insteady of $\ket{\textrm{MAX}}$), where all ``crossing" configurations are linearly dependent on the ``non-crossing" configurations~\cite{moudgalya2018exact}, that doesn't seem to be true here, at least for 4 sites. 
\section{Multiple intervals and multiparticle excitations}
\label{app:multiple}
In the main text, we mostly discuss the evolution of the Renyi entropies for the half line region in one spatial dimension.
To distinguish the entanglement dynamics in chaotic systems from that in integrable systems like the quasiparticle model~\cite{calabrese}, it is important to also consider the evolution of the Renyi entropies for multiple intervals~\cite{leichenauer, spread}.
In this appendix, we first briefly review the implications of the membrane picture for regions consisting of one or more intervals from~\cite{huse}.
We then explain how in the large $q$ limit of the GUE model~(Section \ref{sec:large_q}), these results can be obtained from the structure of the multiparticle excitations of $A_0$. A similar picture should hold away from this special limit, as we have shown that the lowest excitations of $P_4$ are quite generally given by well-defined gapped quasiparticles.

\begin{figure*}[!h]
    \centering
    \includegraphics[width=0.9\textwidth]{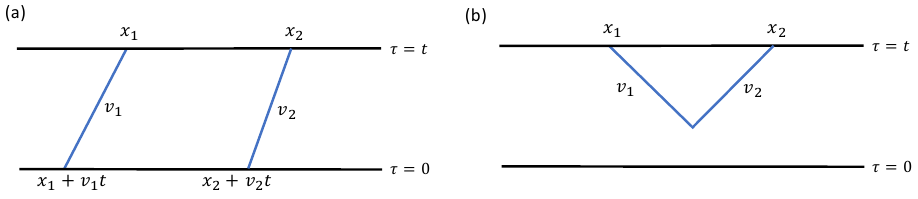}
    \caption{Two possible membrane configurations for a single interval.}
    \label{fig:one_int}
\end{figure*}

Consider the evolution of the entanglement entropy for a single interval.
For simplicity, consider the case where the initial state is a pure product state.
In applying the membrane formula for the Renyi entropy $S_n$ to a finite interval $[x_1,x_2]$, we need to consider the minimum over all possible velocities in the two configurations (a) and (b) in Fig. \ref{fig:one_int}.
From the configurations in (a), we get:
\begin{align}
C_a = & ~ t\, \text{min}_{v_1, v_2} [s_{\rm eq}(\sE_n(v_1)+\sE_n(v_2)) ]   =  ~2  t ~s_{\rm eq} v_{E, n} \, , \quad  v_{E,n} = \sE_n(0) \, .  \label{45}  
\end{align}
From the configuration in (b), since we must have $(|v_1| + |v_2|)t = x_2-x_1$ for the membranes to join together, 
\be \label{46}
C_b = \text{min}_{v_1,v_2} \, s_{\rm eq} \, \le[ (\sE_n(v_1) + \sE_n(v_2)) \frac{(x_2-x_1)}{|v_1|+|v_2|}  \ri] =  s_{\rm eq} (x_2 -x_1),
\ee
where we have used the constraints \eqref{const} to see that the quantity in \eqref{46} is minimized for $|v_1| = |v_2| = v_B$.
Minimizing between $C_a$ and $C_b$, we see a linear growth of entanglement entropy followed by saturation at the thermal value:  
\begin{align} 
S_{n}^{\rm int}([x_1, x_2],t) = \begin{cases}
 2t s_{\rm eq} v_{E, n}  & t < \frac{(x_2 - x_1)}{2v_{E, s}}  \\
  (x_2-x_1)s_{\rm eq} &   t \geq \frac{(x_2 - x_1)}{2v_{E, s}}
\end{cases} \label{interval}
\end{align}

For $R$ consisting of $m$ intervals $[x_1, y_1]\cup[x_2, y_2] \cup \cdots \cup [x_m, y_m]$, we have a minimization over a larger set of configurations, which leads to the following result 
\be 
S_{n, R}(t) = \min_{\gamma \in \sS_m} \le[\sum_{i=1}^m S_n^{\rm int}([x_i, y_{\gamma(i)}], t) \ri]\, .  \label{multi}
\ee 
\eqref{multi} is always non-decreasing as a function of time, in contrast to the evolution of multiple interval entanglement entropy in the integrable quasiparticle model~\cite{leichenauer}.

To derive this formula for the Brownian models of this work, we need understand the structure of the ``multi-particle'' excitations involving more than one domain wall.
We will understand this structure for the simplest case of $A_0$ from \eqref{aleft_def_mt}. 

For an initial state with two or more domain walls, the action of $A_0$ can cause domain walls to annihilate in pairs, and it is useful to divide $A_0$ into parts $A_f$ and $A_c$ as discuss below (\ref{eq:A0divide}).
The action of $A_f$ on the multi-domain wall state $\bra{D_{x_1, ..., x_k}}$,~$x_1<x_2<...<x_n$, is given by  
\begin{align} 
\bra{D_{x_1, ..., x_{k}}} A_f = \sum_{j=1}^{k} \le(\bra{D_{x_1, ..., x_{k}}} - \frac{1}{q} \le(\bra{D_{x_1, ..., x_{j}+1,..., x_{k}}} + \bra{D_{x_1, ..., x_{j}-1,..., x_{k}}}\ri) \ri) \label{443}
\end{align}
when all $|x_i-x_j|>1$.
If any $x_i, x_j$ are such that $|x_i-x_j|=1$, the action is such that we can simply delete the terms with any repeated $x_i$ appearing on the RHS of \eqref{443}. The action of $A_c$ is 
 \be 
\bra{D_{x_1, ..., x_{k}}} A_c = - \frac{2}{q}(\delta_{x_1+1,x_2} \bra{D_{x_3, x_4, ..., x_{k}}} + ... + \delta_{x_{k-1},x_{k}} \bra{D_{x_1, x_2, ..., x_{k-2}}}), \quad k \geq 2 \label{aj}
 \ee
An initial state with zero or 1 domain walls is annihilated by $A_c$.

Due to (\ref{443}), eigenstates of $A_{f}$ take a simple non-interacting, free fermion-like form of (\ref{psin_def_mt}) with energies given by the sum of the one-particle energies in \eqref{plane_waves},  
$E(k_1,..., k_n) = \sum_{j=1}^n E(k_i)$. 
Hence, the total propagator can be written as  
\be 
\braket{D_{x_1, ..., x_n}| e^{-A_ft}| \bar D_{y_1, ..., y_n}} = \prod_{i=1}^n G(x_i, y_i, t) \, .      \label{product_g0}
\ee
where $G(x_i, y_i, t)$ is the single-particle propagator defined in \eqref{g0def}.

Consider the evolution of $S_2$ for a region consisting of $m$ intervals, $R = [x_1, x_2] \cup ... \cup [x_{2m-1}, x_{2m}]$. 
Expanding the evolution under $A_0$ in the interaction picture, treating $A_c$ as the interaction, we get
\begin{align}
& e^{-S_{2, R}(t)_{{\rm large}\, q}} = q^L\braket{D_{x_1, x_2, ..., x_{2m}}| e^{-A_0 t}|\rho_0, e}  \\
& = q^L \braket{D_{x_1,  ..., x_{2m}}| e^{-A_f t}|\rho_0, e} + \int_0^t dt_1  q^L \braket{D_{x_1,  ..., x_{2m}}| e^{-A_f t_1} A_c e^{-A_f (t-t_1)} |\rho_0, e} + ...\nn 
&  + \int_0^t dt_1 \int_0^{t-t_1} dt_2 ... \int_0^{t -\sum_{i=1}^{m-1}t_i} dt_m   \braket{D_{x_1,  ..., x_{2m}}| e^{-A_f t_1} A_c  e^{-A_f t_2} A_c ...  e^{-A_f (t-\sum_{i=1}^{m} t_i)} |\rho_0, e} \label{int_exp}
\end{align}
We can check from \eqref{aj} that after $m+1$ insertions of $A_c$ at any intermediate times, the state $\bra{D_{x_1,..., x_m}}$ necessarily gets annihilated, so that the  interaction picture expansion  truncates at \eqref{int_exp}.
Using \eqref{product_g0}and \eqref{aj}, together with the expression \eqref{436} for the domain wall propagator, we can show that the sum over the $m$ terms in \eqref{int_exp} leads precisely to the minimization of \eqref{multi} in the scaling limit.
\section{Variational approach for low energy excitations of $P_{2n}$} \label{app:var}
In  the variational calculations discussed in this work, our goal can be stated as follows: we select some orthonormal basis of candidate states,  $\{ \ket{\psi_a}\}$, $a = 1, ..., m$ for some $m$. We then consider a general state in the subspace spanned by $\{ \ket{\psi_a}\}$, i.e., $\ket{\psi}= \sum_a c_a \ket{\psi_a}$, and want to minimize the expectation value $\braket{\psi|H|\psi}$ of some Hamiltonian $H$ with respect to the coefficients $c_a$, subject to the normalization condition $\sum_a |c_a|^2=1$.
It is easy to see that this minimization problem is equivalent to finding the lowest energy eigenstate of an ``effective Hamiltonian'' projected to the $\ket{\psi_a}$ subspace: 
\be 
(H^{\rm eff})_{ab} \equiv \braket{\psi_a|H|\psi_b} \,.  
\ee
The smallest eigenvalue of $H^{\rm eff}$ is the minimized expectation value in the given subspace, and the corresponding eigenvector corresponds to the optimal coefficients $c_a$.  
The calculation shown in Appendix \ref{subsec:varcalc} is a trivial $m = 1$ version of this procedure.
For our Hamiltonian $P_4$, we generally look for excitations of the form
\be 
\ket{\psi_k} = \frac{1}{\sqrt{L}}  \sum_x e^{-ikx} \, \ket{\eta} ...  \ket{\eta}_x \ket{\phi}_{x+1, ..., x+\Delta} \ket{ e}_{x+\Delta+1} ... \ket{e} \label{eta_e_modes_app}
\ee 
 It is natural to parameterize the excitations as plane waves due to the translation-invariance of $P_{2n}$. For a system with open boundary conditions, this is a natural ansatz in the thermodynamic limit.
Taking the form of the excitations to be asymptotically $\ket{\eta}..\ket{\eta}$
towards the left and $\ket{e}... \ket{e}$ towards the right ensures that we are in the right sector of the Hilbert space for finding eigenstates of $P_{2n}$ that contribute to the $n$-th Renyi entropy of the left half-line.
It also ensures that the states we consider have vanishing overlap with any of the ground states in the thermodynamic limit, so that the above minimization procedure will give an estimate of the lowest excited energy  with the given momentum, and not the ground state energy. 
Now for any given $\Delta$, we must find an appropriate orthonormal basis of states of the form \eqref{eta_e_modes_app}.
Let us start with the simplest non-trivial example of $\Delta=2$ for $P_4$ of the GUE model, where $\ket{\eta}=q\ket{\down}$, $\ket{e}=q\ket{\up}$.
In this case, the two options $\ket{\phi}_{x+1, x+2} = \ket{\down}_{x+1} \ket{\up}_{x+2}$ and  $\ket{\phi}_{x+1, x+2} = \ket{\up}_{x+1} \ket{\down}_{x+2}$ span the whole relevant Hilbert space -- the other two possibilities are redundant as the the resulting states differ from these cases only by an overall phase in the thermodynamic limit.
However, note that the two states 
\begin{align}
& \ket{\psi_k^1} =\ket{\psi_k^{(\down \up)}} =  \frac{1}{\sqrt{L}}  \sum_{x=1}^{L-1} e^{-ikx} \, \ket{\down} ...  \ket{\down}_x ~[\ket{\down}_{x+1} \ket{\up}_{x+2}]~ \ket{ \up}_{x+3} ... \ket{\up} \label{d3}\\
&\ket{\psi_k^2}=\ket{\psi_k^{( \up\down)}}=\frac{1}{\sqrt{L}}  \sum_{x=1}^{L-1} e^{-ikx} \ket{\down} ...  \ket{\down}_x ~[\ket{\up}_{x+1} \ket{\down}_{x+2}]~ \ket{ \up}_{x+3} ... \ket{\up} \label{d4}
\end{align}
are not orthonormal. 
To find $H^{\rm eff}$ in an orthonormal basis, we use the following two steps:
\begin{enumerate}
\item We find the four  matrix elements $\braket{\psi_k^i|P_4|\psi_k^j}$ directly in the thermodynamic limit. This is easy to do due to the fact that $\bra{ss}(P_4)_{ij} = (P_4)_{ij}\ket{ss} = 0$, where $(P_4)_{ij}$ is the nearest neighbor term in $P_4$.
%
\item We  construct the gram matrix with matrix elements $\braket{\psi_k^i|\psi_k^j}$ semi-analytically for a large value of $L$ and use it to construct an orthonormal basis for the subspace spanned by \eqref{d3}-\eqref{d4}. 
In practice, $L=50$ is sufficient for  the convergence of the $O(1)$ coefficients of $\ket{\psi_k^1}, \ket{\psi_k^2}$ in the orthonormal vectors. 
We then transform $\braket{\psi_k^i|P_4|\psi_k^j}$ found in point 1 to the orthonormal basis to find $H_{\rm eff}$, which is simply a $2 \times 2$ matrix in this case.
\end{enumerate}
We then plot the lowest eigenvalue of $H_{\rm eff}$ obtained for each $k$ to obtain the $\Delta=1$ curves in Fig. \ref{fig:variational_gue_2renyi}. 
The generalization of this procedure to higher $\Delta$ for $P_4$ in the GUE model is straightforward.
For the remaining cases of $P_6$ in the Brownian GUE model and of $P_4$ in the Brownian mixed field Ising model, again we use the same two steps 
Let us describe the $\Delta =1$ case more explicitly in each case. 
For the third Renyi entropy in the GUE model, we need to allow for arbitrary permutations in $\sS_3$.
A convenient way to form an orthonormal basis for this case is as follows.
Let us label the one-site states associated with permutations other than $\ket{e}$ and $\ket{\eta}$ as $\ket{\sigma_a}$, $a$=1, ..., 4.
First, for each $\ket{\sigma_a}$, subtract the components of the state along both $\ket{\eta}$ and $\ket{e}$ to get a state $\ket{\tilde \sigma_a}$.
Then find the orthonormal states $\ket{\tau_a}, a=1, ..., 4$, in the subspace spanned by $\ket{\tilde\sigma_a}$. Then for the $\Delta=1$ case, the relevant set of five states is  
\begin{align} 
&\ket{\psi^{(\eta)}_k}=  \sN(L) \frac{1}{\sqrt{L}}\frac{1}{q^{3L/2}} \sum_{x=1}^L e^{ikx} \ket{\eta}... \ket{\eta}_x \ket{\eta}_{x+1} \ket{e}_{x+2} ... \ket{e} \label{psik_eta}\\
&
\ket{\psi^{(a)}_k}= \frac{1}{\sqrt{L}}\frac{1}{q^{3L/2}}\sum_{x=1}^L e^{ikx} \ket{\eta}... \ket{\eta}_x \ket{\tau_a}_{x+1} \ket{e}_{x+2} ... \ket{e}, \quad a=1, ..., 4 \label{psik_a}
\end{align}
Note that we have not included the state with $\ket{e}$ at $x$, as this is redundant with the choice \eqref{psik_eta}.  The states \eqref{psik_a} are already orthonormal among themselves, and orthogonal to $\ket{\psi_k^{(\eta)}}$.
The normalization factor $\sN(L)$ in $\ket{\psi_k^{(\eta)}}$ has to be computed numerically, and converges to some $O(1)$ constant for large enough $L$.
We can then find the matrix elements of $P_4$ directly in the thermodynamic limit. 

For $P_4$ of \eqref{p2ndef} for the mixed field Ising model couplings in \eqref{halpha_ising}, we need to consider a 16-dimensional Hilbert space at each site.
We can construct an arbitrary one-site orthonormal basis of states $\ket{a}$, $a =1, ..., 14$, which are orthogonal to the permutation subspace spanned by $\ket{e}, \ket{\eta}$.
Then for $\Delta=1$, we can take the orthonormal basis to be 
\begin{align} 
&\ket{\psi^{(\down)}_k}=  \sM(L) \frac{1}{\sqrt{L}}\frac{1}{q^L} \sum_{x=1}^L e^{ikx} \ket{\eta}... \ket{\eta}_x \ket{\eta}_{x+1} \ket{e}_{x+2} ... \ket{e} \label{psik_eta_ising}\\
&
\ket{\psi^{(a)}_k}= \frac{1}{\sqrt{L}} \frac{1}{q^L}\sum_{x=1}^L e^{ikx} \ket{\eta}... \ket{\eta}_x \ket{a}_{x+1} \ket{e}_{x+2} ... \ket{e}, \quad a=1, ..., 14 \label{psik_a_ising}
\end{align}
where again the normalization factor $\sM(L)$ can be computed numerically for finite $L$ until it converges to some $O(1)$ number,  and the only remaining microscopic  inputs needed to compute the matrix elements of $P_4$ in this basis directly in the thermodynamic limit are the two-site matrix elements such as $\bra{a \up} (P_4)_{ij}\ket{b\up}$, $\bra{\down a} (P_4)_{ij} \ket{b\up}$, and so on. 
\section{Arguments using diagrammatic approach in the Brownian GUE model} \label{app:competition}

\subsection{Diagrams in interaction picture}
\label{sec:int_picture}

In Sec. \ref{sec:finite_q} of the main text, we used the structure of the eigenstates for any finite $q$ to argue that the domain wall propagator, defined as 
\be 
G(x,y, t) = \braket{D_x|e^{-At}|\bar D_y}  \label{fullprop}
\ee
is given by 
\be 
G(x, y, t)= \int_{-\pi}^{\pi} \frac{dk}{2\pi} \, e^{-(E(k) + i k v)t}  = e^{-s_{\rm eq} \sE(v) t}, \quad v=\frac{x-y}{t}\label{fullprop_e}
\ee 
where $E(k)$ is the exact dispersion relation at finite $q$, which we obtained numerically.  In this appendix, we will use perturbation theory to understand the structure of the diagrams coming from $A_1$ in \eqref{aleft_def_mt} that correct the large $q$ propagator from $A_0$, 
\be 
G_0(x, y, t) = \braket{D_x|e^{-A_0t}|\bar D_y} =\int_{-\pi}^{\pi} \frac{dk}{2\pi} \, e^{-(E_0(k) + i k v)t}   = e^{- s_{\rm eq}\sE_0(v) t}.  \label{bareprop} 
\ee
The resummation of all such diagrams should in principle lead to the exact result \eqref{fullprop_e}.
Note that we are now using ``0" subscripts to denote $G$, $E$, and $\sE$ for $A_0$ which we found in Sec. \ref{sec:large_q}, to distinguish them from the exact quantities appearing in \eqref{fullprop}.
Quantitatively, the perturbation theory approach is much weaker than the numerical techniques discussed in the main text.
In particular, the corrections that we obtain order-by-order in $1/q$ do not lead to membrane tensions that satisfy the constraints \eqref{const} well at finite $q$.
Moreover, this discussion only applies to $q\geq3$, as the zeroth order Hamiltonian $A_0$  is gapless for $q=2$.
However, one  advantage of discussing the diagrams is that they allow us to better explain the connection of our results to the diagrammatic approach  used for circuit models in~\cite{zhou_nahum}.
The  diagrammatic approach in the interaction picture may also be more directly generalizable to higher dimensions than than the  dispersion relation approach discussed in the main text.

\begin{figure*}[!h]
    \centering
\includegraphics[width=0.7\textwidth]{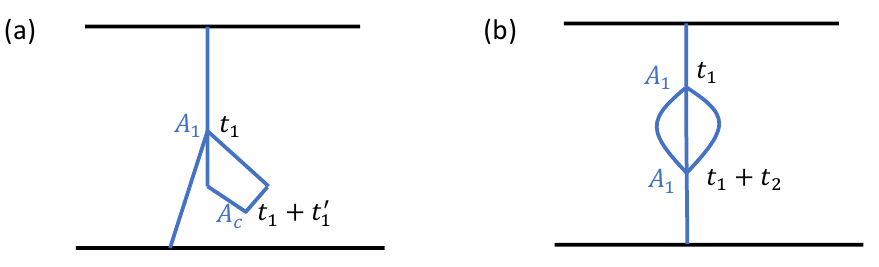}
    \caption{Interaction picture diagrams coming from \eqref{549}.}
    \label{fig:int_diagrams}
\end{figure*}

Let us use the following expansion of the Euclidean time-evolution operator: 
\be 
e^{-P_4t} = e^{-A_0 t} - \int_0^t dt_1  e^{-A_0t_1} A_1 e^{-A_0(t-t_1)} + \int_0^t dt_1 \int_0^{t-t_1} dt_2 \, e^{-A_0t_1} A_1 e^{-A_0t_2} A_1 e^{-A_0(t-t_1-t_2)} +  ... \label{549}
\ee
Putting this expansion into $\braket{D_x|e^{-P_4t}|\bar D_y}$, 
the first term  gives the bare propagator $G_0(x,y,t)$ from the previous section.
Recall that the action of $A_1$ sends a state with one domain wall to a state with three domain walls. 
For the second term in \eqref{549} to contribute to the domain wall propagator, two of the three domain walls created by the action of $A_1$ must annihilate at an intermediate time $t_1+t_1'$ between $t_1$ and $t$ due to the action of $A_c$ defined in \eqref{aj}, see Fig. \ref{fig:int_diagrams}~(a).
The resulting contribution to the domain wall propagator is 
\be 
G(x, y, t) \supset \int_0^t dt_1 \int_0^{t-t_1} dt_1' \braket{D_x| e^{-A_ft_1} A_1 e^{-A_f t_1'} A_c e^{-A_f(t-t_1-t_1')}| \bar D_y} \, . \label{int_insert}
\ee

\begin{figure}[!h]
    \centering
    \includegraphics[width=0.3\textwidth]{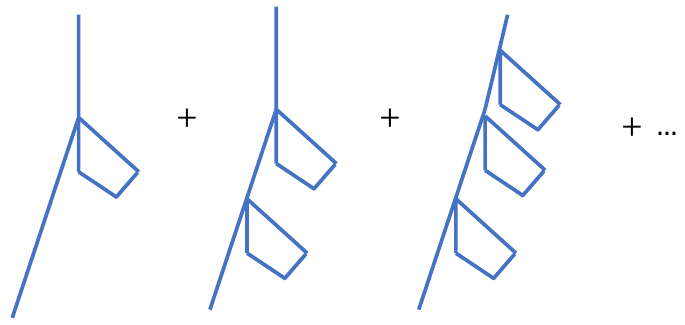}
    \caption{The set of diagrams which leads to exponentiation of corrections from Fig. \ref{fig:int_diagrams} (a).}
    \label{fig:energy_resum}
\end{figure}

By inserting resolutions of the identity in the energy eigenbasis of $A_f$ before and after $A_1$ and $A_c$, and doing the integrals over $t_1'$ and $t_1$,
\begin{align}
& G(x, y, t) =\int dk e^{-ik(x-y) - E_0(k) t}  \nn
&\quad \quad \quad + \sum_{k, k_1, k_2, k_3} e^{-i k(x-y) - E_0(k) t} \frac{\braket{\psi_k| A_1| \bar \psi_{k_1, k_2, k_3}}  \braket{\psi_{k_1, k_2, k_3} | A_c | \bar\psi_{k}} }{E_0(k) - E_0(k_1, k_2, k_3)} \le(\frac{1 - e^{-(E_0(k_1, k_2, k_3)- E_0(k)) t}}{E_0(k_1, k_2, k_3)-E_0(k)}  - t\ri) +\ ...  \nn 
&\quad \quad \quad \quad\approx\int dk e^{-ik(x-y) - E_0(k) t}  ( 1 - F_1(k) t + ... ) , \label{g_int} \\ 
&F_1(k) =  \sum_{k_1, k_2, k_3} \frac{\braket{\psi_k| A_1| \bar \psi_{k_1, k_2, k_3}}  \braket{\psi_{k_1, k_2, k_3} | A_c | \bar \psi_{k}}}{E_0(k)- E_0(k_1, k_2, k_3)}  \label{F_1} 
\end{align}
where $\bra{\psi_{k_1, k_2, k_3}}$ was defined in \eqref{psin_def_mt}, and $\ket{\bar \psi_{k_1, k_2, k_3}}$ is defined such that 
\be 
\braket{\psi_{k_1, k_2, k_3}| \bar \psi_{k_1, k_2, k_3}} =1, \quad \braket{\psi_{p_1, p_2 ...  p_n}| \bar \psi_{k_1, k_2, k_3}} =0\quad \text{for $\{ p_1, ..., p_n \} \neq \{k_1, k_2, k_3\}$}
\ee
In the final expression \eqref{g_int}, we have kept only the leading linear-in-$t$ term in the large $t$ limit, ignoring the $O(1)$ and exponentially suppressed terms.   
This leading contribution comes 
 from the regime in the diagram Fig.~\ref{fig:int_diagrams}~(a) where $t_1'$ is $O(1)$. (The proportionality to $t$ in the second term of \eqref{g_int} comes from the fact that the small bubble of $O(1)$ size  $t_1'$ can be placed at any value of $t_1$ from 0 to $t$.) 
By summing over the leading contributions at large $t$ from the series of diagrams in Fig. \ref{fig:energy_resum}, the correction in \eqref{g_int} exponentiates to 
\be 
G(x, y, t) = \int dk e^{-ik(x-y) -(E_0(k) + F_1(k))t}
\ee

Similarly, one can check that the leading correction at large $t$ from the diagram in Fig. \ref{fig:int_diagrams}~(b), which comes from the third term in \eqref{549}, is  
\be 
G(x, y, t) \supset - \int dk e^{-ik(x-y)} e^{-E_0(k)t} (F_2(k) \,  t) , \quad  F_2(k) =  \sum_{k_1, k_2, k_3} \frac{\braket{\psi_k| A_1| \bar \psi_{k_1, k_2, k_3}}  \braket{\psi_{k_1, k_2, k_3} | A_1 | \bar \psi_{k}}}{E_0(k)-E_0(k_1, k_2, k_3)} 
\ee
and exponentiates to give a correction $F_2(k)$ in the energy.
In all, these ``renormalize" the bare dispersion $E_0(k)$ to $E(k)$, where
\be 
E(k) = E_0(k) + F_1(k) + F_2(k) + ...
\ee
By extending the same reasoning to higher order terms in the expansion \eqref{549}, we can see that the exact dispersion relation at finite $q$ is a resummation of the full set of diagrams in the interaction picture where we have departures from the one domain wall subspace into the multiple domain wall subspace for an $O(1)$ amount of time at intermediate times.
These are precisely analogous to the diagrams considered for circuit and Floquet models in \cite{zhou_nahum}. Note that the corrections $F_1(k)$ and $F_2(k)$ to the dispersion relation $E_0(k)$ found above  can also be independently derived by using a version of Schrodinger picture perturbation theory adapted to the non-Hermitian starting point $A_0$. 

In the next subsection, we will use a thickened line to denote the exact propagator of \eqref{fullprop_e} at finite $q$, which should now be understood as a sum over the following set of diagrams for $O(1)$ departures from the one domain wall subspace:
\begin{figure}[!h]
    \centering
    \includegraphics[width=0.5\linewidth]{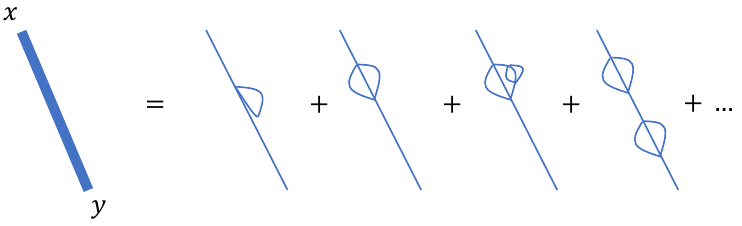}
    
\end{figure}

\subsection{Argument that configurations where the domain wall splits are subleading for any initial state}
\label{sec:competition}
Under the action of the full Hamiltonian $P_4$ in \eqref{aleft_def_mt}, a single domain wall state $\bra{D_x}$ can evolve to a state with an arbitrary number of domain walls, so that \eqref{entlarge} for the large $q$ limit should be replaced with the following expression for the exact evolution at finite $q$: 
\begin{align}
e^{-S_2(x,t)} =& 
 \sum_{y}\braket{D_x|e^{-P_4t}|\bar D_y}\braket{D_y|\rho_0, e}  +  \sum_{y_1,y_2,y_3}\braket{D_x|e^{-P_4t}|\bar D_{y_1, y_2, y_3}}\braket{D_{y_1,y_2,y_3}|\rho_0, e} \nn 
 &+ \sum_{y_1,...,y_5}\braket{D_x|e^{-P_4 t}|\bar D_{y_1, ..., y_5}}\braket{D_{y_1,...y_5}|\rho_0, e} + ... \label{pos_exp}
\end{align}
where we have  inserted a resolution of the identity in the subspace of odd numbers of domain walls to the right of $e^{-At}$.
Note that for instance $\braket{D_{y_1, y_2, y_3}|\rho_0,e} = e^{-S_2([-\infty, y_1]\cup [y_2, y_3], t=0)}$. 

\begin{figure}[t]
    \centering
    \includegraphics[width=0.8\textwidth]{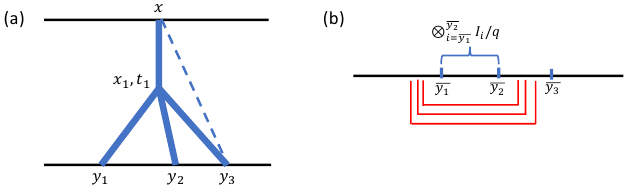}
    \caption{(a) shows a configuration where the domain wall splits into three, and (b) shows an initial state for which (a) with $y_1, y_2, y_3 = \bar y_1, \bar y_2, \bar y_3$ minimizes the initial entropy contribution compared to all other configurations where the domain wall splits into three.}
    \label{fig:line_choices}
\end{figure}

Using the interaction picture diagrams of the previous section, 
$\braket{D_x|e^{-P_4t}|\bar D_{y_1, y_2, y_3}}$ can be expressed as follows in terms of the exact propagator $G(x-y, t')$, again up to an overall prefactor which does not have an exponential dependence on $t$:
\begin{align}
\braket{D_x|e^{-P_4t}|\bar D_{y_1, y_2, y_3}}= \int_0^t dt_1 \sum_{x_1} G(x-x_1, t_1) G(x_1-y_1, t-t_1) G(x_1-y_2, t-t_1) G(x_1-y_3, t-t_1) \, . \label{3split} 
\end{align}
This gives the following contribution to \eqref{pos_exp}, which can be represented by a sum over diagrams shown in Fig. \ref{fig:line_choices}~(a)~(for now, let us ignore the dashed line in the figure):
\be 
e^{-S_2(x, t)} \supset -\frac{1}{q^2}\sum_{y_1, y_2, y_3} \sum_{t_1, x_1} e^{- t_1s_{\rm eq}\sE\le(\frac{x_1-x}{t_1}\ri) - (t-t_1) s_{\rm eq} \le[ \sE\le(\frac{y_1-x_1}{t-t_1}\ri) + \sE\le(\frac{y_2-x_1}{t-t_1}\ri) + \sE\le(\frac{y_3-x_1}{t-t_1}\ri)\ri] } e^{-S_2([-\infty, y_1]\cup[y_2, y_3],t=0)}  \label{split_domain} 
\ee

If we have a pure product initial state $\rho_0$ such that the initial value of $S_2$ for any region is zero, then it is immediately clear that any such  contributions where the domain wall splits are exponentially suppressed in time compared to the first term in \eqref{pos_exp}. Even taking into account the contribution from an arbitrary  initial state, we can use the convexity of the membrane tensions obtained   in Fig. \ref{fig:mem_tensions} to show that the first term of \eqref{pos_exp} dominates, as follows.

Let us fix some arbitrary set of positions $\bar y_1, \bar y_2, \bar y_3$. 
Consider an initial state that is maximally mixed in the region $[\bar y_1, \bar y_2]$, and has maximal entanglement between the degrees of freedom in $[\bar y_2, \bar y_3]$ and the degrees of freedom to the left of $\bar y_1$. See Fig. \ref{fig:line_choices} (b). For this state, the initial entropy contribution $S_2([-\infty, y_1]\cup[y_2, y_3],t=0)$ in \eqref{split_domain} is minimized for $y_1, y_2, y_3 = \bar y_1, \bar y_2, \bar y_3$.

Let us now consider the total potential contribution to the $S_2(x, t)$ from the diagram in Fig. \ref{fig:line_choices}~(a) with $y_1, y_2, y_3 = \bar y_1 \bar y_2, \bar y_3$ for this initial state: 
\be 
S_{\rm solid} = s_{\rm eq} \sE\le(\frac{x_1-x}{t_1}\ri) t_1 +  s_{\rm eq} \le[\sE\le(\frac{y_1-x_1}{t-t_1}\ri) + \sE\le(\frac{y_2-x_1}{t-t_1}\ri)+ \sE\le(\frac{y_3-x_1}{t-t_1}\ri)  \ri] (t-t_1)
\ee
Let us now compare this contribution to the single domain wall configuration shown with the dashed line [coming from the first term of \eqref{pos_exp}], which contributes 
\be 
S_{\rm dashed} = s_{\rm eq} \,  \sE\le(\frac{y_3-x}{t}\ri) t + s_{\rm eq} (y_2-y_1)
\ee
Now using the convexity of $\sE(v)$, 
\be 
\sE\le(\frac{x_1-x}{t_1}\ri) t_1  + \sE\le(\frac{y_3-x_1}{t-t_1}\ri) (t-t_1) \geq \sE\le(\frac{y_3-x}{t}\ri)  t 
\ee
and by similar steps to those leading  to \eqref{46}, by making use of the constraints \eqref{const}, 
\be 
\sE\le(\frac{y_1-x_1}{t-t_1}\ri)(t-t_1) + \sE\le(\frac{y_2-x_1}{t-t_1}\ri)(t-t_1) \geq y_2 - y_1 \, . 
\ee
We therefore find that $S_{\rm dashed}$
always wins in the minimization problem in the scaling limit. It is clear that if $S_{\rm dashed}$ wins even for this choice of initial state, it will also be the dominant contribution for any other choice of initial state.

By a simple extension of this argument, all other diagrams coming from the third and remaining terms in \eqref{pos_exp} are subleading compared to the first term of \eqref{pos_exp} in the scaling limit.

\section{Operator growth interpretation of domain wall splitting}
\label{app:op_growth}
Consider a basis of operators $O_{\alpha}$ for a single site, $\alpha = 1, ..., q^2$, such that $O_1 = I$, and satisfying $\frac{1}{q}\Tr[O_{\alpha}^{\dagger} O_{\beta}] = \delta_{\alpha\beta}$.
By taking tensor products of these operators, we can also construct a basis $\ket{O_a}, a=1, ..., q^{2L}$ of operators for the full system, satisfying 
$\frac{1}{q^{L}}\Tr[O_{a}^{\dagger} O_{b}] = \delta_{ab} \, .$  

Let us introduce the two-copy states $\ket{O_{\alpha}} $  associated with these operators, such that $\braket{ij|O_{\alpha}}= (O_{\alpha})_{ij}$.
In terms of these states, the four-copy spin states defined in \eqref{spin_def}-\eqref{bar_spin_def} can be shown to be written as 
\begin{align}
&\ket{\up}  = \frac{1}{q}\ket{I}\ket{I}, \quad \ket{\down} = \frac{1}{q^2} \sum_{\alpha=1}^{q^2} \ket{O_{\alpha}}\ket{O_{\alpha}^{\dagger}}, \label{updown}\\
& \ket{\bar \up} = \frac{1}{q}\le(\ket{I}\ket{I} - \frac{1}{q^2-1} 
\sum_{\alpha=2}^{q^2} \ket{O_{\alpha}}\ket{O_{\alpha}^{\dagger}} \ri) , \;\; \ket{\bar \down} = \frac{1}{q^2-1} \sum_{\alpha=2}^{q^2} \ket{O_{\alpha}}\ket{O_{\alpha}^{\dagger}} \label{barupdown}
\end{align}

It is natural to introduce the notion of a probability that some initial operator $A$ in the full system evolves to some final operator $B$ under the unitary evolution $U(t)$~\cite{roberts,  syk, nahum_randomcircuit, rak_randomcircuit}:
\be 
P(A \to B, t) = \le|\frac{1}{q^L} \Tr[B^{\dagger}U(t) A U(t)^{\dagger} ]\ri|^2  = \frac{1}{q^{2L}} \bra{B}\bra{B^{\dagger}}(U(t) \otimes U(t)^{\ast})^{\otimes2} \ket{A}\ket{A^{\dagger}}
\ee
Using \eqref{updown}, for any initial operator $A$ and some position $x$ in the system, we define $P(A, x, t)$ as 
\be 
P(A, x, t) \equiv \sum_{O_{a}\text{ ending to the left of }x} P(A \to O_a, t) = \frac{1}{q^{\frac{L}{2}-x}} \, \bra{D_x} (U(t) \otimes U(t)^{\ast})^{\otimes2} \ket{A}\ket{A^{\dagger}} 
\ee
where ``$O_a$ ending to the left of $x$'' means that the $O_a$ is equal to the identity for all sites to the right of $x$.
This is the total probability that the operator $A$ evolves to an operator that has support only on the left of $x$.
Now consider two different kinds of initial operators $A$ from the basis $O_a$ ending at some point $y_2$ (meaning that $A$ has a non-trivial operator $O_{\alpha}\neq I$ at $y_2$, but is equal to the identity everywhere to the right of $y_2$). One type of operator has non-trivial operators  at all sites to the left of $y_2$, while the other is equal to the identity at some $y_1<y_2$ and nontrivial at all other $y<y_2$. Consider the quantity   $P_1-P_2$, where 
\begin{equation}
P_1 = \braket{P(A, x, t)}_{\text{$A$ ending at $y_2$,  identity  at $y_1$}}, \;\;\;P_2 = \braket{P(A, x, t)}_{\text{$A$ ending at $y_2$,   non-trivial at $y_1$}} 
\end{equation}
where $\braket{}$ denotes the average over all basis operators $O_a$ with the property described in the subscript.
Using \eqref{barupdown}, 
\be 
P_1 - P_2 = q^{x-y_2-1} \bra{D_x}(U(t) \otimes U(t)^{\ast})^{\otimes2} \ket{\bar \down ...\bar \down_{y_1-1}\bar\up_{y_1}\bar \down_{y_1+1}... \bar \down_{y_2}\up_{y_2+1...} \up} 
\ee
In random unitary circuits, since a single domain wall does not split, we have $P_1=P_2$, indicating that the evolution of the endpoint of an operator is independent of its internal structure, which is consistent with a maximally random time-evolution. In the Brownian GUE model, $P_1-P_2$ is highly suppressed at late times, but non-zero, indicating that there is some slight sensitivity of the operator growth to the internal structure. However, such effects do not modify qualitative features of the dynamics such as the validity of the membrane picture for the second Renyi entropy.  
%
\end{document}